# Integrated photonic neuromorphic computing: device, architecture, chip, algorithm

Shuiying Xiang[a)], Chengyang Yu, Yizhi Wang, Xintao Zeng, Yuna Zhang, Dianzhuang Zheng, Xinran Niu, Haowen Zhao, Hanxu Zhou, Yanan Han, Xingxing Guo, Yahui Zhang, Yue Hao

AFFILIATIONS
State Key Laboratory of Integrated Service Networks, Xidian University, Xian 710071, China
[a)] Authors to whom correspondence should be addressed: syxiang@xidian.edu.cn

## ABSTRACT

Artificial intelligence (AI) has experienced explosive growth in recent years. Especially, the large models have been widely applied in various fields, including natural language processing, image generation, and complex decision-making systems, revolutionizing technological paradigms across multiple industries. Nevertheless, the substantial data processing demands during model training and inference result in the computing power bottleneck. Traditional electronic chips based on the von Neumann architecture struggle to meet the growing demands for computing power and power efficiency amid the continuous development of AI. Photonic neuromorphic computing, an emerging solution in the post-Moore era, exhibits significant development potential. Leveraging the high-speed and large-bandwidth characteristics of photons in signal transmission, as well as the low-power consumption advantages of optical devices, photonic integrated computing chips have the potential to overcome the memory wall and power wall issues of electronic chips. In recent years, remarkable advancements have been made in photonic neuromorphic computing. This article presents a systematic review of the latest research achievements. It focuses on fundamental principles and novel neuromorphic photonic devices, such as photonic neurons and photonic synapses. Additionally, it comprehensively summarizes the network architectures and photonic integrated neuromorphic chips, as well as the optimization algorithms of photonic neural networks. In addition, combining with the current status and challenges of this field, this article conducts an in-depth discussion on the future development trends of photonic neuromorphic computing in the directions of device integration, algorithm collaborative optimization, and application scenario expansion, providing a reference for subsequent research in the field of photonic neuromorphic computing.

## I. INTRODUCTION

In the current era of rapid information technology development, the exponential growth of data has increasingly exposed the limitations of traditional computing architectures when handling large-scale data processing. Under the von Neumann architecture, the physical separation between the processor and memory severely restricts the optimization of computing efficiency and energy consumption. With the slowdown of Moore's Law, the conventional approach of enhancing computing power by reducing hardware feature sizes has become increasingly challenging. Traditional computing technologies face two major challenges of physical limits and rising energy consumption.

Neuromorphic computing, as a highly promising novel computing paradigm in the post-Moore era, draws on the information processing mechanism of the biological brain. Taking neurons and synapses as basic units, it simulates the structure and information processing process of biological neural networks, achieving the integration of efficient computing hardware and algorithms, and is expected to break through the bottleneck of traditional computing architectures. Among various neuromorphic computing approaches, photonic neuromorphic computing stands out due to its unique advantages of photons, such as ultra-high speed, large bandwidth, and multi-dimensionality, emerging as a crucial solution to the current computing dilemma.[1] Leveraging photonic signals for communication and processing instead of electronic signals, photonic neuromorphic computing effectively reduces energy loss, boosts computing speed and bandwidth, and provides fast, parallel, and adaptive processing capabilities for artificial intelligence (AI) and machine learning applications.[2]

This review systematically explores the field of photonic neuromorphic computing (Fig. 1). It includes several key aspects, including photonic linear computing devices (microring resonator (MRR), Mach-Zehnder Interferometer (MZI), phase-change material (PCM), and semiconductor optical amplifier (SOA)), photonic nonlinear computing devices (photonic nonlinear activation and photonic spiking neurons), photonic neural network (PNN) architectures (multi-layer perception (MLP), convolutional neural network (CNN), spiking neural network (SNN), reservoir computing (RC), reinforcement learning (RL), etc), and PNN training algorithms. A comprehensive analysis of their underlying principles, recent advancements, and existing challenges is presented, aiming to offer valuable insights for the future development of photonic neuromorphic computing.

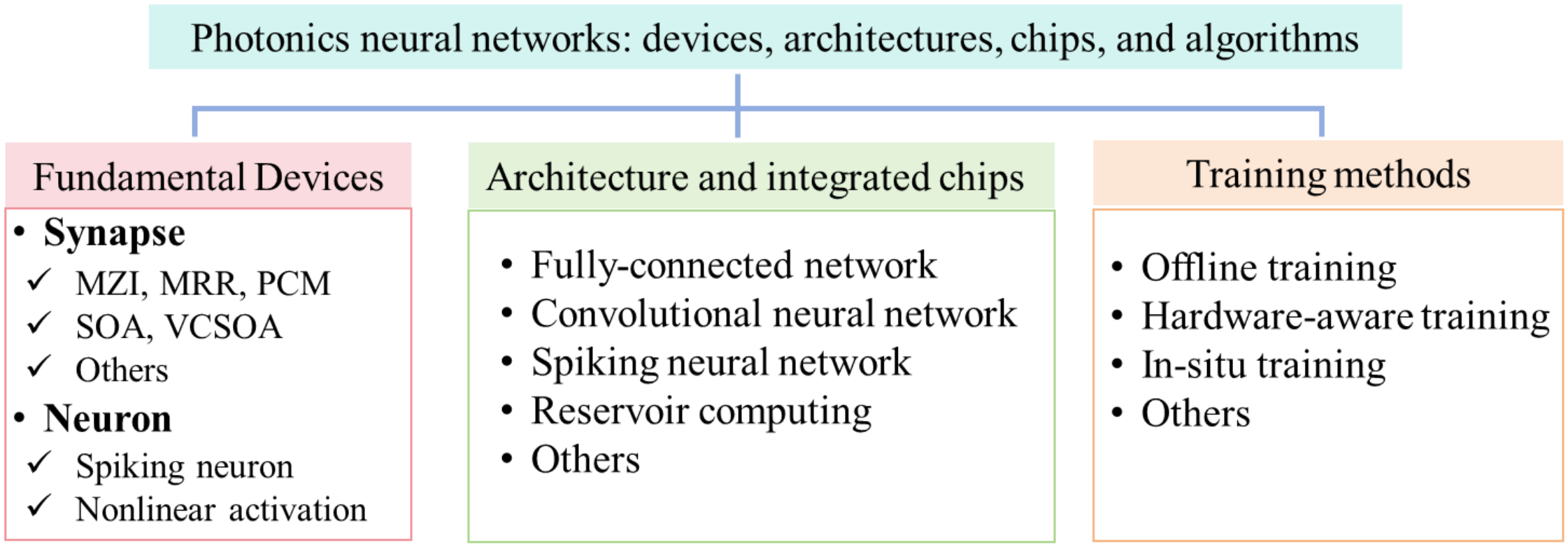

**FIG. 1.** Main contents of photonics neuromorphic computing.

## II. FUNDAMENTAL DEVICES OF PHOTONICS NEUROMORPHIC COMPUTING

In biological neural systems, information is primarily encoded and transmitted through the interplay between the soma and synapses. In artificial neural networks (ANNs), synapses perform linear weighted operations, while the soma is modeled by a nonlinear function (Fig. 2), making both components fundamental to network functionality. Similarly, in PNNs, linear photonic synapses and nonlinear photonic neurons are key to overall system performance. This section presents an overview of linear photonic synapses and nonlinear photonic neurons.

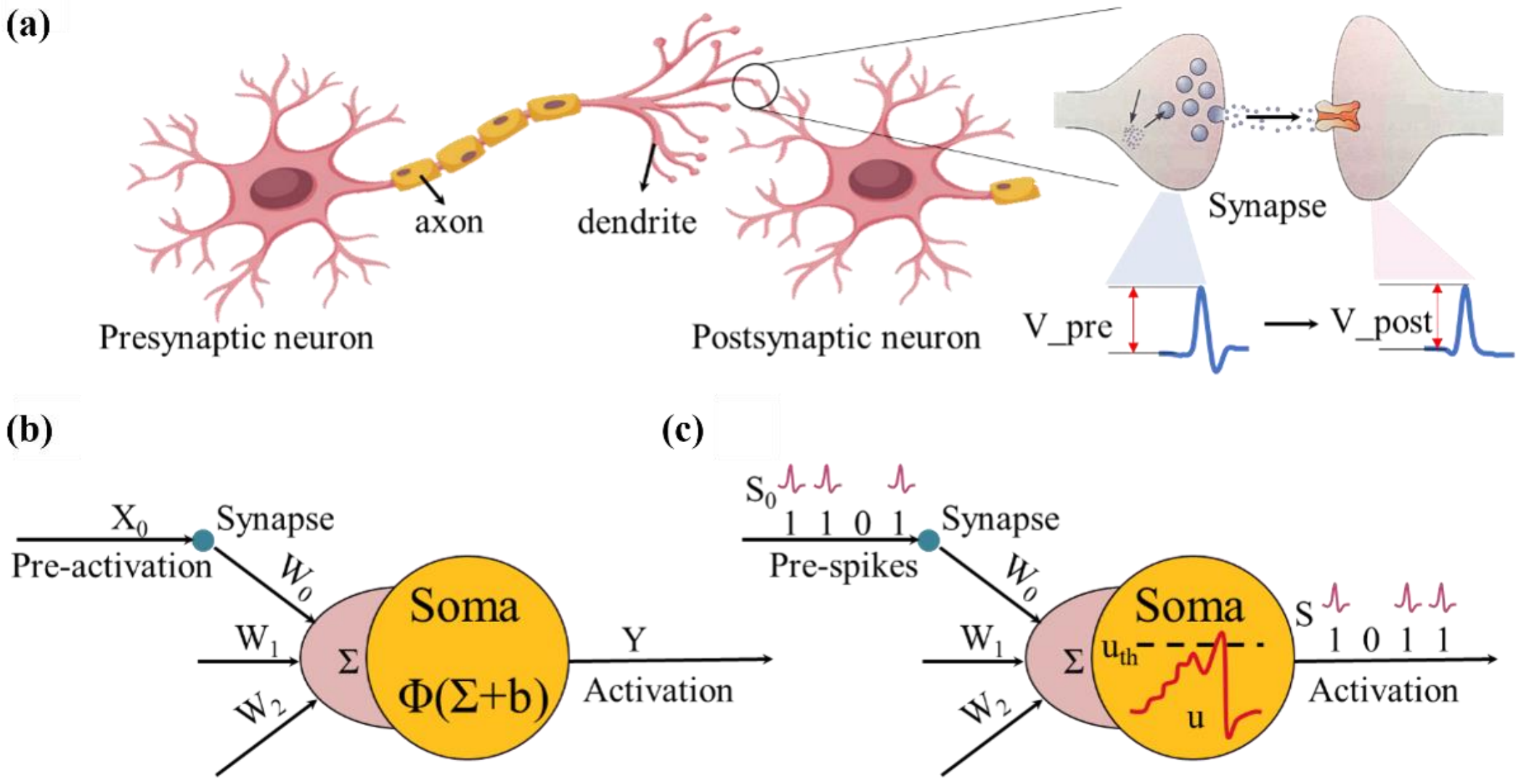

**FIG. 2.** (a) Signal transmission pathway of a biological neuron. (b) Mathematical model of a neuron. (c) Mathematical model of a spiking neuron.

### A. LINEAR PHOTONIC SYNAPSE

In photonic neuromorphic computing systems, photonic synapses serve as a core component, responsible for emulating the weight storage and computational functions of biological synapses. They undertake linear computational tasks such as matrix operations and weighted summation. Therefore, the development of efficient photonic synaptic devices is crucial for advancing PNNs. Currently, various photonic synapse schemes have been proposed, which can be broadly categorized into three types based on their operating principles. The first type includes photonic synapses based on optical interference/resonance or material property, such as MZIs, MRRs, and PCMs. The second type consists of photonic synapses based on optical gain modulation, mainly including SOAs and vertical-cavity SOAs (VCSOAs). The third category encompasses other types of photonic synapses. In the following, recent achievements corresponding to the three types of photonic synapses will be discussed respectively.

#### 1. Photonic synapses based on optical interference/resonance or material property

Photonic synapses, based on optical interference/resonance and material property, harness light coherence and the tunability of dielectric materials to enable synaptic weight modulation and storage. This type of synapse mainly includes MZI, MRR, and PCM. By adjusting parameters such as phase, transmittance, or refractive index, these synapses emulate the update and retention of synaptic weights. Table I summarizes representative research on photonic synapses based on MZIs, MRRs and PCMs.

**TABLE I.** Photonic synapses based on optical interference/resonance and material property.

| Year & Author | Technology Type | Implementation Method | Key Contribution |
|---|---|---|---|
| 2016<br>A. Ribeiro[4] | MZI | 4×4 universal linear circuit | Extinction ratio: > 45 dB<br>Crosstalk: < -20 dB |
| 2017<br>A. Annoni[6] | MZI | Self-configuring mode unscrambler | Crosstalk: < -20 dB |
| 2017<br>D. Pérez[7] | MZI | Waveguide mesh photonic processor | 20+ functionalities<br>FSR: 9.2-18.4 GHz |
| 2023<br>L. Pei[11] | MZI | Photonic neural processing unit | Accuracy: > 98%<br>Computing power: > 100 TOPS |
| 2023<br>B. Wu[12] | MZI | Simplified MZI mesh | Real-valued matrix with $N^2$ phase shifters |
| 2024<br>J. Lin[15] | MZI | QR decomposition-based optical neural network | Reduced MZI count, enhanced robustness |
| 2012<br>L. Yang[16] | MRR | CMOS-compatible MRR | High-speed matrix operations:<br>$8\times10^7$ MAC/s |
| 2016<br>A. N. Tait[21] | MRR | Microring weight banks | Quantitative description of independent weighting |
| 2017<br>A. N. Tait[22] | MRR | Broadcast-and-weight protocol | Dynamic weight accuracy:<br>4.1 + 1(sign) bits |
| 2020<br>C. Huang[23] | MRR | Microring weight bank control | Accuracy and precision: 7 bits |
| 2022<br>J. Cheng[26] | MRR | Microring array | Large-scale matrix computation (16×16) |
| 2024<br>E. C. Blow[29] | MRR | Broadcast and weight silicon PNN | 3 dB weighting bandwidth: 4.7 GHz |
| 2017<br>Z. Cheng[31] | PCM | PCM combined with integrated silicon nitride waveguides | Fully integrated all-photonic synapse |
| 2019<br>J. Feldmann[33] | PCM+MRR | All-optical spiking neurosynaptic network | Supervised/unsupervised learning |
| 2021<br>J. Feldmann[34] | PCM + microcomb | Photonic tensor core | $10^{12}$ MAC operations /<br>tera-MACs per second |
| 2023<br>W. Zhou[39] | PCM | Photonic-electronic dot-product engine | Weight encoding: 4 bit<br>Modulation depth: 1.7nJ/dB |
| 2023<br>Y. Zhang[40] | PCM+MRR | Photonic SNN | MRR+PCM based spiking neuron |
| 2024<br>A. H. A. Nohoji[42] | PCM | Photonic crystal waveguide intersection | Crosstalk: < -60 dB<br>Q-factor: 900 |

In photonic neuromorphic computing, MZI modulates optical intensity through interference, enabling key operations such as weight modulation and matrix computation. Furthermore, MZI-based architectures can be cascaded or arranged in mesh topologies to enable programmable linear operations and enhance computational capabilities. In 1994, M. Reck et al. proposed a recursive algorithm that decomposes any arbitrary N×N unitary matrix into a series of two-dimensional beam splitter transformations, enabling the implementation of arbitrary discrete unitary operators in optical experiments.[3] This algorithm utilizes MZI as the fundamental building block and employs a triangular mesh configuration to perform linear computations in multidimensional spaces. In 2016, A. Ribeiro et al. designed a 4×4 universal linear photonic circuit based on an MZI network and thermo-optic phase shifters. The system enabled arbitrary linear transformations via electronic control and software feedback.[4] By applying adaptive training algorithms to tune the MZI parameters, the circuit demonstrated beam coupling and switching matrix functions, providing experimental verification for dynamic reconfiguration of photonic synapses. Besides, W. R. Clements et al. introduced an optimized universal multiport interferometer architecture. Compared to the traditional Reck structure, this design halved the optical depth and significantly improved loss tolerance, offering an efficient design strategy for large-scale integrated photonic synaptic networks.[5] In 2017, A. Annoni et al. demonstrated self-configuration of an MZI mesh on a silicon photonic chip, as shown in Fig. 3(a). By integrating contactless photonic probes, they achieved real-time demultiplexing and information recovery of multimode optical fields.[6] D. Pérez et al. proposed a programmable photonic processor core based on a hexagonal waveguide mesh, as shown in Fig. 3(b), capable of performing over 20 distinct linear operations using MZI units.[7] In 2019, S. Pai et al. proposed a general-purpose photonic processing architecture based on MZI units arranged in a rectangular mesh, enabling arbitrary unitary matrix transformations.[8] G. Cong et al. achieved arbitrary reconfiguration of silicon photonic circuits using MZI networks, realizing 6-bit photonic digital-to-analog conversion.[9] In 2020, F. Shokraneh et al. proposed an MZI-based diamond mesh architecture for PNNs, as shown in Fig. 3(c).[10] In 2023, L. Pei et al. proposed a joint device-architecture-algorithm co-design method for implementing a photonic neural processing unit.[11] B. Wu et al. developed a simplified MZI mesh to perform real-valued optical matrix-vector multiplication by configuring phase shifters to construct real-valued optical matrices.[12] In addtion, G. Giannougiannis et al. implemented high-fidelity linear transformations using MZIs.[13] In 2024, A. Shafiee et al. analyzed the impact of loss and crosstalk in coherent PNNs (CPNNs) based on MZI arrays.[14] J. Lin et al. proposed a PNN architecture using MZIs, where QR decomposition was employed to implement linear computation. Compared to SVD-based methods, this approach requires fewer MZI units.[15]

MRRs, consisting of ring-shaped waveguides, exhibit optical resonance that enhances signals at specific wavelengths or frequencies. With radii of only a few micrometers, MRRs are highly scalable and can be integrated into compact arrays. These characteristics make MRRs ideal candidates for on-chip photonic synapses.[16-30] In 2012, L. Yang et al. proposed an on-chip silicon MRR array for performing linear matrix-vector multiplication. [16] As shown in Fig. 3(d), the transmission of the MRRs was adjusted to represent the matrix elements, and wavelength-division multiplexing (WDM) was employed to achieve parallel weighted summation of optical signals. In 2014, A. N. Tait et al. proposed the broadcast-and-weight architecture.[17] As shown in Fig. 3(e), MRRs were employed as tunable filters to perform dynamic weighted summation of WDM signals. Subsequently, they experimentally demonstrated the application of WDM-based weighted addition in principal component analysis,[18] achieved continuous calibration of single-channel MRR weights,[19] four-channel MRR weight bank,[20] and 8-channel MRR weight bank.[21] In 2017, they demonstrated a mathematical isomorphism of a continuous-time recurrent neural network (CTRNN), showing a 24-node photonic CTRNN that achieved a 294-fold acceleration over central processing units in solving differential equations. Figure 3(f) depicts a micrograph of the MRR weight bank.[22] In 2020, P. Y. Ma et al. implemented photonic independent component analysis for unknown signals using an on-chip MRR weight bank.[24] In 2021, S. Xu et al. proposed an optical CNN architecture, in which weights were loaded through wavelength-selective coupling by the MRRs.[25] In 2022, J. Cheng et al. extended the MRR weight bank to the complex domain and large-scale matrix operations, by employing matrix decomposition and partitioning.[26] W. Zhang et al. developed a 9-bit MRRs weight by adopting a dithering control scheme to compensate for environmental drift and inter-channel cross talk.[27] In 2024, E. C. Blow et al. evaluated the application of low-Q factor MRRs in broadband optical weighting. Experiments demonstrated that by adjusting the MRR radius and thermal tuning mechanisms, a 3-dB bandwidth of up to 4.7 GHz could be achieved while maintaining high weight accuracy.[29] In the same year, D. Jin et al. proposed a general modeling approach based on the scattering matrix method. The method decomposes devices into fundamental components, such as directional couplers and connecting waveguides, or simplifies them into standalone modules like MRRs, enabling precise modeling of both unidirectional and bidirectional optical devices. [30]

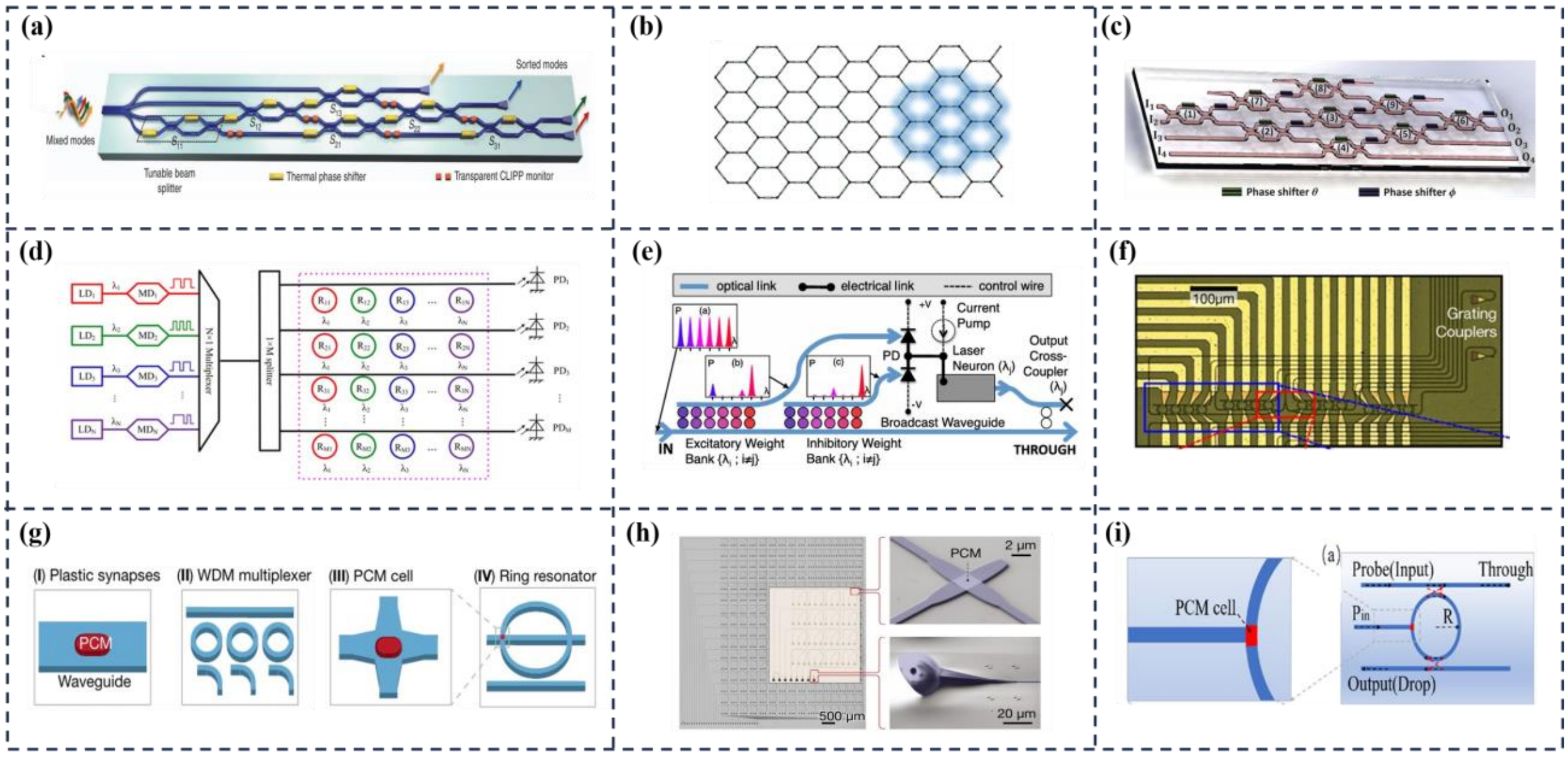

**FIG.3.** (a) Waveguide mesh of cascaded MZIs.[6] (b) Schematic of the hexagonal waveguide mesh.[7] (c) Layout of the 4×4 diamond MZIs.[10] (d) Schematic of the on-chip optical matrix-vector multiplier. [16] (e) Processing-network node coupled to a broadcast waveguide.[17] (f) Micrograph of broadcast-and-weight network.[22] (g) Photonic synaptic structures based on PCM.[33] (h) Optical micrograph of a fabricated 16×16 crossbar.[34] (i) GST-based weighting cell.[40]

PCMs can reversibly switch between crystalline and amorphous states under electrical or optical excitation, modulating optical transmission properties for encoding information and emulating neural and synaptic functions. These properties make PCMs as a key material for photonic synapse. [31-42] In 2017, Z. Cheng et al. proposed an on-chip photonic synapse based on PCM, where tapered waveguides and discrete $Ge_2Sb_2Te_5$ (GST) cells enabled linear weight modulation. The weight adjustment was achieved by controlling the number of pulses with fixed power and duration.[31] In 2019, I. Chakraborty et al. further proposed a photonic SNN computing unit, in which GST was embedded into MRRs to enable parallel dot-product operations.[32] Besides, J. Feldmann et al. demonstrated an all-optical spiking neurosynaptic network, using WDM and PCM for linear weight summation, as shown in Fig. 3(g).[33] In 2021, they further developed an integrated photonic tensor core, utilizing PCM arrays and optical frequency combs to enable parallel convolution processing, as shown in Fig. 3(h).[34] In 2020, M. Miscuglio et al. introduced a photonic tensor core based on PCM, enabling 4-bit precision linear operations.[35] In 2021, Y. Zhang et al. presented an optical synapse device based on directional couplers, using distributed discrete GST islands along the waveguide to tune the optical field distribution for linear weight updates.[36] C. Wu et al. exploited the refractive index contrast between amorphous and crystalline GST states to control modal contrast with up to 64 levels. This contrast is used to represent the matrix elements, with 6-bit resolution and both positive and negative values, to perform matrix–vector multiplication computation in neural network algorithms.[37] In 2022, W. Zhou et al. introduced GST-based waveguide memory devices capable of linear operations via optical or electrical programming.[38] They further applied this PCM-based linear computation to image processing tasks, achieving 87% classification accuracy on the MNIST dataset.[39] In 2023, Y. Zhang et al. introduced a photonic SNN based on MRR and GST (Fig. 3(i)) to complete a pattern recognition task for 12 clockwise directions.[40] In 2024, A. Lugnan et al. proposed a self-adaptive PNN using the non-volatile characteristics of GST.[41] A. H. A. Nohoji et al. proposed a photonic crystal cross-waveguide structure based on PCMs for implementing linear computations.[42]

## 2. Photonic synapses based on optical gain modulation

Photonic synapses based on optical gain modulation utilize gain media such as SOAs and VCSOAs to simulate synaptic weight modulation by controlling the optical amplification process. These synaptic structures can provide both amplification and attenuation of optical signals, enabling programmable weight updates. Table II summarizes representative works of photonic synapses based on SOAs and VCSOAs. In 2013, M. P. Fok et al. experimentally implemented spike-timing-dependent plasticity (STDP) using SOAs and electro-absorption modulators (EAMs), achieving adaptive control of synaptic weights, as shown in Fig. 4(a).[43] They further applied SOAs to the measurement of the angle of arrival of a microwave signal[44] and realized both supervised and unsupervised learning algorithms.[45] In 2015, Q. Ren et al. demonstrated adaptive control of STDP window height and width by adjusting the injection current of an SOA, as shown in Fig. 4(b).[46] In 2016, Q. Li et al. implemented an anti-STDP learning mechanism based on a single SOA.[47] In 2018, .S. Xiang et al. developed a photonic synaptic computing model based on VCSOAs, as shown in Fig. 4(c),[48] and later experimentally verified a photonic STDP scheme using VCSOAs.[49] They further proposed a plastic photonic synapse with a self-feedback loop, achieving all-optical synaptic plasticity through the dynamic gain of VCSOAs, as shown in Fig. 4(d).[50] Additionally, in 2024, they further constructed a PNN by combining the linear weighting capability of SOAs, demonstrating excellent performance in associative learning and image classification tasks.[51–53] In 2020, B. Shi et al. proposed a SOA-based photonic cross-connect chip. As shown in Fig. 4(e), the gain characteristics of SOAs were used as a weight matrix, with injection currents modulated to perform weighted summation of input signals.[54] In 2021, J. A. Alanis proposed a tunable photonic synapse based on VCSOAs, as shown in Fig. 4(f), which was capable of dynamically adjusting weights with a precision of 11.6 bit.[55] In 2022, T. Tian et al. further studied weight-dependent STDP based on VCSOAs. [56]

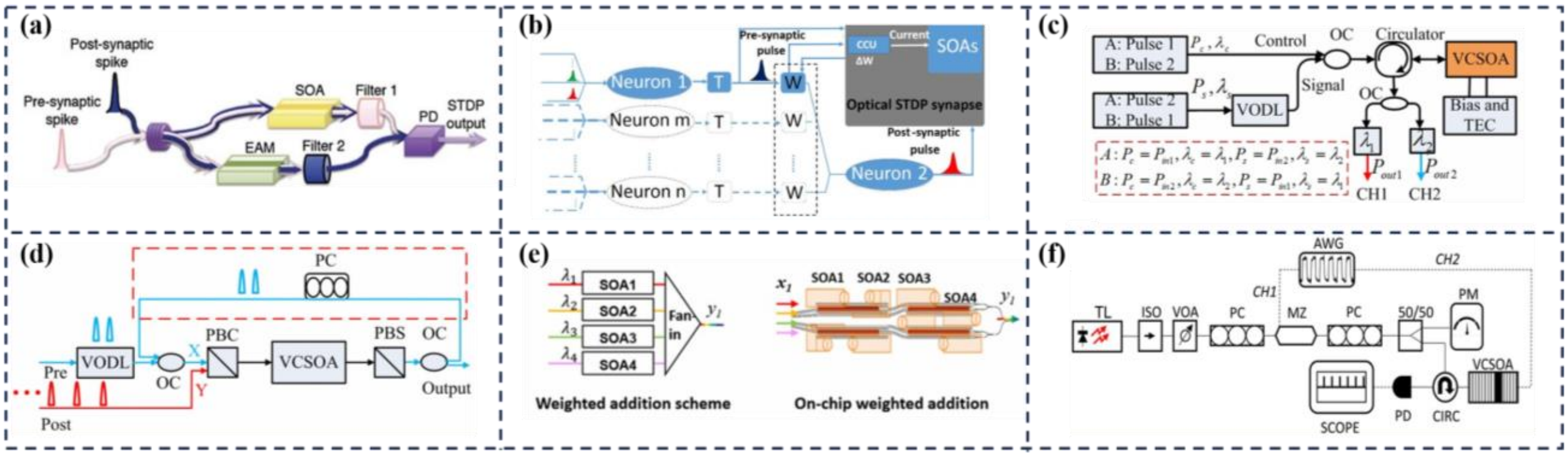

**FIG. 4.** (a) Optical implementation of STDP.[43] (b) Two photonic neurons connected with an optical STDP synapse.[46] (c) Schematic diagram of the proposed photonic STDP based on VCSOA.[48] (d) The schematic diagram of plastic photonic synapse performed synaptic plasticity.[50] (e) Scheme of the weighted addition within one neuron.[54] (f) Schematic description of the experimental setup for VCSOA-based photonic synapse.[55]

**TABLE II.** Photonic synapses based on optical gain modulation.

| Year & Author | Technology Type | Implementation Method | Key Contribution |
|---|---|---|---|
| 2013 M. P. Fok[43] | SOA+EAM | Optical STDP circuit | Adaptive feedback control |
| 2015 Q. Ren[46] | SOA | Weight-dependent STDP with reward modulation | Weight-dependent learning window |
| 2016 Q. Li[47] | SOA | XGM-based STDP circuit | Pulse-width: 80 ps |
| 2018 S. Xiang[48] | VCSOA | Numerical wavelength-dependent STDP | Low-power computational model |
| 2020 B. Shi[54] | SOA | InP cross-connect chip | NRMSE: <0.08 Dynamic range: 27 dB |
| 2021 J. A. Alanis[55] | VCSOA | VCSOA-based synapse | Precision: 11.6 bits Speed: ns rates |
| 2022 T. Tian[56] | VCSOA | VCSOA-based STDP learning window | Stability-competition weight adjustment |
| 2022 Z. Song[49] | VCSOA | Dual-polarization STDP scheme | Low-power polarization-multiplexed STDP |
| 2023 Y. Zhang[50] | VCSOA | Plastic photonic synapse | All-optical synaptic plasticity |
| 2024 D. Zheng[51] | SOA+DFB-SA | Full-function Pavlov associative learning PNN | SOA chip: 1000×2×250 μm³ power consumption: ~80 mW |

### 3. Other types of photonic synapses

Apart from the aforementioned types, there are also other forms of photonic synapses. Table III summarizes key developments in other types of photonic synapses. These architectures achieve more efficient and precise synaptic modulation through coherent detection, interferometric measurement, and complex-valued weight computation. In 2021, S. Xu et al. proposed a silicon-based optical coherent dot-product chip for performing complex deep learning regression tasks, as shown in Fig. 5(a). [57] In 2023, N. Youngblood proposed a large-scale matrix multiplication architecture based on coherent photonic crossbar arrays. The dot-product unit cell, as shown in Fig. 5(b), utilizes homodyne detection and time-division multiplexing techniques to enable efficient computation.[58] In 2024, Zhu et al. developed a universal photonic matrix processor that combines a coherent multi-dimensional photonic core with error management strategies, supporting high-precision matrix operations. Figure 5(c) shows the optical path design for 2×2 matrix-vector multiplication.[59] M. Moralis-Pegios et al. proposed a silicon photonic coherent crossbar (Xbar), as shown in Fig. 5(d), which achieved high-fidelity linear operations using EAMs and thermo-optic phase shifters. [60] B. Dong et al. leveraged partially coherent light to enhance the parallelism of photonic tensor cores. Figure 5(e) presents a schematic of matrix-vector multiplication for an N-dimensional input vector using partially coherent light.[61] S. R. Kari et al. designed a time-multiplexed coherent dot-product unit cell (DPUC), as presented in Fig. 5(f) which supports complex-valued computations and performs dot product of two 64-element vectors.[62]

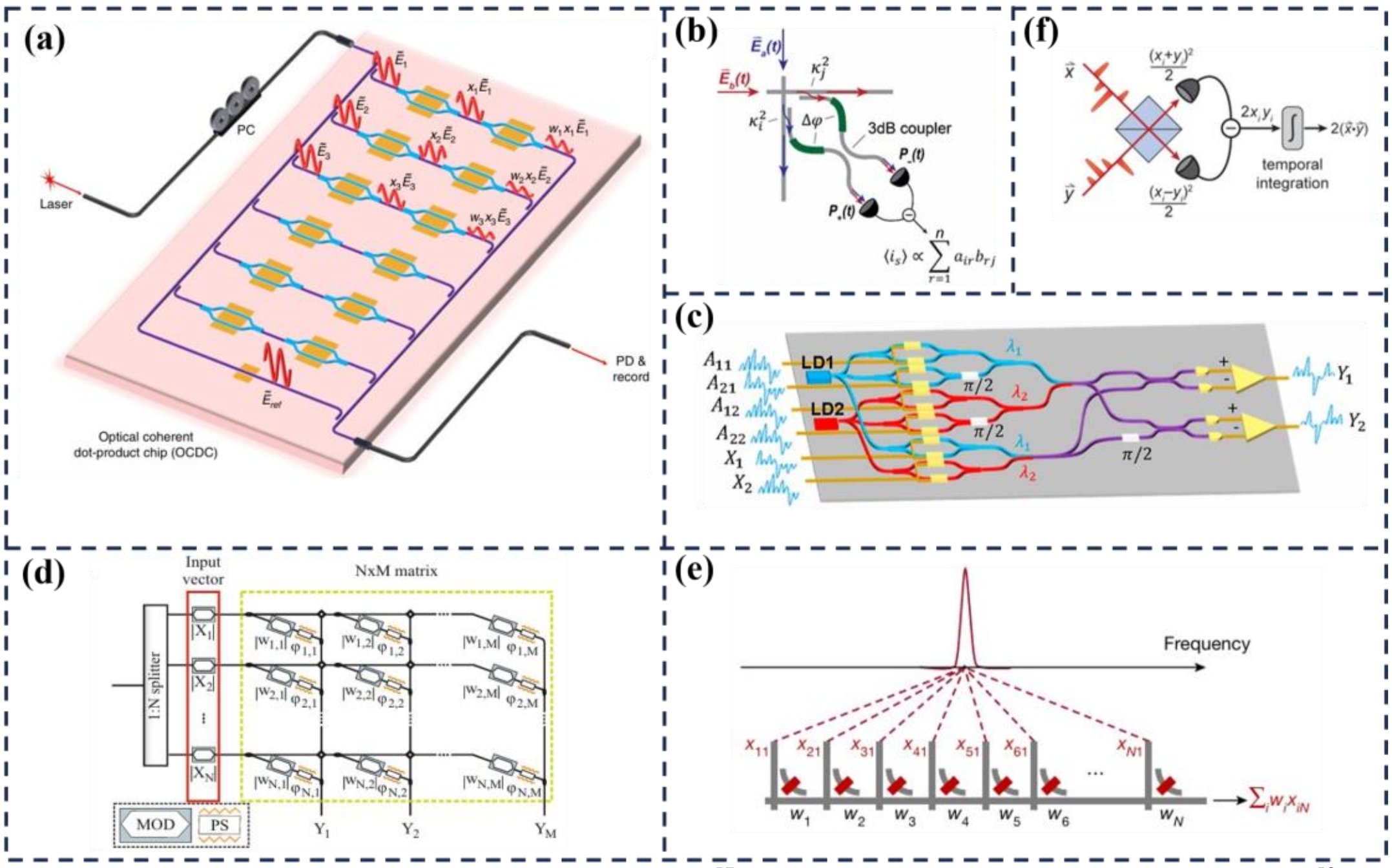

**FIG. 5.** (a) Schematic of the optical coherent dot-product chip.[57] (b) Schematic of the dot-product unit cell.[58] (c) Schematic of a photonic core supporting 2×2 matrix-vector multiplication.[59] (d) The N×M Crossbar architecture operating as a linear operator.[60] (e) Concept of partial-coherence-enhanced parallelized photonic computing.[61] (f) Schematic of time-multiplexed coherent dot-product.[62]

**TABLE III.** Other types of photonic synapses.

| Year & Author | Technology Type | Implementation Method | Key Contribution |
|---|---|---|---|
| 2021 S. Xu[57] | Coherent dot-product chip | Optical Coherent Dot-product Chip | Full real-valued domain deep learning regression |
| 2023 N. Youngblood[58] | Coherent photonic crossbar array | Photonic Matrix-Matrix Multiplier (MMM) | Peak computational speed: ~98 TeraOPs |
| 2024 Z. Zhu[59] | Coherent photonic matrix processor | Photonic Matrix Processing Unit (MPU) | General-purpose matrix processing for scientific computing |
| 2024 M. Moralis-Pegios[60] | Xbar architecture | Silicon photonic coherent crossbar | Fidelity: 99.997% $\pm 0.002$ |
| 2024 B. Dong[61] | Partially coherent tensor core | EAM-based photonic tensor core | Parkinson gait accuracy: 92.2% MNIST accuracy: 92.4% |
| 2024 S. R. Kari[62] | Coherent DPUC | Integrated silicon photonic DPUC array | RMSE: 0.09 Precision: 3.8 bits |

Linear photonic synapses are expected to continue advancing toward higher integration density, lower energy consumption, and broader scalability. Future developments will likely benefit from synergistic innovations across the three main categories: interference- and material-based designs, gain-modulated devices, and emerging coherent processing architectures. By combining novel materials, compact structures, and intelligent control schemes, future photonic synapses will not only improve computational precision and robustness, but also evolve toward higher practicality.

## B. NONLINEAR PHOTONIC NEURON

Nonlinear photonic neurons are the core components of PNNs for achieving complex functions. The implementation of nonlinear photonic neurons primarily relies on optical nonlinear materials and specialized optical devices, and this field is still in its early developmental stages. Current research on photonic nonlinear computing mainly focuses on two major directions: photonic nonlinear continuous-value activation neurons and photonic nonlinear spiking activation neurons.

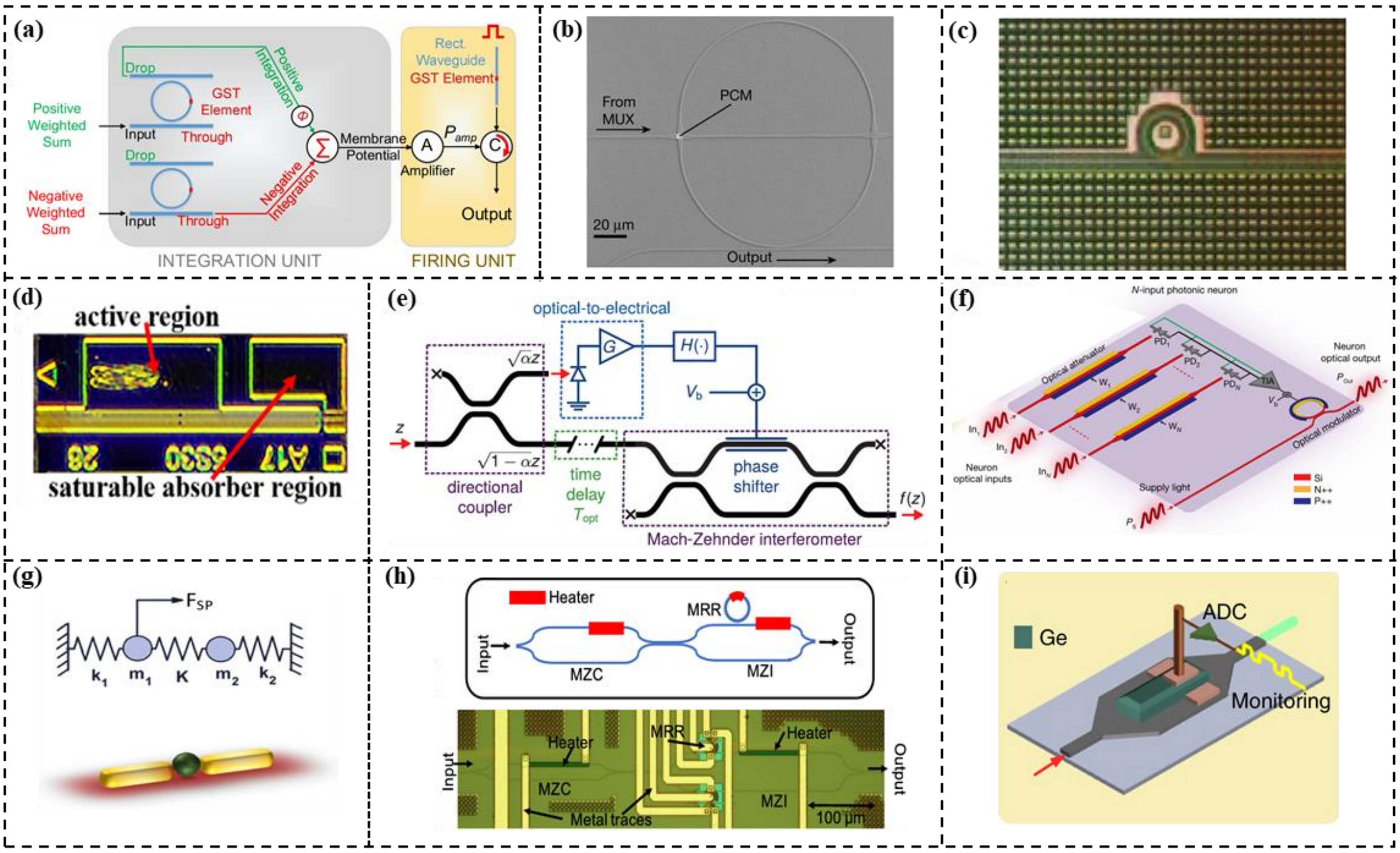

**FIG. 6.** Nonlinear photonic neurons. (a) Schematic of a bipolar integrate and fire neuron based on GST-Embedded Ring resonator devices.[99] (b) Scanning electron micrograph of a ring resonator used to implement the activation function.[33] (c) Microscope image of the silicon microring spiking neuron.[96] (d) The micrograph picture of the fabricated DFB-SA laser chip.[71] (e) Schematic of the activation function which achieves a nonlinear response by converting a small portion of the optical input.[103] (f) Schematic of photonic-electronic neuron.[104] (g) Electromagnetically induced transparency.[110] (h) Reconfigurable all-optical nonlinear activation functions based on a cavity-loaded MZI.[106] (i) Nonlinear germanium-silicon photodiode for activation.[114]

Nonlinear photonic spiking neurons are typically implemented using semiconductor lasers, MRRs, and PCMs. Table IV summarizes representative studies of nonlinear photonic spiking neurons. [63-100]By leveraging the optical injection dynamics of lasers, polarization switching effects, and the modulation properties of saturable absorbers, researchers can successfully mimic biological neuronal behaviors, including excitatory responses, inhibitory responses, and spiking activity. In 2010, A. Hurtado et al. proposed to emulate the fundamental functions of biological neurons using the polarization switching effect in vertical-cavity surface-emitting lasers (VCSELs), achieving excitatory and inhibitory responses through optical injection.[63] In 2014, B. J. Shastri et al. introduced a laser model based on graphene saturable absorbers, demonstrating its application in spike processing.[64-66] Since 2016, S.Xiang et al reported spiking dynamics based on VCSEL and VCSELs with saturable absorbers (VCSEL-SA) [67-78], including spiking rate encoding based on VCSEL[67], polarization-multiplexed spike encoding based on VCSEL-SA[76], XOR with a single VCSEL[68], and binary convolution and image edge detection based on VCSEL. [69] In 2023, S. Xiang et al proposed and fabricated a photonic spiking neuron chip based on a Fabry-Pérot (FP) laser with saturable absorber (FP-SA), and demonstrated the nonlinear neuron-like dynamics, including threshold, temporal integration, and refractory period. They further realized the mapping of SNN algorithm to the FP-SA chip for hardware-software collaborative computing.[86] Besides, they also reported a photonic spiking neuron based on a FP laser, utilizing its spike rate encoding characteristics to construct a high-speed obstacle avoidance system.[70] They further proposed and fabricated a photonic spiking neuron chip based on a distributed feedback laser with saturable absorber (DFB-SA). As illustrated in Fig. 6(d), neuron-like dynamical characteristics under both single-wavelength and multi-wavelength incoherent optical injection conditions were demonstrated.[71] Additionally, various other spiking neuron implementations based on: VCSEL-SA,[72-78] micropillar lasers,[79-83] FP-SA,[84-88] DFB lasers, [71,89-93] have been reported.

Meanwhile, MRR-based photonic spiking neurons have gained increasing attention due to their advantages in large-scale integration and low power consumption. In 2012, T. Van Vaerenbergh et al. systematically demonstrated the feasibility of silicon-based MRRs as photonic spiking neurons.[94] Since 2020, J. Xiang et al. have proposed all-optical spiking neurons (as illustrated in Fig. 6(c))[95,96] and electrically driven spiking neurons[97] based on silicon MRRs, confirming their ability to replicate typical spiking neuron behaviors. More recently, graphene-silicon heterogeneously integrated MRRs have also been explored.[98] Furthermore, the integration of PCMs has expanded the implementation strategies for photonic spiking neurons, offering novel approaches to emulate integrate-and-fire behaviors. In 2018, I. Chakraborty et al. developed an all-optical spiking neuron model using a $Ge_2Sb_2Te_5$ (GST)-embedded MRR, as illustrated in Fig. 6(a), where optical pulses triggered the crystalline-to-amorphous transition in GST to emulate biological integrate-and-fire dynamics.[99] In 2019, J. Feldmann et al. employed PCM-MRR-based neurons in photonic SNNs for nonlinear activation, as illustrated in Fig. 6(b), modulating GST's amorphization degree via weighted input pulses to control probe pulse transmission.[33] In 2024, Q. Zhang et al. proposed a thermodynamic leaky integrate-and-fire (TLIF) neuron model based on an electrically reconfigurable GST optical switch.[100]

**TABLE IV.** Photonic nonlinear spiking activation neurons.

| Year & Author | Activation Functions | Type | Technology | Programmability |
|---|---|---|---|---|
| 2010 A. Hurtado[63] | N/A | All optical | VCSEL | No |
| 2012 T. Van Vaerenbergh[94] | N/A | All optical | MRR | No |
| 2014 F. Selmi[79] | N/A | All optical | Micropillar laser with saturable absorber | No |
| 2018 I. Chakraborty[99] | IF | All optical | MRR + PCM | No |
| 2019 J. Feldmann[33] | ReLU | All optical | MRR + PCM | No |
| 2022 J. Xiang[96] | N/A | All-optical | MRR | No |
| 2023 S. Xiang[86] | LIF | All optical | FP-SA laser | No |
| 2023 S. Gao[70] | N/A | All optical | FP laser | No |
| 2024 Y. Zhang[71] | LIF | All optical | DFB-SA laser | No |
| 2024 Q. Zhang[100] | TLIF | Optoelectronic | PCM | No |

Photonic nonlinear continuous-value activation neurons are primarily implemented through various optical nonlinear effects and devices, including EAMs, electro-optic modulators (EOMs), SOAs, and novel functional materials. Table V summarizes representative studies of photonic nonlinear continuous-value activation neurons.[101-111] Among these, EAMs leverage the electro-absorption effect, where an applied voltage alters the material's optical power absorption characteristics, thereby modulating light signal intensity to achieve nonlinearity. In 2019, J. K. George et al. modeled the nonlinear transfer functions of five different EAM types, analyzing and comparing their performance in PNNs.[101] They also proposed an indium tin oxide (ITO)-based EAM monolithically integrated with a silicon waveguide.[102] In 2019, I. A. D. Williamson et al. reported a reconfigurable nonlinear activation function based on the electro-optic effect, as illustrated in Fig. 6(e). Their approach involved converting a fraction of the input optical signal into an electrical signal, which was then applied to an EOM to control the intensity of the remaining optical signal, achieving an opto-electro-optic nonlinear conversion.[103] In 2022, Z. Xu et al. proposed a programmable nonlinear accelerator based on non-volatile opto-resistive RAM switches. They exploited the opto-resistive RAM's high-to-low resistance transition to realize nonlinear functionality.[104] F. Ashtiani et al. demonstrated a ReLU activation function using the electro-optic effect in a PN-junction MRR modulator, as illustrated in Fig. 6(f).[105] All-optical nonlinearity can also be achieved by leveraging the free-carrier dispersion effect and thermo-optic effect in MRRs. In 2020, A. Jha et al. implemented multiple nonlinear functions by loading an MRR onto a MZI and exploiting free-carrier dispersion, as illustrated in Fig. 6(h).[106] In 2022, Z. Fu et al. proposed a programmable, low-loss all-optical activation device comprising a MRR integrated with a GST thin film. This design leverages the intrinsic nonlinear properties of the silicon MRR while utilizing the phase-change characteristics of GST to enable dynamic programmability of nonlinear activation functions.[107] SOAs generate nonlinear responses through gain saturation and cross-gain modulation (XGM) during optical interactions. In 2019, G. Mourgias-Alexandris et al. realized nonlinear activation using an SOA-MZI and an SOA-XGM gate.[108] In 2020, B. Shi et al. employed an SOA-based wavelength converter for nonlinear activation, where XGM converted multi-wavelength signals into a single-wavelength output.[109] Beyond these approaches, laser-nonlinear material interactions enable all-optical control over light propagation. In 2018, M. Miscuglio et al. demonstrated nonlinear activation using a nanophotonic structure, as illustrated in Fig. 6(g), where plasmonic resonance in metal nanoparticles interacted with excitonic transitions in quantum dots, modulating transmission.[110] In 2024, C. Chen et al. reported an on-chip integrated all-optical nonlinear activation device based on 2D $MoTe_2$ and optical waveguides, utilizing its saturable and reverse-saturable absorption effects.[111] Additional innovative methods include: exploiting the Kramers-Kronig relations between optical amplitude and phase for nonlinear activation,[112] novel device architectures like germanium-silicon hybrid structures (as illustrated in Fig. 6(i))[113-116] and graphene/silicon hetero-integration.[117] These advancements significantly expand the possibilities for realizing continuous-value nonlinear activation functions in PNNs.

**TABLE V.** Photonic nonlinear continuous-value activation neurons.

| Year & Author | Activation Functions | Type | Technology | Programmability |
|---|---|---|---|---|
| 2018 M. Miscuglio[110] | N/A | All optical | Nanophotonic structures | No |
| 2019 G. Mourgias-Alexandris[108] | Logistic sigmoid | All optical | SOA-MZI + SOA-XGM | No |
| 2019 I A D Williamson[103] | ReLU | Optoelectronic | MZI | Yes |
| 2019 R. Amin[100] | N/A | Optoelectronic | ITO-based EAM | No |
| 2020 A. Jha[106] | Sigmoid, Radial Basis, Clamped ReLU, Softplus | All optical | MZI + MRR | Yes |
| 2022 Z. Fu[107] | ReLU, ELU, Softplus, Radial Basis | All optical | MRR + PCM | Yes |
| 2022 F. Ashtiani[105] | ReLU | Optoelectronic | MRR | No |
| 2022 Y. Shi[114] | N/A | All optical | Ge-Si PD | No |
| 2022 Z. Xu[104] | N/A | Optoelectronic | ORS + MZI + PCM | Yes |
| 2023 Y. Tian[112] | ReLU, Softplus | All optical | Kramers-Kronig activation | Yes |
| 2024 C. Chen[111] | N/A | All optical | $MoTe_2$/OWG | No |

Nonlinear photonic neurons emulate biological neurons through precise control of optical nonlinearities and device physics. Current research mainly focuses on two areas: spiking and continuous-value activation. These areas are based on different physical mechanisms and device architectures, and offer unique advantages for neuromorphic computing. At present, the research on nonlinear photonic neurons is still in a stage of rapid development. As materials science, device design and optoelectronic integration technology advance, nonlinear photonic neurons will facilitate the construction of multi-layer PNNs, contributing to realize sophisticated functions and higher performance.

## III. PNN ARCHITECTURE AND CHIPS

ANNs are foundational to intelligent computing, with architectures such as fully connected ANNs, SNNs, CNNs, and reservoir computing (RC), each excelling in different tasks—from data integration and biological emulation to image processing and temporal data handling (Fig. 7). This section explores recent advances in photonic implementations of ANNs, CNNs, SNNs, diffractive networks, and RC, and examines how photonic technologies are reshaping neural computing.

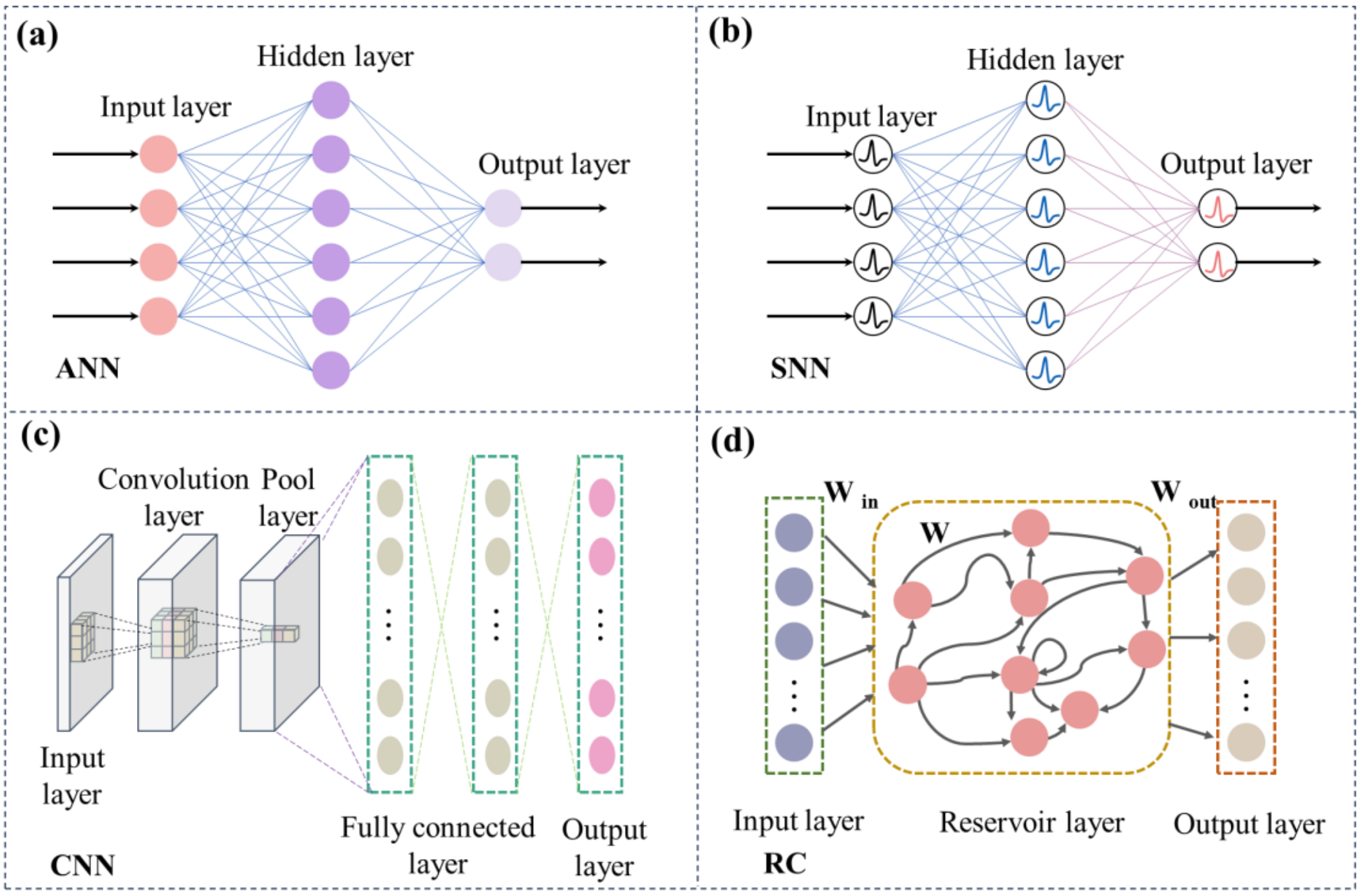

**FIG. 7.** Schematic diagrams of (a) fully connected ANN, (b) SNN, (c) CNN, and (d) RC.

### A. PHOTONIC FULLY-CONNECTED NETWORK

A fully-connected network (FCN) is a basic neural network architecture. While FCNs based on traditional electronic hardware have achieved remarkable progress, they struggle with processing speed and energy efficiency as data volumes explode and application scenarios grow more complex. Photonic FCN exhibits advantages of high-speed parallel transmission, ultra-wide bandwidth, and low energy consumption, and has attracted lots of attention in recent years[115-128]. The relevant progress of photonic FCN is shown in Table VI. In 2017, Y. Shen et al. proposed a PNN architecture based on coherent nanophotonic circuits. A FCN was realized with 56 MZIs, as shown in Fig. 8 (a).[119] In 2020, C. Huang et al. demonstrated a PNN for fiber nonlinearity compensation in long-haul transmission systems. In an experiment over a 10,080-km trans-Pacific link, the PNN achieved a Q-factor improvement of 0.51 dB, just 0.06 dB lower than numerical simulations, proving the feasibility of PNNs for optical fiber transmission applications.[120] In 2021, they also designed a WDM-based PNN architecture, which adopted MRRs for weight matrix operations, and balanced photodetectors for signal summation and nonlinear activation (Fig. 8(b)), and achieved 0.60 dB Q-factor improvement for fiber nonlinearity compensation.[121] In the same year, H. Zhang et al. proposed an optical neural chip capable of implementing complex-valued neural networks. By encoding information in the phase and amplitude of light and exploiting optical interference, the chip performed complex-valued arithmetic operations, significantly enhancing computational speed and energy efficiency, as shown in Fig. 8(c).[123]

In 2022, G. Mourgias-Alexandris et al. proposed a noise-resilient, high-speed photonic deep learning architecture. Based on coherent silicon photonics, it achieves a computation rate of 10 billion multiply-and-accumulate operations per second per axon (10GMAC/sec/axon). By using a noise-aware training model, its noise resilience was enhanced, as shown in Fig. 8(d).[124,125] C. Feng et al. designed a compact butterfly-style silicon photonic-electronic neural chip for hardware-efficient deep learning. It reduced the use of optical components and energy consumption by limiting the universality of weight representation.[126] S. Ohno et al. demonstrated a prototype chip of a 4 × 4 MRR crossbar array for on-chip inference and training of PNN by directly mapping target matrix elements to the transmittance of MRRs. Moreover, it enabled on-chip backpropagation through the transpose matrix operation of the MRR crossbar array, accelerating the training of the PNN, as shown in Fig. 8(e).[127] F. Ashtiani et al. proposed an end-to-end photonic deep neural network for image classification. It achieved sub-nanosecond image classification by directly processing optical waves, as shown in Fig. 8(g).[105] Y. Shi et al. proposed a chip-based photonic neuron using nonlinear germanium-silicon (Ge-Si) photodiodes, and built a self-monitored nonlinear PNN, as shown in Fig. 8(f).[114]

**TABLE VI.** Evolution and performance analysis of photonic FCN.

| Year & Author | Technology Type | Implementation Method | Key Contribution |
| --- | --- | --- | --- |
| 2017 Y. Shen[119] | Coherent MZI mesh | 4×4 MZI mesh | Vowel recognition accuracy:76.7% |
| 2021 C. Huang[121] | Silicon PNN for compensating fiber nonlinearity | 4×2 PNN with WDM, MRR and BPD | 0.60 dB Q-factor improvement over 10,080 km |
| 2021 H. Zhang[123] | MZI-based complex-valued PNN | 6×6 MZI mesh and coherent detection | Iris accuracy: 97.4%<br>Circle and spiral accuracy: 98%<br>MNIST accuracy: 90.5% |
| 2022 G. Mourgias-Alexandris[124] | Silicon coherent PNN | 4 fan-in dual IQ-modulator | Compute speed: 10GMAC/s; MNIST accuracy: >98% |
| 2022 C. Feng[126] | Butterfly-style photonic-electronic neural chip | 4 port PNN with phase shifters, directional couplers, waveguide crossings, and MZI attenuator | 7× fewer trainable optical components;<br>Energy efficiency: ~9.5 TOPS/W;<br>Compute density: ~225 TOPS/$mm^2$;<br>MNIST accuracy: 94.16% |
| 2022 S. Ohno[127] | MRR for on-chip inference and training | 4 ×4 MRR crossbar array | Energy efficiency: 15 TOPS/W;<br>Iris accuracy: 93% |
| 2022 F. Ashtiani[105] | End-to-end deep PNN | PIN attenuator, GeSi PD, and micro-ring modulator | End- to-end latency: 570 ps<br>Energy efficiency: 345 fJ/OP;<br>Compute density: 3.5 TOPS/$mm^2$;<br>Accuracy: 93.8% (two-class), 89.8% (four-class) |
| 2022 Y. Shi[114] | GeSi PDs for nonlinear self-monitored PNN | 4×4 MZI mesh and GeSi PD | Energy efficiency: ~0.27 pJ/OP;<br>Compute density: 1.92 TFLOPS;<br>MNIST accuracy :97.3% |
| 2023 H. Zhang[128] | Complex-valued PNN for molecular prediction | MZI network with 8 modes and 56 phase shifters | Coefficients of determination: 0.9325 (molecular property prediction) |
| 2024 Z. Xu[131] | Distributed computing architecture Taichi | Diffractive-interference hybrid photonic chiplet, and 8×8 MZI mesh | Energy efficiency: 160 TOPS/W;<br>Compute density: 878.90 TMACS/$mm^2$;<br>1623-category Omniglot accuracy: 91.89% |

In 2023, H. Zhang et al. utilized photonic chip technology for molecular property prediction, showcasing the potential of PNNs in computational chemistry. They utilized photonic chips to implement complex-valued neural networks and employed a multi-task regression learning algorithm to predict quantum mechanical properties of molecules. This study paved the way for future large-scale molecular property prediction and material design.[128] In 2024, T. Xu et al. reported a method for control-free and efficient PNNs via hardware-aware training and pruning. By shifting neural network weights to noise-insensitive areas, this method remarkably enhanced the robustness and energy efficiency of PNNs.[130] In the same year, Z. Xu et al. proposed and fabricated a large-scale photonic computing chip Taichi based on reconfigurable MZI arrays and on-chip diffractive units. Based on an integrated diffractive-interference hybrid design and a general distributed computing architecture, it achieved an energy efficiency of 160 tera-operations per second per watt (160-TOPS/W), as shown in Fig. 8(h).[131]

PNNs have the potential to overcome the limitations of traditional electronic architectures in terms of speed, energy efficiency and scalability. As photonic technologies advance, integrating components such as MZIs, MRRs and nonlinear photodiodes onto silicon chips enables high-speed, low-latency processing and real-time learning capabilities. Recent advancements demonstrate a clear trajectory towards chip-scale integration, noise-resilient training, and hybrid diffractive-interference architectures. These developments collectively pave the way for energy-efficient and high-throughput photonic computing platforms. Moreover, the ability to implement complex-valued computations and multitask learning on photonic chips significantly enhances the versatility of these systems for applications in communication, computational chemistry, and AI. Moving forward, significant progress in hardware-aware optimization, distributed computing frameworks, and large-scale photonic integration will be essential for the development of general-purpose, intelligent photonic processors.

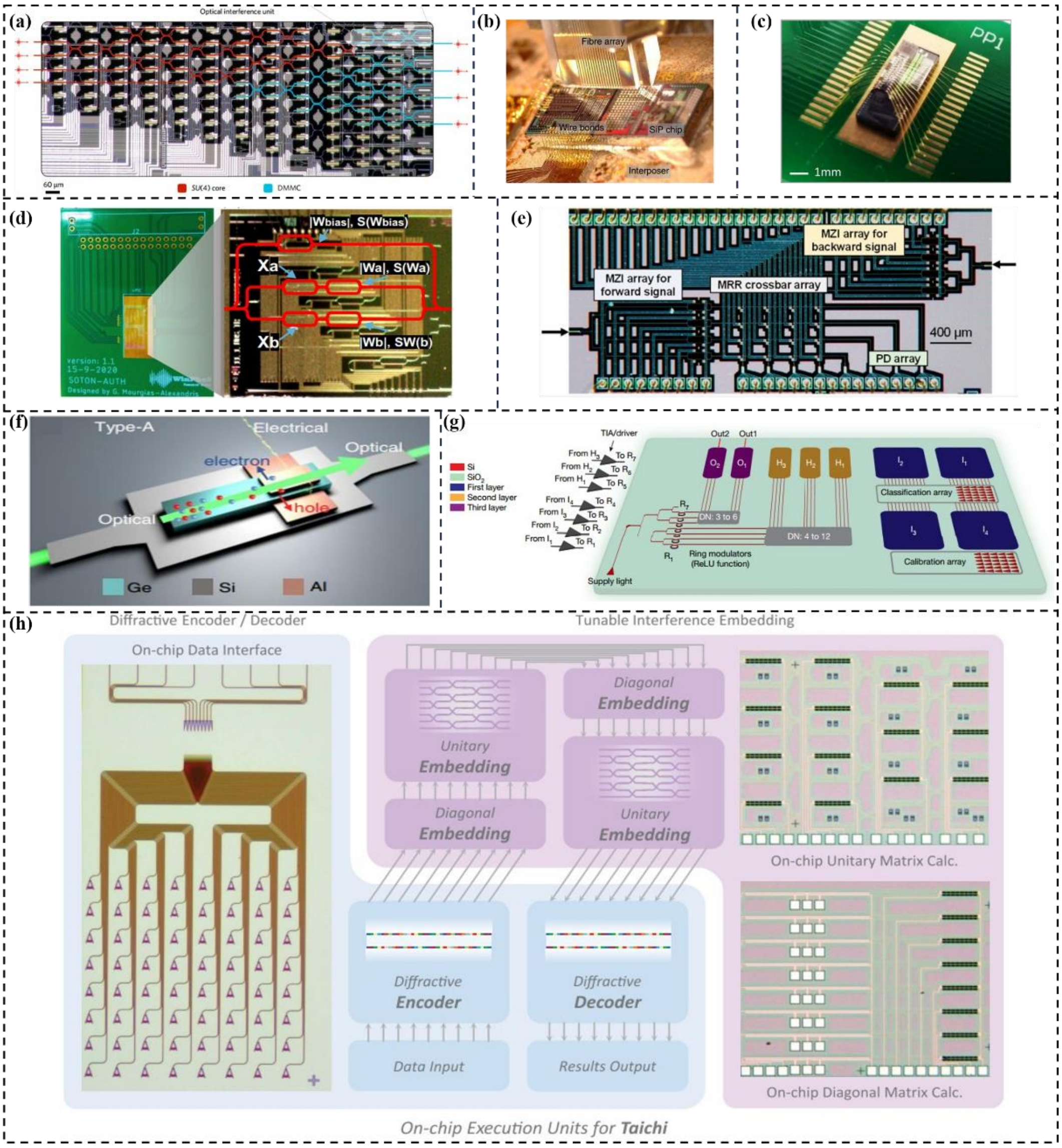

**FIG. 8.** (a) Optical micrograph illustration of the experimentally demonstrated OIU.[119] (b) Chip packaging and optical coupling setup.[121] (c) Fabrication and packaging of silicon photonic chip.[123] (d) Silicon photonic neuron.[124] (e) Plan-view image of the fabricated MRR crossbar array.[127] (f) The structure and schematic of the proposed AONU (Type-A). Many carriers are accumulated in the non-electrode part of the Ge film, which enhances the nonlinear interaction with light. At the tail end, the carrier movement forms the photocurrent that served as a monitoring signal. The yellow wave ray represents the data flow of the electrical monitoring signals.[114] (g) The top-level block diagram of the PDNN chip. Two 5 × 6 arrays of grating couplers are used as the input pixel array.[105] (h) Taichi chiplets design. Diffractive-based encoder and decoder are applied for input data perception, and the MZI array serves as feature embeddings. Conv., convolutional layer; NL., nonlinear layer; Pool., pooling layer; Elec., electronic; Opto., optical; Calc., calculation.[131]

## B. PHOTONIC CONVOLUTIONAL NETWORK

CNNs have achieved remarkable success in various applications such as computer vision and speech processing. However, traditional electronic implementations are increasingly constrained by power and speed limitations as computational demands grow. Photonic CNNs based on waveguide interconnects have advanced rapidly. Current implementations can be primarily classified according to their core devices, including MRRs, MZIs, PCMs, phase shifters, and arrayed waveguide gratings (AWGs). Table VII summarizes representative studies of photonic CNNs consisting of these fundamental devices.

### 1. MRR-based CNNs

MRR-based synapses represent a key class of photonic synapses. When cascaded into arrays, they enable incoherent matrix multiplication, with each MRR individually tunable and its transmission coefficient precisely controlled to match specific computational requirements. In photonic integrated circuits, MRRs offer advantages such as compact size, high tunability, and compatibility with WDM technologies, making them well-suited for flexible weight modulation in photonic CNNs. Furthermore, the Kerr effect in high-Q MRRs enables parametric oscillations under optical pumping, generating new frequencies through four-wave mixing. These frequencies form evenly spaced comb lines, producing optical frequency combs that can serve as multi-wavelength sources for convolution operations.

In 2014, A. N. Tait *et al.* achieved high-precision, scalable weight control for CNN computations by combining the proposed on-chip optical broadcast-and-weight architecture with MRRs and WDM.[17] In 2018, A. Mehrabian *et al.* developed a photonic CNN accelerator using MRR weight banks and WDM, achieving highly parallel operations with theoretical speeds thousands of times faster than electronic counterparts.[132] In 2020, V. Bangari *et al.* introduced the digital electronic and analog photonic (DEAP) CNN architecture, realizing a 2.8–14× speedup and ~25% energy savings over GPUs, with 97.6% accuracy on the MNIST dataset (Fig. 9(a)).[133] D. J. Moss *et al.* demonstrated a universal optical vector convolution accelerator combining MRR-generated microcombs with electro-optic Mach–Zehnder modulators (MZMs), achieving 11.322 TOPS through wavelength, temporal, and spatial multiplexing (Fig. 9(b)).[134] W. Zou *et al.* presented an integrated photonic tensor flow processor that processed high-order tensors directly in the optical domain, avoiding digital duplication and reaching 480 GOP/s throughput and 588 GOP/s/mm² computing density, with scaling potential beyond 1 TOPS/mm².[135] In 2023, J. Dong et al. combined MRR weight banks with microcomb-based multi-wavelength sources, achieving 8-bit weight precision, 51.2 TOPS throughput, 4.18 TOPS/W energy efficiency, and 78.5% accuracy on the RAF-DB emotion dataset.[136] J. E. Bowers *et al.* developed a microcomb-driven chip-based photonic processing unit with 9-bit precision and 1.04 TOPS/mm² compute density.[137] In 2025, X. Xu *et al.* introduced an optical tensor convolution accelerator that used multidimensional multiplexing and triple modulation paths to surpass single-path limits, achieving over 3 TOPS for 3×3×3 convolutions with reduced memory consumption.[138] They also leveraged wavelength-synthesizing and time-wavelength interleaving to perform complex-valued convolutions exceeding 2 TOPS.[139]

**TABLE VII.** Representative research on photonic CNNs.

| Year & Author | Technology Type | Implementation Method | Key Contribution |
|---|---|---|---|
| 2017 J. K. George[140] | MZI | Convolution based on FFT | Compute speed: ~$10^3$ speedup over state-of-the-art GPUs |
| 2020 V. Bangari[133] | MRR | Digital electronic and analog photonic CNN | Compute speed: 2.8 to 14 times faster than GPUs; Energy efficiency: 25% less than GPUs; MNIST accuracy: 97.6% |
| 2021 D. J. Moss[134] | MRR and EOM | Time-wavelength-space interleaving technology | Compute speed: 11.3 TOPS; MNIST accuracy: 88% |
| 2021 H. Bhaskaran[34] | MRR and PCM | Parallelized in-memory computing | Compute speed: 2 tera-MACs/s; Energy efficiency: 17 fJ/MAC; MNIST accuracy: 95.3% |
| 2022 Y. Tian[141] | MZI | Real-value matrix representation | MZI amounts: O($N\log_2 N$); MNIST accuracy: ~99.3% |
| 2023 J. E. Bowers[137] | MRR | Time-wavelength plane stretching | Compute density: 1.04 TOPS/mm$^2$; Energy efficiency: 2.38 TOPS/W; MNIST accuracy: 96.6% |
| 2023 H. Bhaskaran[39] | PCM | In-memory dot-product engine | Energy consumption: 1.7nJ/dB; MNIST accuracy: 87%; Fashion-MNIST accuracy: 86% |
| 2024 D. Yi[148] | AWG and MZI | Eliminating repetitive multiplication | Weight precision: 8-bit; MNIST accuracy: 96% |
| 2024 J. Dong[149] | AWG and MZM | Inherent routing principles | Compute density: 8.53 TOPs/mm$^2$; MNIST accuracy: 91.9% |
| 2025 C. Pappas[150] | AWGR and EOM | Time-space-wavelength multiplexed architectures | Compute speed: 163.8 TOPS; DDoS detection: 0.799 Cohen's kappa score; MNIST accuracy: 93.35% |

### 2. MZI-based CNNs

The widely-adopted MZI-based synapses, which can be configured in mesh topologies to perform linear operations, are well-suited for implementing convolution computations in photonic CNNs.

In 2017, J. K. George et al. employed the MZI mesh for fast Fourier transform (FFT) and achieved an all-optical CNN, improving the computation speed by approximately $10^3$ compared to the most advanced GPUs of the time.[140] In 2020, F. Shokraneh *et al.* proposed a programmable MZI optical processor based on a diamond mesh structure, enabling various sizes of PNNs. Compared to the triangular mesh, the diamond mesh showed higher robustness to phase errors and loss tolerance, offering better scalability for MZI-based optical processors.[10] In 2022, Y. Tian *et al.* introduced a photonic matrix architecture that used the real part of a non-universal N×N unitary MZI mesh to represent real-valued matrices. This method was applied to CNNs and successfully completed the MNIST dataset classification in experiments. This architecture significantly reduced the number of MZIs in the network while preserving the learning capabilities of the MZI mesh.[141]

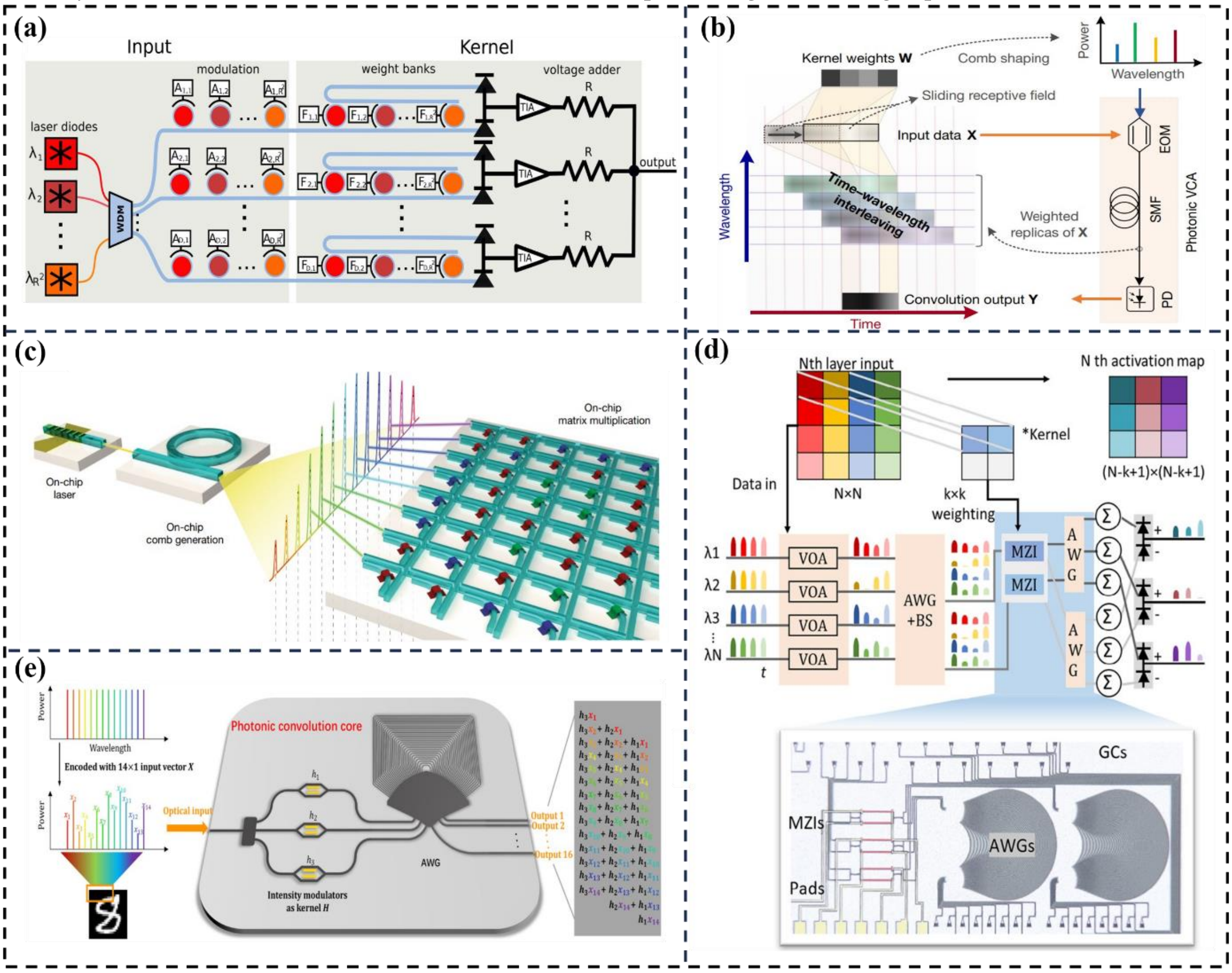

**FIG. 9.** Representative research on photonic CNNs. (a) DEAP-CNN hardware architecture based on MRRs.[133] (b) Optical convolutional accelerator based on optical frequency combs.[134] (c) Fully integrated photonic convolutional architecture based on PCM.[34] (d) Parallel optical convolution core.[148] (e) Photonic convolution core with AWG chip.[149]

### 3. PCM-based CNNs

With fast response, non-volatility, and excellent compatibility with optoelectronic devices, PCM-based synapses hold great promise for silicon photonic integration. By combining silicon photonics with PCM-based synapses, both photonic memory and computation can be integrated on a single platform, enabling efficient, fully optical processing. This approach is pivotal for building large-scale, high-speed, and low-power photonic CNNs, advancing their application in AI and beyond.

In 2011, C. D. Wright et al. experimentally demonstrated a PCM-based processor capable of basic arithmetic operations—addition, subtraction, multiplication, and division—with simultaneous result storage. This foundational work significantly propelled PCM-enabled neuromorphic photonic computing.[142] In 2021, H. Bhaskaran et al. introduced a photonic tensor core by combining PCMs with integrated microcombs and WDM, enabling parallel convolution at speeds approaching $10^{15}$ MAC operations per second (Fig. 9(c)).[34] In 2023, they further developed a non-volatile, electronically reprogrammable PCM memory cell that achieved 4-bit weight encoding, 1.7 nJ/dB crystallization energy, and 158.5% switching contrast—offering high scalability for large CNNs and achieving effective performance in image and pattern recognition tasks.[39] B. Dong et al. proposed a PCM-based photonic tensor core leveraging an added radio-frequency (RF) dimension for multiplexing across space, wavelength, and RF domains. This design enabled a 100-way parallelism—two orders of magnitude beyond conventional architectures—and achieved 93.5% accuracy in a CNN-based sudden cardiac death risk detection task.[143]

### 4. Phase shifter-based CNNs

Phase shifters, which operate via the thermo-optic effect, are essential components of MZMs. When current flows through the micro-heaters of a phase shifter, the waveguide temperature changes, altering the refractive index due to its temperature dependence. This enables precise phase modulation of optical signals, which, combined with interference principles, allows for intensity modulation and thus optical weighting.

In 2021, W. Zou et al. proposed a PNN based on a silicon photonic coherent dot-product chip, where phase shifters were used for both weighted operations and adaptive calibration. This system enabled real-valued computation across arbitrary scales of convolution and matrix multiplication through multiplexing strategies, achieving image reconstruction accuracy comparable to 32-bit digital systems.[57] In 2023, M. Li et al. introduced a low-loss silicon nitride optical convolution processing unit based on multimode interference (MMI). By integrating WDM and thermo-optic phase shifters, they developed a compact and linearly scalable parallel convolution architecture. The chip achieved 5-bit MAC precision, offering a high-density (12.74 TMACs/mm²) and energy-efficient (4.84 pJ/MAC) solution for large-scale optical CNNs.[144] In 2025, they further proposed a reconfigurable optical CNN utilizing a novel data encoding scheme across wavelength, temporal, and spatial domains. This design achieved a 92.86% data utilization rate and a theoretical computing throughput of up to 10.51 TOPS.[145]

#### 5. AWG-Based CNNs

AWGs are key elements of parallel computation in optical CNNs, thanks to their wavelength-selective properties and high integration density. AWGs operate by leveraging fixed optical path differences and interference among waveguides, directing different wavelengths to specific output ports, thereby supporting wavelength multiplexing and routing. These characteristics supports the high parallelism and scalability of AWG-based optical CNNs, facilitating efficient multitask processing.

In 2023, B. Shi et al. proposed an on-chip parallel photonic convolution scheme using a cross-connected architecture and cyclic AWGs. This design introduced a cyclic wavelength domain and combined it with spatial and free spectral range dimensions to enable parallel convolution. A convolution core based on an AWG–SOA–AWG structure was implemented, achieving 2.56 TOPS and an energy efficiency of 3.75 pJ/bit.[146] In 2024, X. Dong et al. proposed a wavelength-routing convolution scheme using AWGR's unique sliding characteristics to perform sliding window operations in the wavelength–spatial domain, thus avoiding decomposition into large numbers of MAC operations. Compared to optical matrix–vector multiplier (MVM)-based methods, this approach offers higher scalability, simplicity, and speed.[147] D. Yi et al. integrated AWGs with MZI meshes to develop an optical convolution processor that eliminated redundant multiplications via spectral filtering, validated with an 8-bit resolution proof-of-concept experiment (Fig. 9(d)).[148] J. Dong et al. designed an integrated architecture that performed both sliding kernel operations and summation within AWGs, enabling M×N MAC operations using only M+N units in a single clock cycle, thus minimizing resource redundancy (Fig. 9(e)).[149] In 2025, C. Pappas et al. introduced an optical accelerator based on time–space–wavelength multiplexing with AWGRs. Their architecture featured dedicated matrix-by-matrix and matrix-by-tensor engines, achieving a peak computing performance of 163.8 TOPS.[150]

Photonic CNNs have evolved across diverse device platforms, each leveraging distinct physical mechanisms. MRRs offer scalable, WDM-compatible weight modulation, particularly when integrated with optical frequency combs. MZIs enable programmable matrix operations, while PCMs provide non-volatile, reconfigurable memory-compute integration. Phase shifters and AWGs add flexibility through dynamic phase control and high-density spectral routing. Despite these advances, there are still key challenges to overcome, such as achieving scalable and precise weight control, tolerance to fabrication and thermal variability, and seamlessly integrating optical nonlinearities and memory.

### C. PHOTONIC SPIKING NETWORK

SNNs, first introduced by Wolfgang Maass as the third generation of ANN models,[151] incorporate spiking neurons that closely resemble their biological counterparts. Unlike traditional neural networks, SNNs encode neural information through spikes that capture temporal dynamics, enabling the processing of complex spatiotemporal patterns. Consequently, SNNs offer several advantages, including enhanced biological plausibility, suitability for hardware implementation, improved energy efficiency, and increased computational power. Although microelectronics-based SNNs have made significant progress, they encounter inherent limitations regarding energy consumption and processing speed. In contrast, photonics offers a promising alternative for information processing, leveraging their inherent advantages such as high speed, large bandwidth, and low power consumption. As early as 2014, N. Alexander et al. proposed a parallel photonic neural interconnect architecture known as "broadcast-and-weight" and a reconfigurable processing-network node, demonstrating an on-chip photonic SNN protocol.[17] In recent years, significant research efforts have been devoted to the on-chip implementation of photonic SNNs, with representative works summarized in TABLE VIII.

#### 1. On-chip photonic SNN models and architectures

In recent years, significant achievements have been made in the development of photonic SNN architectures based on MRRs.[32,98-100,152] In 2018, I. Chakraborty et al. proposed a spiking neuron based on GST-embedded MRRs. These neurons can be integrated with on-chip synapses to create a fully photonic SNN inference framework, offering the potential for ultrafast computation and a wide operational bandwidth.[99] In 2019, they further developed non-volatile synaptic arrays using GST-based MRRs and integrated them with spiking neurons, as shown in Fig. 10(a), and achieved a classification accuracy of 97.85% on the MNIST dataset.[32] In 2022, S. Xiang et al. proposed a photonic SNN architecture and system-level computational model based on MRRs, employing supervised learning algorithms for spike sequence learning tasks. Their work highlighted the significance of general and robust algorithms to address the limitations of photonic hardware and to promote the practical application of photonic SNNs.[152] More recently, in 2024, N. Jiang et al. introduced a photonic computing primitive for integrated SNNs based on add-drop ring microresonators (ADRMRs) and electrically reconfigurable PCM photonic switches. This approach enables both spiking response and synaptic plasticity, providing theoretical support for MRR-based photonic SNN chip development.[100]

In addition to MRR-based approaches, significant research efforts have been directed toward photonic SNN architectures based on VCSELs.[50,74,153-162] As shown in Fig. 10(b) and Fig. 10(c), since 2019, S. Xiang et al. have proposed a photonic SNN computational model utilizing VCSELs and VCSOAs. This model is capable of reproducing biologically inspired neural dynamics and spike-timing-dependent plasticity (STDP). A self-consistent unified model of neurons and synapses within an all-optical SNN framework was developed to support this architecture. In this model, synaptic plasticity is emulated based on the dynamic response of VCSOAs under dual optical pulse injection. They further demonstrated unsupervised learning for first-spike-based pattern recognition tasks,[50,153] and extended the model to supervised learning schemes for achieving pattern classification.[154]In addition, several other research groups have also reported photonic SNN architectures based on VCSELs.[74,158-162] In 2022, A. Hurtado et al. demonstrated a photonic SNN constructed using a single VCSEL, as illustrated in Fig. 10(d), achieving 97% classification accuracy on the Iris dataset. In 2024, N. Li et al. proposed a simple and effective data encoding scheme for photonic SNNs using VCSELs-SA. This approach enabled successful implementation of pattern recognition, facial expression recognition, and Iris dataset classification, achieving an accuracy of 94.67%.[74]

**TABLE VIII.** Representative works about advancing on-chip implementations of photonic SNN.

| Year & Author | Technology type | Implementation Method | Key Contribution |
|---|---|---|---|
| 2019 I. Chakraborty[32] | PCM and MRR array | Photonic SNN with 16 GST and 16 MRR | Energy consumption: 12.5 fJ/synapse, 5 pJ/neuron; MNIST accuracy: 97.85% |
| 2019 J. Feldmann[33] | PCM and MRR | 4 neurons and 60 all-optical synapses | Supervised and unsupervised learning; 15-pixel images classification |
| 2019 S. Xiang[153] | VCSEL | VCSEL-based neuron, VCSOA for photonic STDP | Unsupervised learning; First spike timing recognition |
| 2020 S. Xiang[154] | VCSEL | VCSEL-based neuron, VCSOA for photonic STDP | Spike sequence learning; 10-class classification |
| 2021 T. Inagaki[170] | DOPO | Class-I/II spiking neurons fiber ring cavity network | Mode switching Synchronization control 150-node Ising problem |
| 2023 A. Hurtado[158] | VCSEL | GHz-rate photonic SNN | Iris accuracy: >97%; Speed: GHz |
| 2023 S. Xiang[166] | VCSEL-SA and MZI | 4×4 MZI array and 4 VCSELs-SA | "0-3" recognition:100%; 400×10 photonic SNN for digit recognition task |
| 2023 S. Xiang[86] | FP-SA | Time-multiplexed spike encoding, hardware-software collaboration | Energy consumption: 7.329 fJ/spike; Spiking response rate: 3.3GHz |
| 2023 S. Xiang[173] | FP-SA | Two electrodes as excitatory/inhibitory dendrites | Frequency encoding range: 1.43–3.34 GHz Iris classification:100% |
| 2023 S. Xiang[175] | DFB-SA | Single chip of photonic spiking neuron | Temporal encoding and rate encoding; Neuron-like response; MNIST accuracy: 92.2% |
| 2023 S. Xiang[92] | Four-channel DFB-SA array | Photonic convolutional SNN with time-multiplexed matrix convolution | Parallel weighting and nonlinear activation; Energy consumption: 19.99 fJ/spike; MNIST accuracy: 87% |
| 2024 N. Jiang[100] | ADRMRs | ADRMRs and PCM | Dual neural dynamics of ADRMRs; MNIST accuracy: 98% |
| 2024 N. Li[74] | VCSEL-SA | Photonic SNN based on VCSEL-SA | Recognized four spiking patterns; Iris classification: 100% |
| 2024 Y. Lee[165] | Co-integrated CMOS and MZI, VCSEL | Optoelectronic neuron with 4×4 MZI mesh and VCSEL | Energy consumption: 1.18 pJ/spike; Iris accuracy: 89.3% |
| 2025 X. Guo[97] | MRR | MRR with p-n junction hybrid spiking CNN | Spike encoding rate: 250 MHz; Energy consumption: 20 pJ/spike; MNIST accuracy: 94.1% |

In 2022, Y. Lee et al. proposed a monolithic optoelectronic SNN hardware design inspired by the Izhikevich model, comprising an event-driven laser spiking neuron integrated with an incoherent MZI network, as shown in Fig. 10(e). Subsequently, they developed an optoelectronic SNN capable of performing handwritten digit classification on the MNIST dataset,[163] and the Iris dataset.[165] In 2022, S. Xiang et al. proposed a hybrid integrated photonic SNN inference framework consisting of MZIs and VCSELs-SA for pattern recognition tasks, as depicted in Fig. 10(f). By employing an improved remote supervised method (ReSuMe) based on a tempotron-like algorithm, they successfully achieved recognition of digits '0–3' and performed optical character recognition (OCR) tasks.[166]

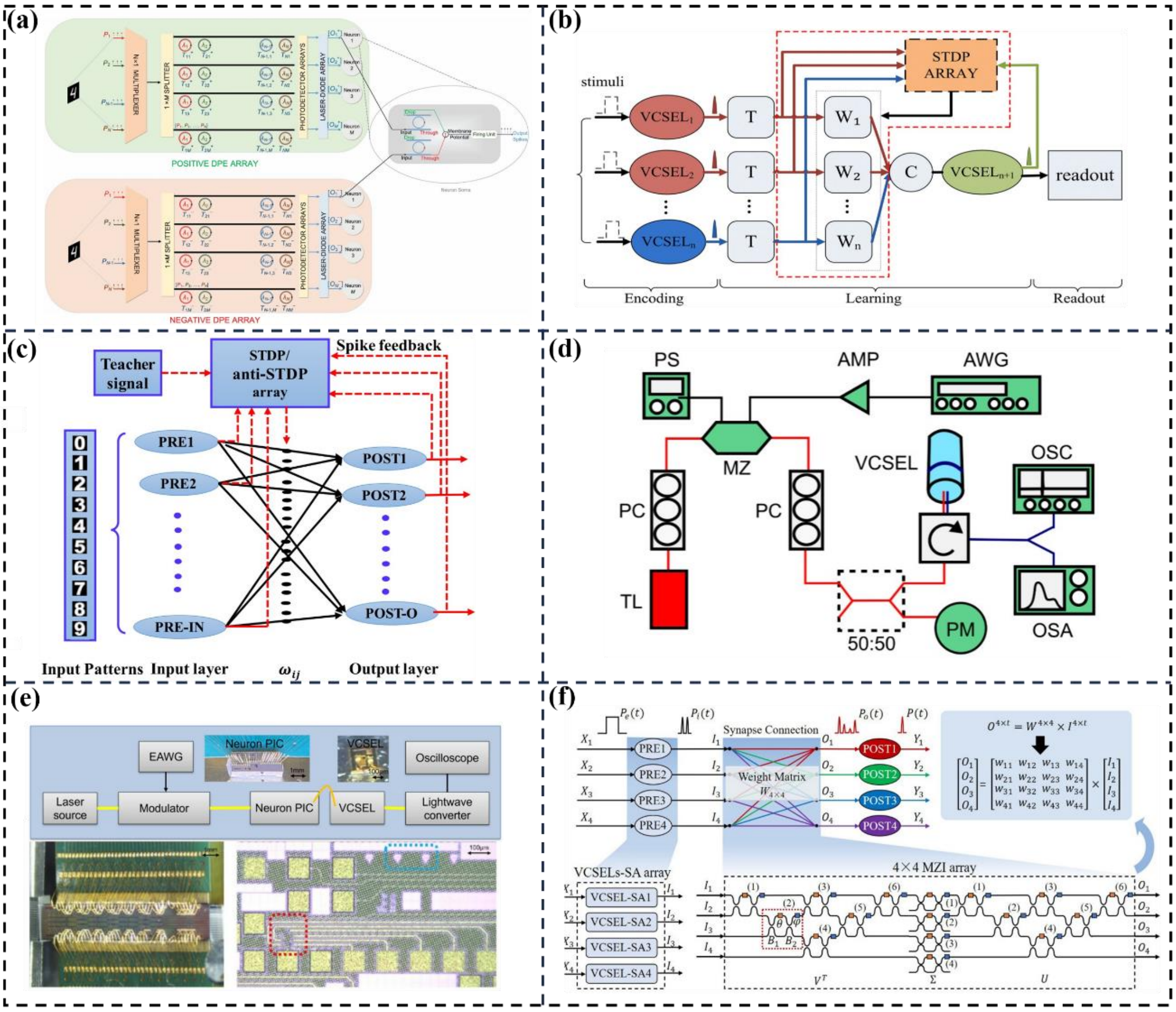

**FIG. 10**. (a) The photonic SNN computing primitive based on MRRs.[32] (b) The photonic SNN consisting of photonic spiking neurons based on VCSELs.[153] (c) The computing primitive of a photonic SNN based on VCSELs for supervised learning.[154] (d) The experimental setup for the spiking reservoir computer/SNN.[158] (e) Neuron spiking dynamics experimental setup and photo of MZI mesh.[165] (f) Schematic diagram of a hybrid-integrated 4 × 4 photonic SNN architecture based on a 4 × 4 MZI array and VCSELs-SA arrays.[166]

For the software-hardware co-design of photonic SNN architectures, in 2021, S. Xiang et al. introduced synaptic delay plasticity and developed an improved supervised learning algorithm tailored for photonic SNNs based on traditional weight training. This approach enabled high classification accuracy (reaching 92%) even with a limited number of optical neurons.[167] Furthermore, in 2023, they proposed a multi-synaptic connection strategy utilizing a delayed-weight co-training version of the ReSuMe algorithm, which significantly enhanced network performance.[91] In the same year, the team demonstrated a hybrid architecture for photonic convolutional SNNs [168] and proposed a complete photonic SNN conversion framework,[169] achieving successful classification across multiple datasets. These works provide a solid theoretical foundation for the practical deployment of photonic SNNs.

**2. On-chip integrated photonic SNN chips**

The photonic SNN integrated chips have also witnessed rapid development, paving the way of widespread application of photonic SNNs in real-world scenarios.

Multiple teams have achieved significant milestones in the development of integrated chips for photonic SNNs. In 2019, J. Feldmann et al. proposed and fabricated an integrated photonic synapse-based spiking neuron system (circuit). As shown in Fig. 11(a). The system supports both supervised and unsupervised learning modes and successfully classified four 15-pixel images, demonstrating its capability for pattern recognition as a prototype AI system.[33] In 2021, T. Inagaki et al. show that photonic spiking neurons implemented with paired nonlinear optical oscillators can be controlled to generate two modes of bio-realistic spiking dynamics by changing optical-pump amplitude, as shown in Fig. 11(b). [170] In 2022, X. Guo et al. reported the electrically-driven spiking neuron based on a silicon microring under the carrier injection working mode, which is shown in Fig. 11(c). By programming time-multiplexed spike representations, photonic spiking convolution based on this MRR is realized for image edge feature detection and yields a classification accuracy of 94.1% on the MNIST.[97] W. Zou et al. further proposed a noise-injection scheme to implement a GHz-rate stochastic photonic spiking neuron (S-PSN).

The firing-probability encoding was experimentally demonstrated and exploited for Bayesian inference with unsupervised learning. In a breast diagnosis task, the stochastic photonic spiking neural network (S-PSNN) can achieve a classification accuracy of 96.6%.[171]

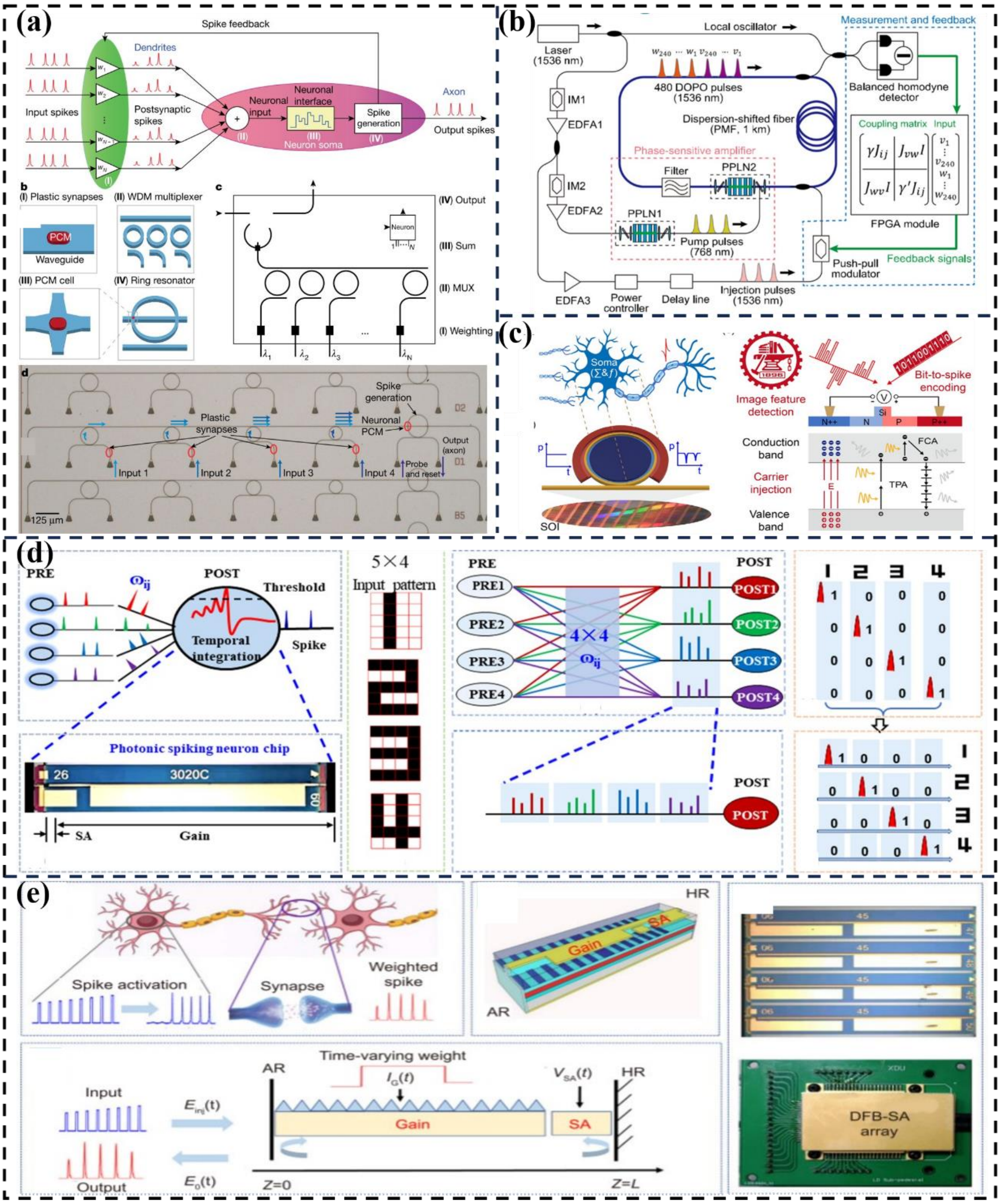

**FIG. 11.** (a) Schematic of the spiking neuron circuits, consisting of several pre-synaptic input neurons and one post-synaptic output neuron connected via PCM synapses.[33] (b) Experimental setup of photonic SNN based on DOPO.[170] (c) Schematic of the silicon microring spiking neuron with an embedded p-n junction.[97] (d) Operational principle of photonic SNN based on FP-SA for pattern recognition.[86] (e) The photonic neuron-synaptic core based on DFB-SA.[92]

On spiking neurons based on laser chips, in 2020, N. Alexander et al. demonstrated that a laser neuron, fabricated in a photonic integrated circuit platform, could function as a processing node in a larger scale spiking neural network. This approach calculated speed and energy efficiencies— 1 TMAC/s per neuron and 260 fJ/MAC, respectively—exceed current microelectronic performance figures, particularly in speed.[172] By using time-multiplexed temporal spike encoding, as shown in Fig. 11(d), S. Xiang et al. proposed a PSNN with FP-SA laser chip are experimentally demonstrated to realize hardware-algorithm collaborative computing, showing the capability to perform classification tasks with a supervised learning algorithm.[86] Furthermore, the team conducted in-depth research on FP-SA laser chip. The frequency encoding and spatiotemporal encoding schemes, the recognition accuracy can be enhanced based on the FP-SA neurons with double dendrites by expanding the weight range, reach the 97.5% accuracy in Iris dataset.[173] Besides, based on the FP-SA laser chip, classification tasks on Iris and WBC datasets were implemented.[87,88] In 2023, based on DFB-SA laser chip, they also benchmarked the handwritten digit classification task with a simple single-layer FCN and achieved a recognition accuracy of 92.2%.[175] They further used a four-channel DFB-SA laser array to implement activation and linear weighting on a single chip, as illustrated in Fig. 11(e), successfully achieving a recognition accuracy of 94.42% on MNIST dataset classification.[92] In 2024, they proposed and experimentally demonstrated the full-function Pavlov associative learning PNN based on DFB-SA laser chip and SOA.[51] In 2025, they proposed a PSNN with DFB-SA laser chip and direct modulated laser to realize 94% accuracy on the MNIST dataset. This PSNN show that the energy efficiency reaches 0.625 pJ/MAC.[93]

Photonic SNN has achieved remarkable advancements in terms of architecture and chip. Novel nonlinear spiking neuron chip with low thresholds, unsupervised and supervised learning algorithms have been developed, as well as the integtated Photonic SNN chips. As a further attempt, photonic SNN will emphasize monolithic or hybrid integration of photonic spiking neurons and synapses, aiming to realize compact, low-power, and ultrafast neuromorphic systems. Collectively, these developments are anticipated to pave the way for large-scale deployment of photonic SNNs in practical applications such as edge computing, intelligent sensing, and reinforcement learning.

## D. DIFFRACTIVE OPTICAL NEURAL NETWORK

Diffractive optical neural networks (DONNs) have emerged in recent years as a novel model of PNNs, leveraging the principles of diffractive optics combined with deep learning. DONNs have already demonstrated remarkable achievements in the domain of free-space optics.[177-180] Compared to their free-space counterparts, on-chip DONNs have attracted significant attention due to their advantages in miniaturization and energy efficiency. As compared to bulk optics, integrated photonics offers a scalable solution in terms of alignment stability and the overall network footprint.[181-187]

**TABLE IX.** The representative works about on-chip DONN.

| Year & Author | Technology type | Implementation Method | Key Contribution |
|---|---|---|---|
| 2020<br>J. Ong[181] | OCNN integrated diffractive optics | Star couplers Fourier-transform-based convolutions | MNIST accuracy: 97.9%<br>F-MNIST accuracy :88.6% |
| 2022<br>D. Jiang[182] | IDNN | Diffractive cells MZI | Energy consumption: 17.5 mW;<br>Iris accuracy: 98.3%;<br>MNIST/F-MNIST accuracy: 89.3%/81.3% |
| 2022<br>S. Zarei[183] | on-chip DONN | WDM SOI platform with 1D metasurfaces | Three optical logic gate<br>Bandwidth: 60 nm |
| 2023<br>H. Chen[185] | on-chip DONN | SOI platform 1D dielectric metasurface | Computation throughput: 81.6 TOPS;<br>Energy consumption: $1.808\times10-4$ J;<br>F-MNIST accuracy: 91.63%;<br>CIFAR-4: 86.25% |
| 2024<br>J. Dong[186] | TDONN | SOI platform<br>Tunable diffractive units<br>Stochastic gradient descent and drop-out mechanism | Computation throughput: 217.6 TOPS;<br>Computing density: 447.7 TOPS/mm2;<br>Energy efficiency: 7.28 TOPS/W;<br>Latency: 30.2 ps;<br>Multimodal test accuracy: 85.7%; |

Representative works on on-chip DONNs are summarized in TABLE IX. In 2020, J. Ong et al. proposed and simulated an optical CNN based on a star coupler (Fig. 12(a)). By combining phase and amplitude masks, they achieved an accuracy of 97.9% on the MNIST dataset.[181] In 2022, D. Jiang et al. designed an integrated DONN using two ultra-compact diffractive units and *N* MZIs, enabling parallel Fourier transforms, convolution operations, and application-specific optical computing, as illustrated in Fig. 12(b). The system achieved classification accuracies of 98.3%, 92.5%, and 83.2% on the Iris, MNIST, and Fashion-MNIST datasets, respectively.[182] S. Zarei et al. reported a DONN capable of performing optical logic operations. Three logic gates (NOT, AND, and OR) were demonstrated within a single DONN operating at a wavelength of 1.55 μm, as shown in Fig. 12(c). Additionally, wavelength-independent operations across seven wavelengths were demonstrated, enabling WDM for parallel computing.[183]

In 2023, H. Chen et al. proposed a passive DONN architecture based on integrated one-dimensional (1D) dielectric metasurfaces, as illustrated in Fig. 12(d). The DONN with one hidden layer (DONN-l1) and with three hidden layers (DONN-l3) achieved classification accuracies of 86.7% and 90.0% on the Iris dataset, respectively. Additionally, DONN-l3 achieved an experimental accuracy of 86% on the MNIST dataset.[184] Based on this architecture, they further implemented classification and regression tasks. The classification tasks on the Fashion-MNIST and CIFAR-4 datasets yielded accuracies of 91.63% and 86.25%, respectively.[185] J. Dong et al. proposed and fabricated a trainable DONN (TDONN) chip, as shown in Fig. 12(e), by integrating electrodes into the hidden layers of the on-chip diffractive optical devices. The TDONN chip consists of one input layer, five hidden layers, and one output layer. The chip successfully performed classification and achieved an accuracy of 85.7% on a multimodal test dataset.[186]

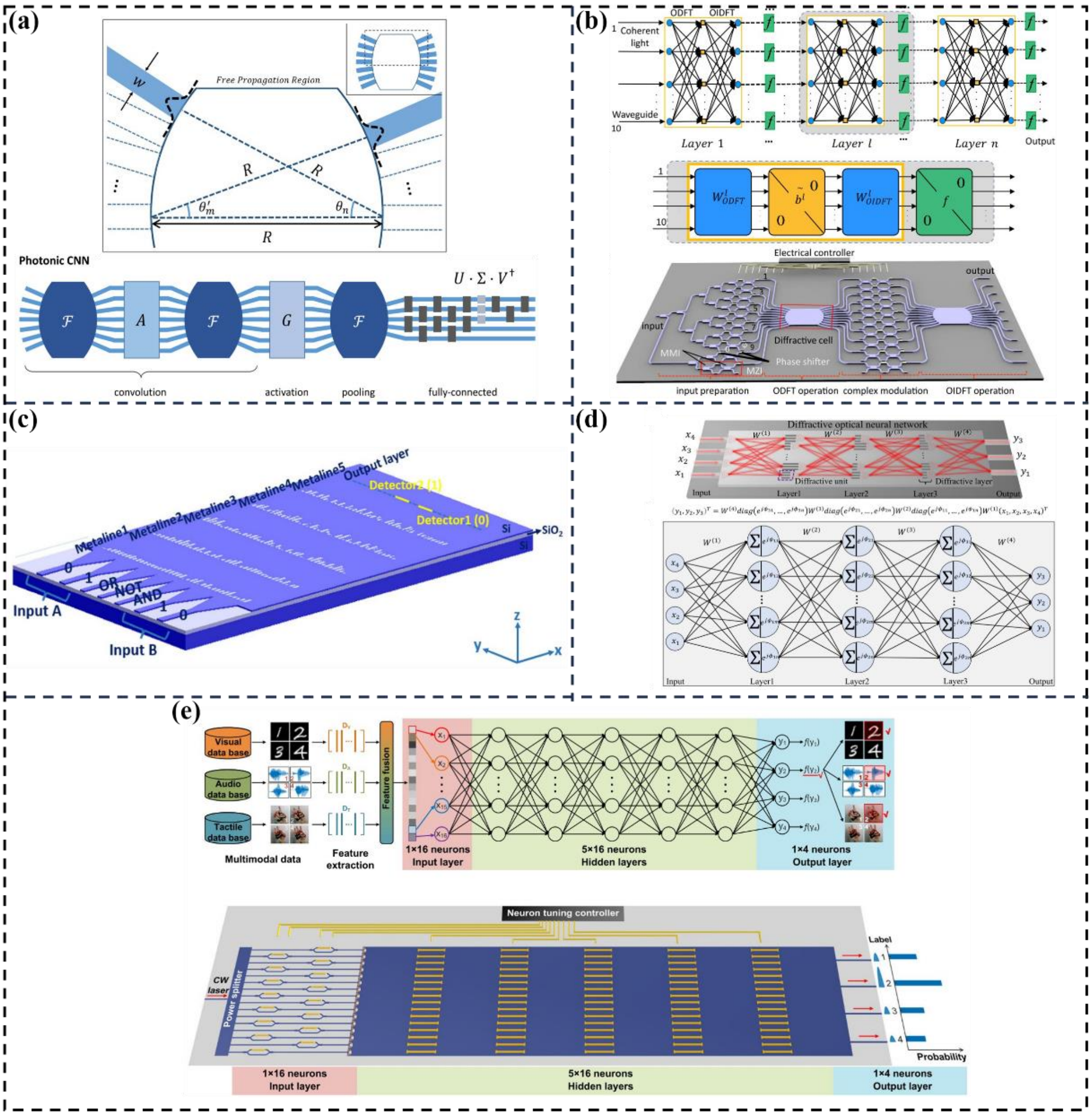

**FIG. 12.** (a) Schematic of N×M star coupler and OCNN based on it.[181] (b) The integrated diffractive optical network based on two ultracompact diffractive cells and MZI.[182] (c) Schematic of on-chip DONN trained to perform optical logic operations AND, NOT and OR.[183] (d) Schematic and logic diagram of on-chip diffractive optical neural network based on a silicon-on-insulator platform.[185] (e) Schematic of on-chip diffractive optics for multimodal deep learning.[186]

Diffractive optical structures are capable of supporting high-density neuron mapping and programmable interconnects at the nanoscale. These capabilities significantly enhance the computational throughput, parallelism, and reconfigurability of DONNs. By incorporating tunable optical materials, electro-optic modulation elements, and trainable diffractive architectures, DONNs can perform real-time optical inference with low power consumption, low latency, and ultrahigh bandwidth. However, realizing the full potential of DONNs still requires addressing several key technical challenges, including training accuracy, system stability, and large-scale on-chip integration.

## E. PHOTONIC RESERVOIR COMPUTING

In 2007, echo state networks (ESN) and liquid state machines (LSM) were defined as reservoir computing (RC) by D. Verstraeten et al., laying the foundation for subsequent developments.[188] RC is divided into three components: the input layer, the reservoir layer, and the output layer. The input and reservoir weights in RC are randomly initialized and remain fixed, while only the output layer weights are learned through training algorithms, such as gradient descent and least squares methods. These RC algorithms are characterized by their low training complexity and minimal computational cost. Both theoretical and experimental studies of RC have advanced rapidly along two main paths: spatially distributed multi-node RC systems and nonlinear reservoir layer with delayed feedback loops.

### 1. Space RC

The development of space RC systems is summarized in Table X.[189-204] With the advancement of photonic on-chip integration technology, space RC has evolved from a complex system involving multiple discrete nonlinear optical devices[189-193] to an integrated silicon photonic reservoir chip.[194-204] In 2008, K. Vandoorne et al. proposed a photonic parallel reservoir system model based on 25 SOAs coupling theoretical models.[190] After that, advancements in device integration and architectural optimization have led to significant milestones, including 81 SOA adjacent node interconnect array models,[189] a passive silicon photonic chip,[191] and a multimode Y-junction passive photonic reservoir system.[192] In 2014, K. Vandoorne et al. demonstrated a passive on-chip silicon photonic reservoir chip, as shown in Fig. 13(a).[191] To address the energy loss caused by the large number of nodes and long delay lines in the original chip, they integrated a multimode Y-junction-based passive photonic reservoir system.[192] In 2015, D. Brunner et al. implemented an RC system with a diffractive coupling architecture comprising an 8 × 8 VCSELs array (Fig. 13(b)).[193]

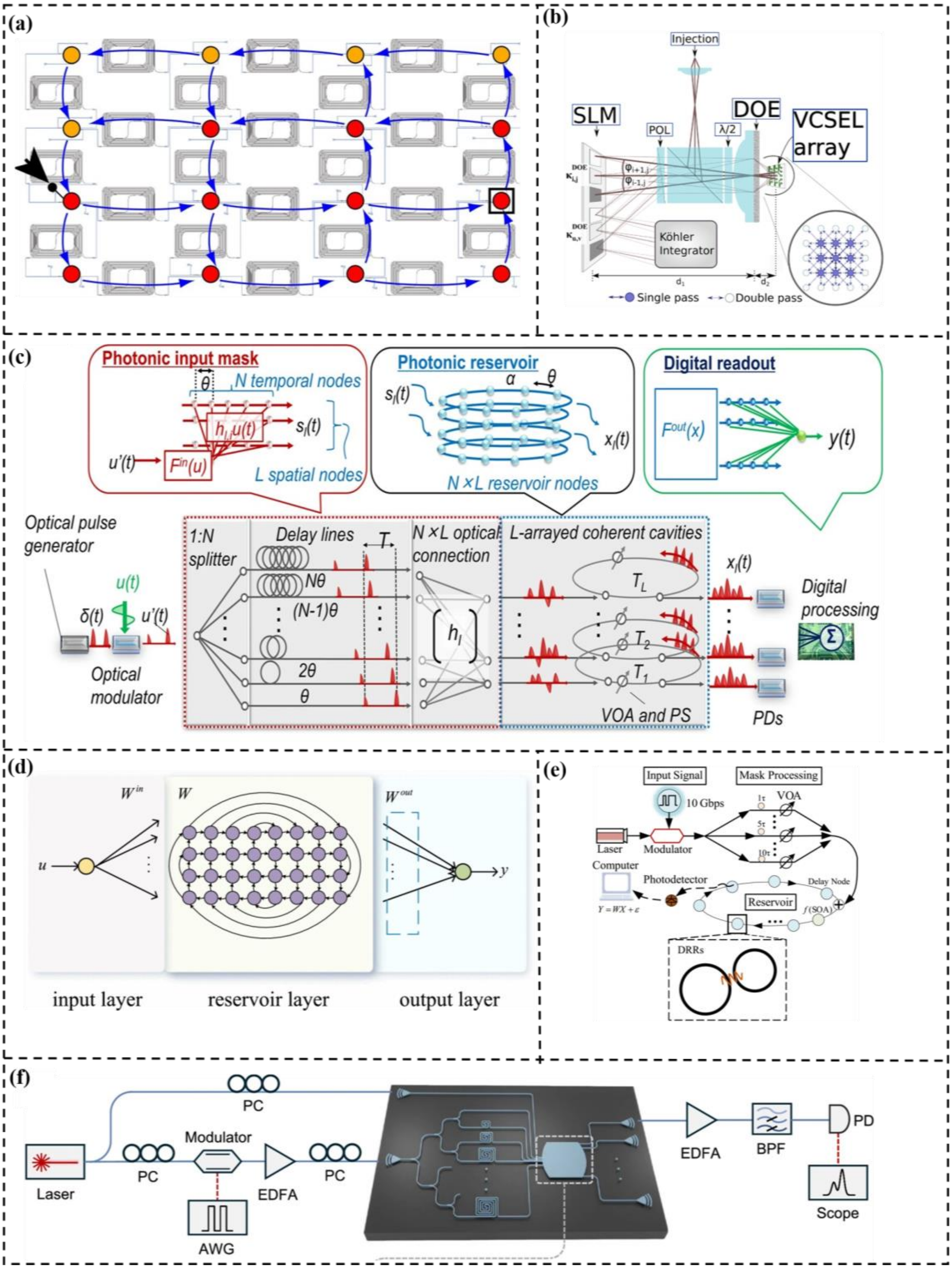

**FIG. 13.** (a) 16-node passive reservoir in 4×4 configuration.[191] (b)The VCSEL array's emission passes a diffractive optical element and is imaged onto a reflective spatial-light modulator.[193] (c) Scalable reservoir computer on coherent photonic processor.[194] (d) An optical channel distortion equalization method based on silicon photonic RC structure with particle swarm optimization (PSO) algorithm.[197] (e) an all-optical reservoir, consisting of integrated double-ring resonators (DRRs) as nodes[200] (f) Conceptual diagram of traditional RC, digital next generation RC (NG-RC), and photonic NG-RC engine.[198]

Since then, there are increasing reports on silicon photonic reservoir chips using integrated double-ring resonators, MZIs, and other components.[194-204] In 2021, M. Nakajima et al. fabricated an on-chip scalable integrated coherent linear photonic reservoir layer, as depicted in Fig. 13(c). [194] In 2022, E. Gooskens et al. utilized 3×3 MMIs as the reservoir layer, enabling 2-bit delayed XOR operations and nonlinear signal equalization.[195] In 2023, L. Pei et al. presented an all-optical equalizer with a rectangular node array based on a 4×8 silicon photonic rectangular reservoir structure (Fig. 13(d)), achieving a three order of magnitude reduction in the bit error rate of 25 Gb/s on-off keying signals.[197] L. Zheng et al. proposed an all-optical reservoir system using cascaded dual-ring resonators as nodes (Fig. 13(e)), which reduced the chip size.[201] In 2024, C. Huang et al. introduced a simplified silicon photonic RC system by using a balanced photodiode (BPD) as the nonlinear device to realize nonlinear mapping of input information (Fig. 13(f)). Experimental results showed that this system can process chaotic time prediction tasks at an information processing speed exceeding 60 GHz.[199]

**TABLE X.** The representative works about space RC.

| Year & Author | Technology& Methods | Task | Results |
|---|---|---|---|
| 2008<br>K. Vandoorne[189] | 25 SOAs | Rectangular and triangular<br>Waveform recognition task | ER=0.02 |
| 2011<br>K. Vandoorne[190] | 9×9 SOAs array | Digital spoken signal recognition | WER=0.06 |
| 2014<br>K. Vandoorne[191] | 16-node passive reservoir | Arbitrary Boolean logic Operations with memory<br>Isolated spoken digit recognition | BER=$10^{-4}$<br>WRE=0.01 |
| 2018<br>A. Katumba[192] | Multimodal Y junctions | 3-bit header recognition | BER=$10^{-3}$ |
| 2021<br>M. Nakajima[194] | Scalable on-chip photonic reservoir | Chaotic time series prediction<br>Handwritten digit recognition | NMSE=0.06<br>TA=91.3% |
| 2023<br>X. Zuo[197] | Silicon photonic RC structure with PSO algorithm | 25 Gb/s on-off keying | BER = $9.15 \times 10^{-5}$ |
| 2023<br>Z. Li[201] | Double-ring resonators | 3-bit and 6-bit packet header recognition tasks | BER=$5\times10^{-4}$ and $9\times10^{-4}$ |
| 2024<br>D. Wang[198] | Delay line and star coupler | Santa Fe time serial prediction<br>Image COVID-19 task | NMSE=0.03<br>Accuracy=92.3% |

**2. Time delay RC**

The time delay RC system employs a nonlinear device combined with a delayed feedback loop structure as nonlinear nodes to form the reservoir layer. Based on the principle of time-division multiplexing, the system samples the feedback loops at equal intervals to obtain the reservoir layer response. This time delay RC scheme effectively reduces hardware costs and the complexity of system implementation. In 2013, D. Brunner et al. constructed time delay RC systems using semiconductor lasers, and performed speech recognition and Santa Fe chaotic time series prediction tasks.[205] Since then, research on time delay RC has increased, primarily focusing on optimizing the system's input layer, reservoir layer, and output layer, as well as expanding its applications, as shown in Table XI.

The input layer design schemes of TD-RC systems demonstrate a developmental trajectory from theoretical simulations to practical systems and from single masks to diverse masks. In 2016, A. Uchida et al. proposed the chaotic signal masking theory, as illustrated in Fig. 14(a),[206] and validated that high-complexity masks can enhance system performance. In 2021, I. Fischer et al. presented a physical generation approach for clockless, sub-nanosecond repeating pattern masks, increasing mask diversity and utilizing them as input-layer mask sequences to optimize reservoir system performance.[207] In 2024, S. Xiang et al. constructed a photonic RC architecture based on a single VCSEL and two cascaded MZMs, achieving simultaneous implementation of the input layer and reservoir layer in the optical domain.[208]

The development of the reservoir layer in time delay RC systems exhibits three major technological trends: parallelization, deep integration, and feedback-free operation. From single-node serial processing to multi-node parallel architectures, from single-layer planar networks to deep loop structures, and from relying on external feedback loops to exploiting device-intrinsic memory, technological innovations have consistently focused on improving information processing density and energy efficiency. In 2015, G. Verschaffelt et al. constructed an all-optical time delay RC system based on a semiconductor ring laser. By leveraging both clockwise and counterclockwise operating modes, they enabled parallel processing of Santa Fe sequence prediction and nonlinear channel equalization tasks.[209] As shown in Fig. 14(b), J. Bueno et al. experimentally validated the photonic time delay RC system in 2017. They achieved a nonlinear prediction accuracy of NMSE = 0.056 for the Mackey-Glass time series task.[210] In 2018, J. Vatin et al. developed a time delay RC system employing a VCSEL with self-feedback as the nonlinear device.[211] G. Q. Xia et al. utilized dual optical feedback and optical injection to simultaneously perturb the nonlinear dynamics output of a semiconductor laser, as shown in Fig. 14(d). Their work demonstrated that the dual-optical-feedback RC system outperforms single-optical-feedback systems.[212] In 2019, S. Xiang et al. experimentally verified that the two polarization modes of a VCSEL could independently handle parallel tasks.[213] They further implemented a four-channel photonic time delay RC system using the two polarization modes of two mutually coupled VCSELs, as illustrated in Fig. 14(c).[214] In 2021, they further proposed a RC system based on a semiconductor nano-laser with dual phasic conjugate feedback.[215] In 2022, C. Wang et al. theoretically validated a reservoir system using a FP quantum dot laser with multiple longitudinal modes. The system leveraged different longitudinal modes of the laser as physical neurons to process multi-channel input signals in parallel. For memory capacity, time series prediction, nonlinear channel equalization, and speech recognition tasks, the parallel system exhibited faster operation and improved performance on multiple benchmark tasks compared to single-channel time delay RC systems with the same number of nodes.[216] In 2023, they further introduced a deep photonic recurrent reservoir system and applied it to a real-world optical signal equalization fiber system. For a four-layer deep reservoir system, the bit error rate was minimized to the order of $10^{-3}$.[217] In 2024, S. Xiang et al. proposed an integrated photonic time delay RC system based on a four channel DFB layers array with optical feedback and injection for pattern recognition tasks. For the Iris flower classification task, the system achieved a 100% recognition accuracy, significantly outperforming single-channel reservoir systems.[218] They also experimentally validated a time delay RC system using a FP laser, as shown in Fig. 14(e).[219] In 2024, L. Y. Zhang et al. introduced a parallel deep tree photonic reservoir computing architecture, which integrates deep reservoir layers with a hierarchical tree structure.[220] They also advanced the field by incorporating imperfect physical models into photonic RC. Importantly, it exhibits strong robustness against inherent imperfections in physical models.[221] In 2024, N. Li et al. proposed a 4-layer deep recurrent time-delay RC system using four injection-locked DFB lasers, as illustrated in Fig. 14(g).[222] By introducing additional time delays in the residual structure, the system integrated 960 interconnected neurons (240 neurons per layer), drastically increasing computational density. This design showcases how residual time delays can enhance memory capacity and nonlinear processing capabilities in deep reservoir architectures, marking a significant milestone in hardware scalability for photonic RC.

**TABLE XI.** Some representative works about time delay RC.

| Year & Author | Technology & Methods | Task | Results |
|---|---|---|---|
| 2016 J. Nakayama[206] | A chaos mask signal | Santa Fe series prediction | NMSE=0.08 |
| 2024 X. X. Guo[208] | Both the input layer and reservoir in optical domain | Santa Fe series Prediction<br>Handwritten digit recognition | NMSE=0.0456<br>WER=0.0667 |
| 2015 R. M Nguimdo[209] | A single-longitudinal mode semiconductor ring laser | Santa Fe series prediction Nonlinear channel equalization | NMSE=0.03<br>SER=$10^{-3}$ |
| 2017 J. Bueno[210] | Experiment of photonic RC | Mackey Glass prediction | NMSE=0.056 |
| 2018 J. Vatin[211] | Photonic RC with VCSEL | Santa Fe series prediction<br>Channel equalization | NMSE=$10^{-3}$<br>SER=$10^{-5}$ |
| 2018 G. Q. Xia[212] | SL with double feedback | Santa Fe series prediction | NMSE=0.03 |
| 2019 X. X. Guo[213] | Parallel time delay RC based on Multi-polarization of VCSEL | Santa-Fe series prediction<br>Waveform recognition | NMSE=0.0266<br>NMSE=0.0167 |
| 2022 J. Y. Tang[216] | Parallel time delay RC with FP-QD laser | Santa Fe series prediction<br>Nonlinear channel equalization<br>Spoken digit recognition | NMSE=0.024<br>SER=0.0179<br>WER=1.44% |
| 2023 Y. W. Shen[217] | Deep photonic RC | Nonlinear fiber compensation | BER=$10^{-3}$ |
| 2024 X. X. Guo[219] | Parallel RC with FP laser | Santa Fe series prediction<br>Nonlinear channel equalization | NMSE=0.013<br>BER=0.001 |
| 2024 X. X. Guo[218] | An integrated photonic time delay RC based on F-DFBs | Iris classification task. | BER=0 |
| 2024 C. D. Zhou[222] | Deep residual time delay RC | Time series prediction<br>Nonlinear channel equalization | NMSE=0.0169<br>SER=$2.6\times10^{-3}$ |
| 2024 L. Zhang[221] | Forecasting RC based on VCSELs with knowledge | Santa Fe series prediction | NMSE=0.0683 |
| 2024 R. Zhang[220] | The parallel deep tree RC | Santa Fe series prediction<br>Nonlinear channel equalization | NMSE= $5\times10^{-3}$<br>SER=$10^{-3}$ |
| 2025 Z. W. Dai[224] | A photonic spiking RC based on DFB-SA laser | Iris classification task | BER=0.044 |
| 2025 C. D. Zhou[223] | Photonic RC with quasi-convolution coding | Time-series prediction<br>Nonlinear channel equalization<br>Memory capacity | NMSE=0.0054<br>SER=0.001<br>MC=3.8818 |
| 2022 J. Y. Jin[225] | An adaptive photonic RC by Kalman filter | Santa Fe series prediction<br>Nonlinear channel equalization | NMSE=0.0028<br>SER=$10^{-4}$ |
| 2023 G. O. Danilenko[226] | SL with tunable bandpass filtered feedback | Memory capacity<br>Computation ability | MC=1<br>CA=0.6 |

To eliminate the reliance on feedback loops and address their drawbacks in on-chip integration, such as hardware implementation challenges and low integration density, feedback-free reservoir architectures have been explored in recent years. These efforts aim to fulfill memory requirements through advanced pre/post-processing algorithms or by leveraging the intrinsic properties of physical devices. In 2025, N. Li et al proposed an efficient convolution-like encoding scheme based on convolutional coding principles. This scheme encodes input information in the input layer to incorporate past information before injecting it into the reservoir layer, enabling the construction of a feedback-free RC system compatible with on-chip integration, as shown in Fig. 14(f).[223] S. Xiang et al. reported a photonic spiking RC system using a DFB-SA laser.[224] Compared with traditional continuous-signal processing, spiking processing requires less energy, contributing to reduce system power consumption. Notably, it omits delay feedback loops by using only DFB-SA lasers as the hardware structure. These achievements indicated that the reservoir layer shifts from single-node dynamic regulation to multi-mode collaborative computing, from linear superposition architectures to hierarchical network designs, and from relying on feedback delays to mining intrinsic system memory.

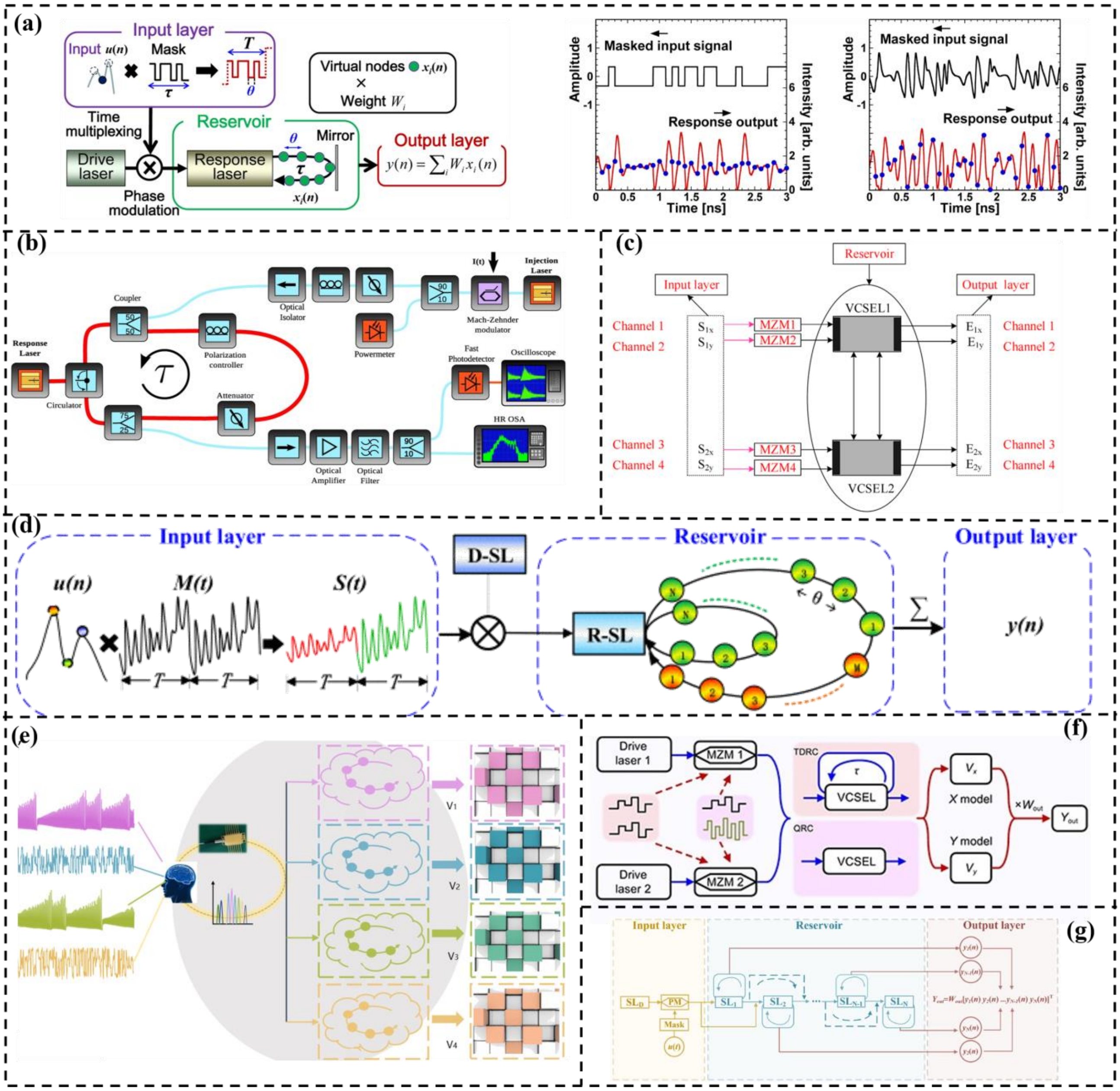

**FIG. 14.** (a) A chaos mask signal in photonic time delay reservoir computing.[206] (b) The experimental set up of photonic time delay reservoir computing.[210] (c) The Four-channels RC based on MDC-VCSELs.[214] (d) The photonic reservoir computing based on VCSEL with double feedback delay.[212] (e) Parallel reservoir computing based on FP laser.[219] (f) QRC which could be enabled to deal with time-related tasks or sequential data without the implementation of FL.[223] (g) The Schematic representation of DR-TDRC.[222]

The optimization of output layer algorithms for time delay RC systems has also attracted a lot of attention. In 2022, N. Jiang et al. proposed the Kalman-filter RC architecture, which recursively updates readout weights using a state-space model. [225] In 2023, G. O. Danilenko et al. demonstrated that filtering the output light of a semiconductor laser and feeding it back into the laser could flatten the eigenvalue spectrum, optimizing the reservoir's memory capacity. [226] Y. Tanaka et al. introduced a self-organizing multi-readout structure in reservoir systems to address limitations in traditional reservoir training and catastrophic forgetting. By allocating training data to multiple readouts and using self-organizing maps for data classification, this method enables each readout to specialize in specific data, improving overall training performance. The model excels in continuous learning tasks and sound recognition.[227]

The application expansion of time delay RC systems has seen a significant transition from benchmark task validation to real-world industry empowerment. In 2021, S. Sackesyn et al. experimentally demonstrated that a waveguide-based photonic reservoir chip can compensate for linear and nonlinear impairments in optical fibers.[228] S. Xiang et al. utilized a time delay RC system based on VCSEL in 2024 to achieve short-term prediction of an optical chaotic system with a three VCSELs coupled network.[229] They further proposed extreme events generated by microcavity semiconductor lasers, using a photonic time delay RC system for prediction.[230]

The prospects of on-chip silicon photonic integrated reservoir chips and time delay RC systems are extremely promising. Owing to their tailored application scenarios and increasingly sophisticated network architectures, photonic time delay RC systems exhibit profound potential for expanding practical applications. From a hardware perspective, the continuous progress in silicon-based photonic integration technology enables the realization of all-optical integrated photonic RC chips. At the algorithmic level, enabling adaptive updates of reservoir output weights to reduce training complexity and shorten convergence time would enhance the system's adaptability to complex application scenarios.

## F. LARGE-SCALE PHOTONIC COMPUTING PROCESSORS

Large-scale photonic computing processors combine the advantages of photonic computation with high computational density to meet the growing computational demands of AI. In 2025, S. R. Ahmed et al. proposed a photonic AI processor that executes advanced AI models, including ResNet3 and BERT20,21, and the Atari deep RL algorithm. This photonic AI processor integrated four 128×128 photonic tensor cores based on MZI networks (Fig. 15(a)). Through 3D heterogeneous integration of vertically stacked photonic cores with 12 nm CMOS digital control chips, this photonic AI processor achieves a computational efficiency of 65.5 TOPS at 78W electronic power.[231] In 2025, S. Hua et al. developed a 64×64 photonic matrix accelerator consisting of more than 16,000 integrated photonic components (Fig. 15(b)). Through 2.5D advanced packaging for electronic-photonic co-design, this photonic accelerator can perform matrix multiply–accumulate operations with high speed up to 1 GHz frequency and low latency as small as 3 ns.[232] Table XII summarizes the main implementation methods and key contributions of these two architectures. However, persistent bottlenecks in optoelectronic interfaces, thermal stability, and manufacturing uniformity demand urgent resolution to transition from lab-scale prototypes to industrial deployment.

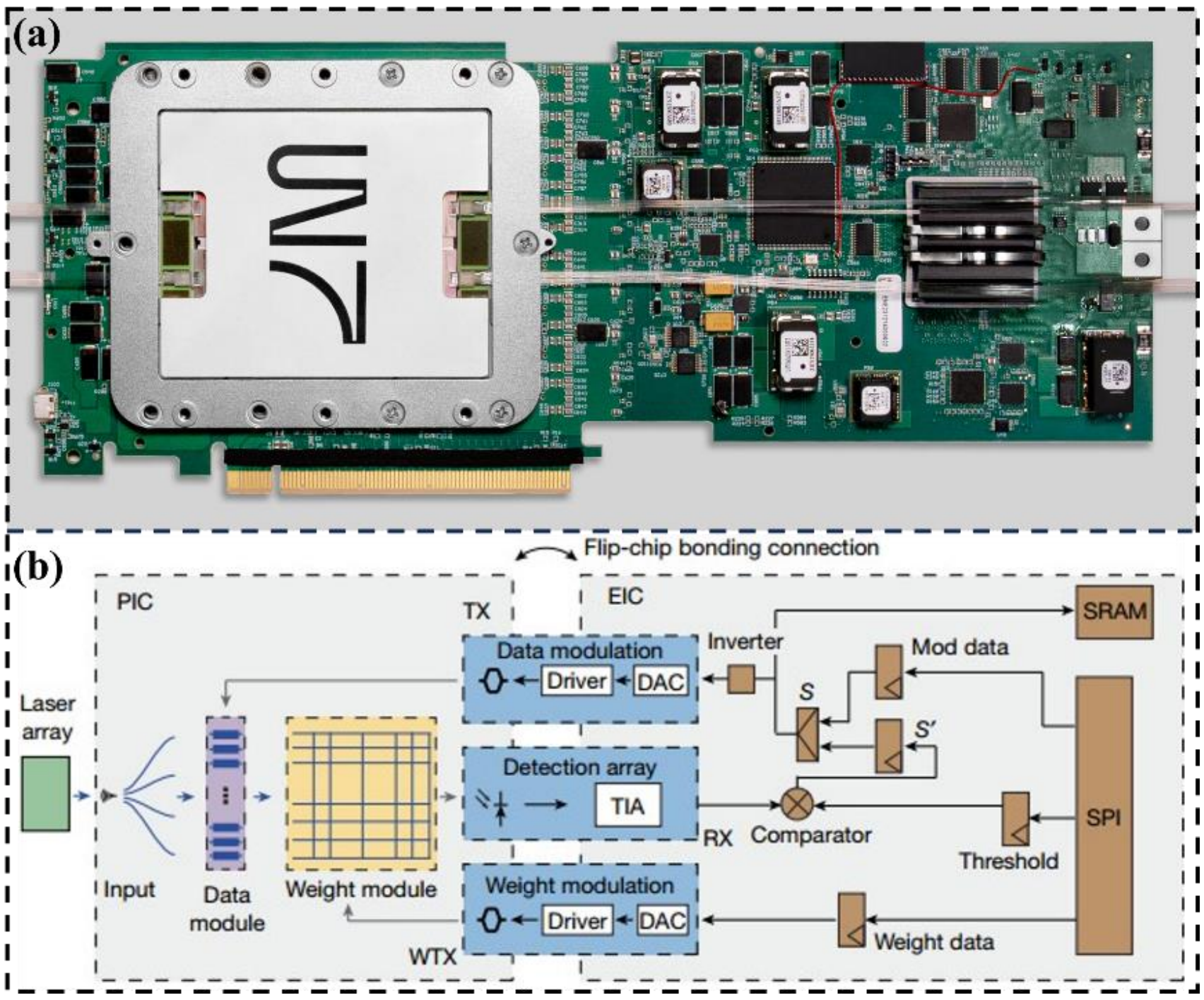

**FIG. 15.** (a) Universal photonic AI acceleration.[231] (b) An integrated large-scale photonic accelerator with ultralow latency.[232]

**TABLE XII.** Large-scale photonic computing processors.

| Year & Author | Technology Type | Implementation Method | Key Contribution |
|---|---|---|---|
| 2025 Ahmed[231] | 3D packaged photonic accelerator | 4 photonic tensor core chip based on 128×128MZI and 2 digital control chip | Energy efficiency of 65.5 TOPS/W; ResNet (CIFAR10:86.4%, ImageNet 79.3-79.7%), BERT-tiny (IMDB: 83.2%, QuAD:12), Atari(Beamrider, Pacman) |
| 2025 S. Hua[232] | 2.5D packaged photonic accelerator | 64 modulator array+64×64 MZI weigth array+64 detector array | Low latency with 3 ns; Energy efficiency of 2.38 TOPS/W; NP-hard Ising problem |

## G. OTHERS

Except for the above mentioned architectures, various novel photonic computing architectures have been reported in recent years. Table XIII shows the main implementation methods and key contributions of other photonic neuromorphic architectures.

In 2023, Z. Chen et al. demonstrated PNN inference using a coherent VCSEL array, where matrix multiplication was directly performed through optical field interference (Fig. 16(a)).[233] S. Afifi et al. proposed a non-coherent silicon photonic Transformer architecture which leverages MZI arrays to represent attention kernels.[234] In 2024, H. Zhu et al. proposed photonic Transformer accelerator by employing tunable MRRs to realize reconfigurable weight mapping (Fig. 16(b)).[235] Hsueh et al. developed a hybrid optoelectronic multi-head attention chip based on optical frequency combs, achieved 27.9 fJ/MAC energy efficiency.[236] H. Sha et al. proposed an optical Transformer based on MZI cascading grids that achieved 96.12% accuracy on the MNIST dataset.[237]

Besides, photonic decision-making systems have demonstrated ultrafast response characteristics in real-time optimization tasks. [238-241] In 2020, Y. Ma et al. implemented a semiconductor laser network with Sagnac ring phase modulation to solve the multi-arm slot machine problem and enhance the security through chaotic time delay signature hiding.[238] In 2023, B. Shen et al. used an optical frequency comb to generate a high-dimensional chaotic entropy source to achieve high accuracy in a multi-channel parallel determination task (Fig. 16(c)).[240]

Photonic associative learning aims to emulate synaptic plasticity in biological systems, enabling efficient pattern association through fully optical means.[242-242]In 1990, M. Ishikawa et al. proposed a photonic associative memory system based on a microchannel spatial light modulator.[242] In 2020, S. Wang et al. implemented synaptic weight modulation by using VCSOA-based STDP rules for associative learning and pattern recall tasks.[243] In 2022, J. Y. S. Tan proposed a photonic Pavlovian learning network based on PCM to implement classical conditioning, at an ultralow power consumption of just 1.8 nJ (Fig. 16(d)).[244,245] In 2024, D. Zheng et al. demonstrated a fully functional associative network based on DFB-SA lasers and SOA, enables associative learning, forgetting functionality, and pattern recall tasks. [51]

**TABLE XIII.** Other photonic neuromorphic architectures.

| Year & Author | Technology Type | Implementation Method | Key Contribution |
|---|---|---|---|
| 2023 Z. Chen[233] | VCSEL-based PNN | 5×5 VCSEL arrays | Energy efficiency: 7 fJ/OP; Compute density: 6TOPS/mm$^2$ MNIST accuracy: (93.1±2.0)% |
| 2024 H. Zhu[235] | Transformer | Dynamic photonic tensor core | >2.6× energy reduction and >12× latency reduction(Compared to prior photonic accelerators) |
| 2023 B. Shen[240] | Photonic decision making system | Micro-ring optical frequency combs | 256-armed bandit problems with correct decision ratio >95% |
| 2022 Z. Cheng[245] | Photonic associative learning | PCMs | Compute density: 118 TOPS/mm$^2$ |
| 2024 J. Ouyang[246] | Photonic solver | 16-channel FFT-mesh MZI array | Computing speed: 1.66 TFLOPS Compute density: 44.4 GFLOP/mm$^2$ Energy efficiency : 0.458 nJ/FLOP |
| 2024 X. Li[250] | Photonic RL | MZI and OCTOPUS hybrid architecture | 56% improvement in efficiency |
| 2025 B.Wu[252] | Photonic Ising machine | Optoelectronic coupled oscillators | The spin evolution time: 150 ns; Roundtrip time :1.71ns |
| 2025 B.Wu[253] | Optical recurrent accelerator | Optical hidden Markov model and an optical recurrent neural network | Two-class classification: 95%; Eight-class classification: 87.7% |

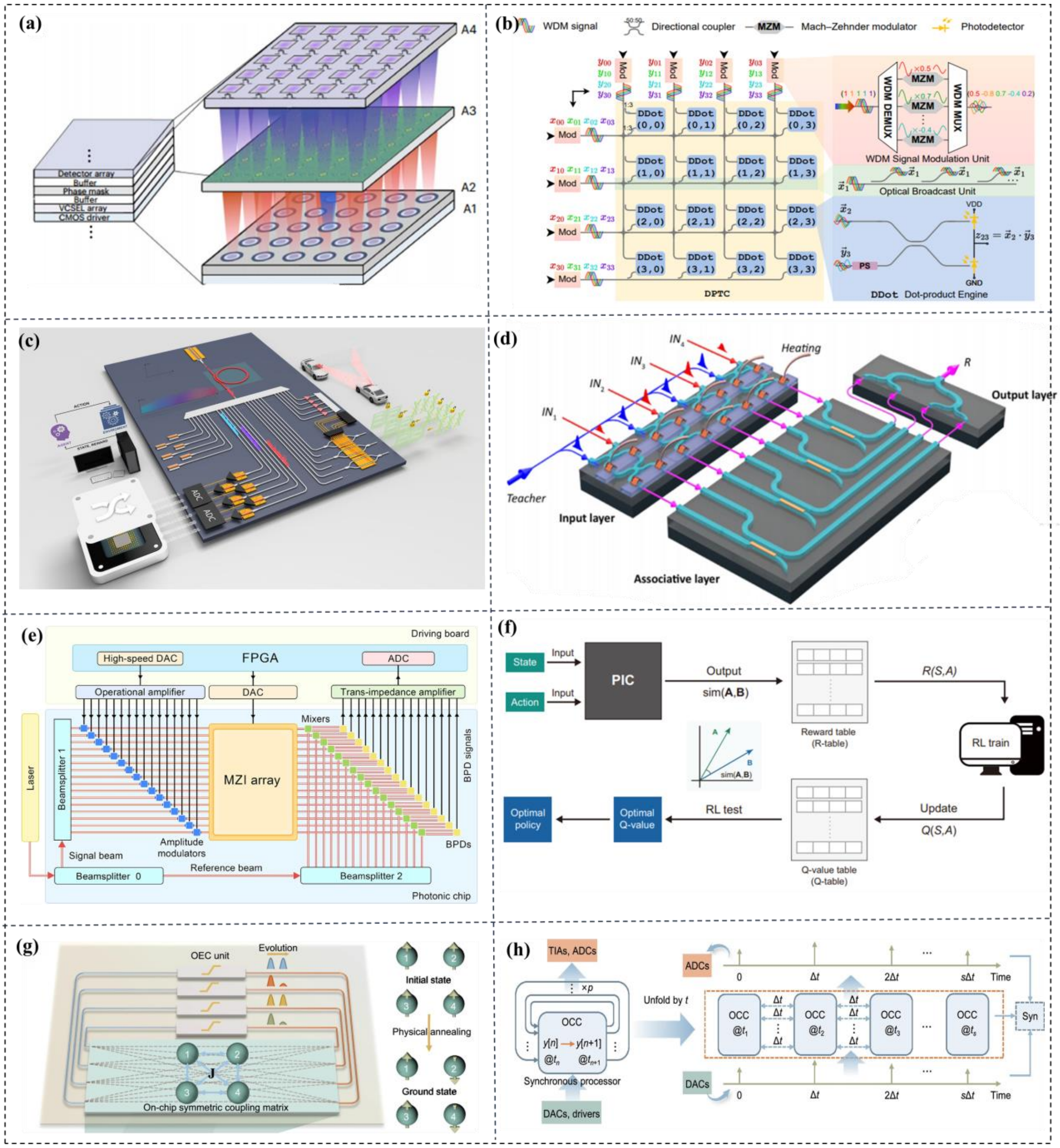

**FIG. 16.** Alternative architectures for PNNs. (a) VCSEL-driven PNN.[233] (b) Photonic transformer accelerators.[235] (c) Photonic decision-making system.[240] (d) Associative learning PNN.[245] (e) Photonic solver based on MZI array.[246] (f) Photonic RL processing architecture.[250] (g) A monolithically integrated optical Ising machine.[252] (h) A monolithically integrated asynchronous optical recurrent accelerator.[253]

In 2025, J. Ouyang et al. proposed a 16-channel silicon-based photonic solver for NP-hard combinatorial optimization problems, using a FFT-mesh MZI array for arbitrary real-valued matrix multiplication (Fig. 16(e)).[246]

Photonic reinforcement learning (RL) systems offer unique advantages by enabling policy updates at the speed of light.[247-251] In 2017, M. Naruse et al. leveraged chaotic laser dynamics to generate stochastic policies that balanced exploration and exploitation, achieving decision latencies as low as 1 ns.[247] In 2023, Z. Yang et al. mapped deep RL architectures onto optically biased neural networks, enabling real-time action-value computation via MZIs mesh, resulting in a 2.5× speedup in convergence during path planning tasks.[249,250] In 2024, X. Li et al. proposed an optical computing of dot-product units (OCTOPUS) based on MZI meshes for high-efficiency RL. Their framework implemented Q-learning to optimize perovskite material synthesis and cliff-walking tasks, achieving a 56% improvement in algorithmic efficiency through photonic simulation of agent-environment interactions (Fig. 16(f)).[251]

Photonic Ising machines have garnered significant attention due to their immense potential in solving combinatorial optimization problems. In 2025, J. Dong et al. proposed a monolithically integrated optical Ising computing scheme based on optoelectronic coupling (OEC) oscillators, achieving the demonstration of an on-chip four-spin Ising solver operating without external electrical assistance. This system incorporates MZIs and OEC nonlinear units, as shown in Fig.16(g). [252] They also proposed an asynchronous computing paradigm for on-chip optical recurrent accelerators based on wavelength encoding, effectively mitigating synchronization challenges (Fig. 16(h)).[253] To demonstrate the flexibility and efficacy of this asynchronous paradigm, they presented two monolithically integrated recurrent models, an optical hidden Markov model and an optical recurrent neural network.

Photonic computing is rapidly evolving from discrete components toward large-scale integrated systems. VCSEL arrays offer high-density light sources, while silicon-based MZI and MRR networks enable programmable optical information processing. The emerging of different photonic network architectures may further pave the way of integrated photonic neuromorphic computing.

# IV. TRAINING METHODS OF PNNS

Training is a crucial step of PNNs, determining the performance of the entire system. Currently, the training methods for PNNs can be divided into two categories: hardware-aware Ex-situ training and on-chip In-situ training. Ex-situ training refers to training conducted with the assistance of a digital computer. It utilizes various hardware-aware training techniques to capture and model the hardware behavior during the training phase, considering various non-ideal effects. In-situ training aims to perform training directly on the chip. This approach maximizes accuracy by directly incorporating the actual behavior of the photonic hardware and on-chip non-idealities into the training process.

**TABLE XIV.** Hardware-aware Ex-situ training methods for PNNs.

| Year & Author | Technology Type | Implementation Method | Key Contribution |
|---|---|---|---|
| 2020<br>J. Gu[254] | Noise-aware quantization | Quantitative training; Group Lasso regularization. | >80% on 3-bit MNIST |
| 2022<br>G.Mourgias-Alexandris[124] | Noise-resilient and high-speed deep learning | Training with Gaussian noise (σ=0.4). | 10 GMAC/s/axon;<br>98% on MNIST |
| 2022<br>M. Kirtas[255] | Quantization-aware training | Quantified error injection training. | >90% on 4-bit MNIST;<br>40% lower MSE on 4-bit FI-2020 |
| 2022<br>M. Kirtas[256] | Normalized post-training quantization | Normalization quantization of Gaussian distribution. | 76.77% on 4-bit CIFAR10 |
| 2022<br>C. Feng[126] | Photonic–electronic neural chip | Hardware-aware training framework. | 7 × component reduction;<br>3.3 × reduction in footprint;<br>5.5 × latency improvement |
| 2022<br>R. Shao[257] | Imprecise components | Gradient-Genetic Hybrid Optimization. | 90.8% on MNIST |
| 2022<br>J. Spall[258] | Hybrid training | Optical system calculation error, digital system calculation gradient | Optical linear: 88.0% ,<br>Hybrid optoelectronic: 92.7% ,<br>Complex optical: 92.6% on MNIST |
| 2024<br>Y. Zhan[259] | Physics-aware analytic-gradient training | Pre-trained differentiable DNN parses the gradient | 30 × faster training and 4.5 × lower energy vs. in-situ |
| 2024<br>T. Xu[130] | Hardware-aware training and pruning | Loss function regularization term incorporation | 95.0% on MNIST;<br>10 × tuning power reduction |
| 2025<br>Y. Wang[260] | Asymmetrical training | Extra forward passes in a digital parallel model | 95.8% on MNIST;<br>87.5% on F-MNIST;<br>85.6% on K-MNIST |

TABLE XIV shows the development overview of Ex-situ training. In 2020, J. Gu et al. proposed a noise-aware quantization scheme to enable PNNs to adapt to low-precision controls and non-ideal environments with phase shifter noises.[254] This scheme achieves low-precision voltage control of PNNs through coarse gradient approximation and unitary projection, and mitigates the corresponding accuracy degradation. In 2022, G. Mourgias-Alexandris et al. combined a noise-tolerant linear neuron architectural scheme with noise-aware training methods based on a coherent silicon integrated circuit, achieving a high-performance photonic deep learning model.[124] This model offers on-chip compute rates per axon that are 6 orders of magnitude higher and classification accuracy that is >7% higher. M. Kirtas et al. proposed a quantization-aware training framework for training photonic deep-learning models with limited precision.[255] This framework can effectively reduce the precision requirements of photonic deep-learning models. They also proposed a Gaussian distribution-aware normalized post-training quantization method to address this issue.[256] R. Shao et al. proposed a two-step ex-situ training scheme.[257] First, the phase configuration is rapidly optimized under ideal conditions through stochastic gradient descent. Then, in combination with the genetic algorithm, the optimal configuration is found while considering the parameter imprecisions in the MZIs. J. Spall et al. proposed a hybrid training framework.[258] The forward propagation is computed in real-time through the optical system, while the backward propagation calculates the error gradient digitally to update the weight matrix. In 2024, Y. Zhan et al. proposed a new hybrid training framework based on the physics-aware analytic-gradient training method to address the training challenges of non-differentiability of PNN chips.[259] In 2025, Y. Wang et al. proposed a novel asymmetrical training method to address the training challenges of encapsulated deep PNNs.[260] This method combines the forward propagation of a digital model with the pseudo-gradient update of a physical system and completes the training relying only on the information of the output layer.

**TABLE XV.** In-situ training methods on the PNNs chips.

| Year & Author | Technology Type | Implementation Method | Key Contribution |
|---|---|---|---|
| 2018 T. W. Hughes[261] | In Situ training | Adjoint variable method | Implement the XOR function |
| 2020 T. Zhou[262] | In Situ optical backpropagation training | Light reciprocity and phase conjunction | 92.19% on MNIST; 2-3 orders lower gradient error vs. elec-training |
| 2020 H. Zhou[263] | Self-configuring and reconfigurable | Modified gradient descent algorithm | Realize a 3×3 optical switch |
| 2021 H. Zhang[264] | On-chip training | Genetic algorithm and chip parameters optimization | Realize the 6×6 cross switch; 93.3% on Iris |
| 2022 M. J. Filipovich[265] | On-chip training | Direct feedback alignment algorithm | 20 TOPS; 96.33% on MNIST |
| 2023 W. Zhang[266] | Online training and pruning | Particle swarm optimization and power-pruning regularization | 100% on Iris; 96.9% on MNIST |
| 2023 S. Pai[267] | In Situ backpropagation | Bidirectional light propagation | 96% on circle dataset; 98% on moon dataset; 97.2% on MNIST |
| 2024 Y. Wan[268] | On-chip training | SPGD algorithm and phase shifters optimization | 6×6 optical switching |
| 2024 Z. Xue[178] | Fully forward mode training | Gradient descent based on forward-propagated fields | 92.5% on F-MNIST |
| 2025 J. Spall[269] | End-to-end optical backpropagation Training | Saturable absorbers for activation | 100% on Rings dataset; 98.5% on XOR dataset; 99.0% on Arches dataset |

TABLE XV shows the development overview of In-situ training. In 2018, T. W. Hughes et al. theoretically proposed the adjoint variable method to derive the photonic analogue of the backpropagation algorithm,[261] enabling highly efficient in-situ training of PNNs. This method implements on-chip backpropagation by interfering the adjoint field with the forward field, and directly measures the gradient information as an in-situ intensity measurement. In 2020, T. Zhou et al. proposed an optical error backpropagation algorithm for the in-situ training of linear and nonlinear DONNs.[262] H. Zhou et al. proposed a self-configuring and reconfigurable optical signal processor based on silicon photonics, which is capable of achieving fully automatic and multifunctional photonic signal processing.[263] In 2021, H. Zhang et al. proposed an on-chip training method based on the genetic algorithm for the efficient optimization of programmable PNNs.[264] In 2022, M. J. Filipovich et al. proposed a parallel and efficient training architecture for deep neural networks using the direct feedback alignment algorithm.[265] This algorithm directly propagates the error of the output layer to each hidden layer through a fixed random feedback matrix, avoiding the inter-layer dependence of backpropagation and significantly enhancing parallelism. In 2023, W. Zhang et al. proposed an online training and power optimization method for PNNs.[266] They used a gradient-free online training framework based on particle swarm optimization and incorporated an additional regularization term into the loss function to account for power consumption, thus achieving pruning of PNNs. In addition, S. Pai et al. constructed a three-layer, four-port silicon PNN chip with programmable phase shifters and optical power monitoring.[267] They experimentally demonstrated the in-situ backpropagation by interfering the forward- and backward-propagating light to measure the backpropagated gradients of the phase-shifter voltages. In 2024, Y. Wan et al. proposed an efficient training method for on-chip optical processors based on the stochastic parallel gradient descent (SPGD) algorithm.[268] Z. Xue et al. proposed a fully forward mode (FFM) learning for the efficient training of PNNs.[178] By leveraging the spatial symmetry of light propagation and Lorentz reciprocity, it enables end-to-end training directly within the physical system. In 2025, J. Spall et al. proposed an all-optical backpropagation method, achieving end-to-end optical computing for PNNs.[269] The nonlinear saturation and linear transmission characteristics of saturable absorbers are employed to approximate the derivative of the activation function during backward propagation.

At present, both ex-situ training and in-situ training methods have achieved remarkable results in terms of model accuracy and controllability. Ex-situ training has focused on to decrease the effect of noise and low-bit quantization through the integration of multiple algorithms and hardware-aware strategies. In situ training has advanced from the introduction of basic algorithms to the exploration of feedback mechanisms, online training, and all-optical methods. These research advancements continuously enrich the training methodologies for PNNs. In the future, the development of training methods for PNNs requires comprehensive consideration of multiple aspects, including algorithm efficiency, model robustness, and hardware adaptability.

# V. CHALLENGES

Although integrated photonic neuromorphic computing has developed rapidly, it still faces key challenges in aspects such as low-threshold photonic nonlinear computing, the scale of integrated chips, the compatibility of opto-electronic collaboration and software-hardware collaboration, and unclear application scenarios.

Firstly, the low-threshold photonic nonlinear computing is still in its infancy. In complex computing tasks, nonlinear operations are the core for realizing functions such as activation functions of neural networks and pattern recognition. However, traditional optoelectronic devices are mostly based on the principles of linear optics, and there are technical bottlenecks in achieving efficient and controllable nonlinear optical effects. Existing optical nonlinearities have problems such as low modulation efficiency, excessively high optical power, and poor compatibility with existing semiconductor integration processes, which limit the ability of integrated photonic chips to solve complex AI tasks.

Secondly, the development of large-scale photonic integration and packaging is hindered. On the one hand, the manufacturing process of optoelectronic devices is complex. Key components such as light sources, modulators, and detectors often rely on different material systems to achieve their optimal performances, and there are significant differences in manufacturing processes, making it difficult to achieve monolithic high-density integration. On the other hand, with the increase in integration density, the problem of crosstalk of optical signals within the chip intensifies. The coupling between waveguides leads to signal distortion, and the performance uniformity between different chips is poor, seriously affecting the computing accuracy and stability. In addition, the current advanced packaging technology for photonic integrated chips is not yet mature, which greatly restricts the large-scale industrial application of integrated photonic neuromorphic computing chips.

Thirdly, the optoelectronic collaboration mechanism urgently needs to be improved. Integrated photonic neuromorphic computing requires the collaboration of photonic computing and electronic computing to take full use of their respective advantages. However, in reality, there are problems such as signal loss, delay, and noise interference in optoelectronic conversion, which reduce the overall performance of the system. The parallel and high-speed characteristics of photonic computing and the logical control advantages of electronic computing are difficult to integrate at the architectural level, and there is a lack of a mature optoelectronic collaboration strategy. At the same time, the design and research and development costs of the optoelectronic collaboration system are high, and the cycle is long.

Fourthly, the software-hardware collaboration adaptability is poor. Existing programming languages and development tools are mostly oriented towards traditional electronic computing, making it difficult to fully unleash the multi-dimensional and highly parallel processing advantages of integrated photonic neuromorphic computing hardware. In addition, there is a huge gap between the scale of neural networks and the photonic computing chips scale, and the software-hardware adaptability is insufficient. Traditional algorithms are difficult to achieve optimal performance on the photonic hardware. Therefore, it is urgent to develop dedicated software tools and algorithms that are suitable for the hardware characteristics of integrated photonic neuromorphic computing chips to achieve better software-hardware collaboration.

Fifthly, the advantageous application scenarios are not clear. In the traditional computing field, integrated photonic neuromorphic computing is difficult to break through the balance between cost and performance. In the fields of AI, although there is great theoretical potential, restricted by technical maturity and cost, it is difficult to replace the existing mature electronic computing system. Therefore, it is urgent to explore advantageous application scenarios that can give full play to the unique advantages of integrated optical technology to promote the practical application process of integrated photonic neuromorphic computing.

# VI. OUTLOOK

Photonics neuromorphic computing, with its inherent advantages of high parallelism, low latency, and low power consumption, has gradually emerged as a key technological solution to break through the computing power bottleneck in the post-Moore era. Remarkable progress has been achieved in integrated chips, architectures, and algorithms, demonstrating enormous development potential and application prospects. In the field of photonic linear computing devices, components such as MRR, MZI, and PCM provide crucial building blocks for the construction of photonic computing chips, leveraging their unique optical properties and physical mechanisms.[270] In the fields of photonic nonlinear computing devices, photonic nonlinear activation and photonic spiking neurons have successfully emulated the nonlinear behaviors and spike processing mechanisms of biological neurons, enabling more efficient and intelligent neuromorphic computing. Moreover, continuous optimization of PNN training algorithms have enabled their adaptation to the characteristics of photonic computing, achieving application in fields such as image recognition and natural language processing. In the future, photonics neuromorphic computing is expected to achieve breakthroughs in multiple aspects, as shown in Fig.17.

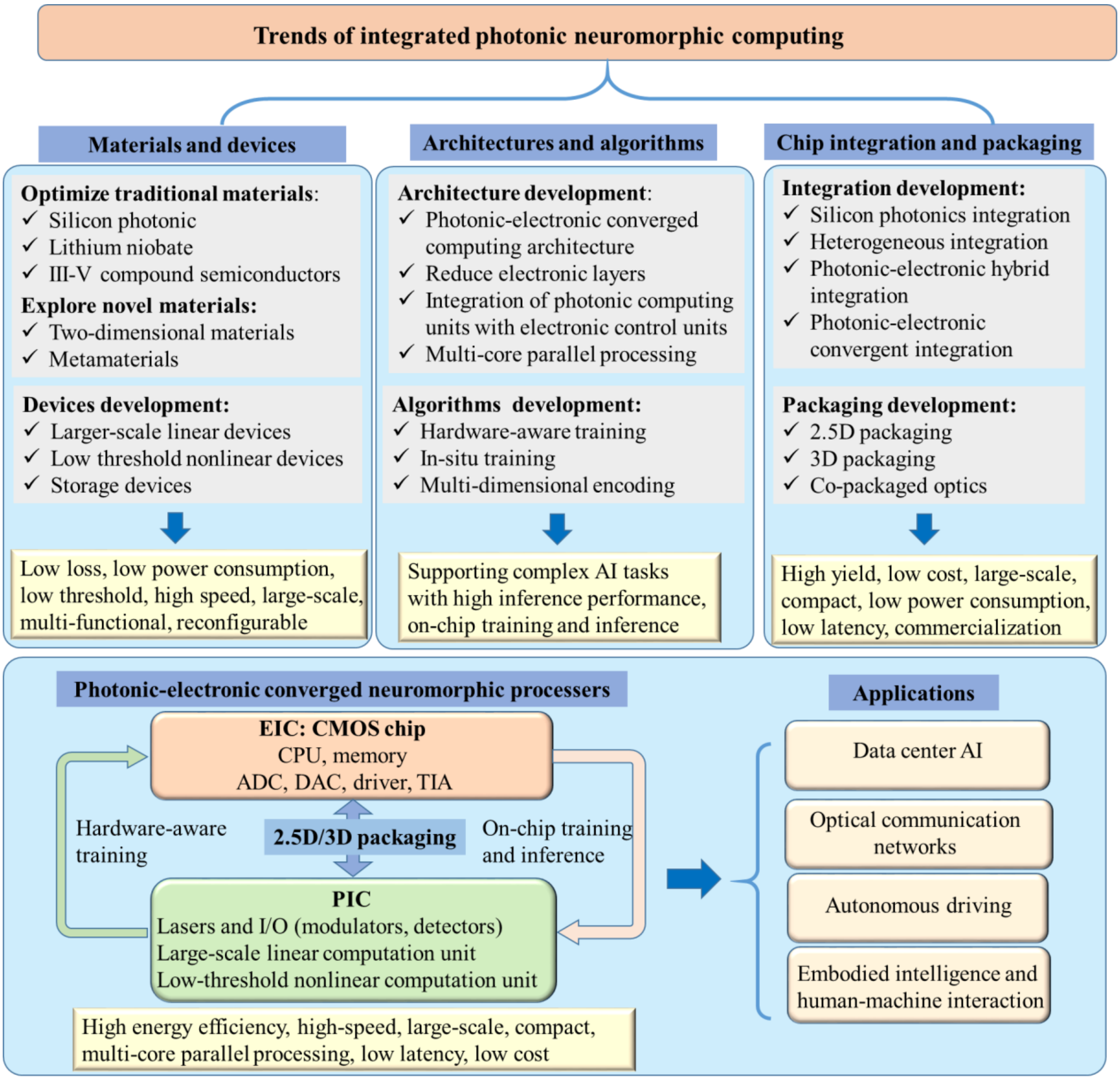

**FIG. 17.** Future development trends of integrated photonic neuromorphic computing.

**(1) Development of novel photonic neuromorphic computing materials and devices**

Materials serve as the cornerstone of photonic neuromorphic computing. Currently, although traditional silicon-based optoelectronic materials have achieved certain progress in integration density and performance, the development of novel materials is crucial to meet the stringent requirements of higher-performance photonic computing systems for high data rates, low losses, and low thresholds.[271,272] On one hand, continuous efforts are being made to optimize traditional materials. By improving the preparation processes and device designs of silicon-based materials, it is expected to further reduce transmission losses and enhance integration density and performance. Additionally, traditional photonic materials such as lithium niobate are constantly expanding their applications, playing significant roles in devices like modulators. III-V compound semiconductors, with their high-gain characteristics, continuously enhance the performance of key devices such as light sources and optical amplifiers. On the other hand, exploration of novel materials is desired. The emergence of new photonic materials, such as two-dimensional materials and metamaterials, has brought new possibilities to photonic computing. These materials not only possess excellent optical properties but also operate across a broader range of wavelengths, thereby expanding the application scenarios. For example, two-dimensional materials can be used to fabricate more compact and energy-efficient photonic devices, while metamaterials, through artificially designed microstructures, enable precise control of light propagation characteristics, holding promise for the realization of high-performance photonic computing chips. In the future, the development of photonic computing will rely on the collaborative integration of traditional and novel materials to give full play to their respective advantages, achieving more efficient and stable photonic computing systems.

In the field of device research and development, silicon photonic devices have become one of the important development directions in photonic computing due to their high compatibility with existing semiconductor manufacturing processes. Significant progress has been made in silicon photonic devices such as high-speed optoelectronic modulators and high-sensitivity photodetectors. These devices take full advantage of existing semiconductor manufacturing processes, excelling in cost control and integration density improvement, laying a solid foundation for the commercialization of photonic computing. With the continuous advancement of materials science and manufacturing processes, novel photonic devices are evolving rapidly towards higher integration density, lower power consumption, and faster response speeds.

For nonlinear optical devices, researchers are actively exploring various novel nonlinear optical materials and devices. By enhancing the intensity and efficiency of nonlinear effects, reducing the operating thresholds of optical devices, and minimizing energy consumption, low-threshold all-optical nonlinear neurons can be realized. This approach circumvents the energy consumption and latency issues associated with optoelectronic conversion, fully unleashing the high-speed parallel advantages of photonic computing.[273]

In the field of storage, PCM realizes optical storage based on the phase-change properties of materials. With the advantages of high-speed read-write capabilities and non-volatility, PCM opens up new paths for improving the performance and expanding the functions of photonic integrated devices for both computing and storage.[274,275]

From the iterative upgrading of traditional materials to the exploration and application of novel materials, and from the commercialization of silicon photonic devices to the technological breakthroughs of nonlinear optical devices and optical storage devices, each innovation injects new vitality into photonic computing technology. In the future, with continuous efforts in the fields of materials and devices, photonic computing will evolve towards low power consumption, high integration, and multifunctional, providing powerful computing support for fields such as AI and big data processing.

**(2) Innovation in photonic computing architectures and algorithms**

Algorithms and architectures are the core elements for fully exploiting the advantages of photonic computing. Currently, although photonic computing theoretically possesses powerful parallel computing capabilities, the algorithms and architectures compatible with it are still in the development stage.

In the field of algorithms, traditional computing paradigms are insufficient to fully unleash the potential of photons.[2] In the future, it is essential to achieve collaborative design of algorithms and photonic neuromorphic hardware based on the physical characteristics of optical devices, incorporating physical processes such as light propagation, interference, diffraction, resonance, and nonlinear dynamics into algorithmic.[276] Additionally, developing new in-situ training algorithms for neural networks in the optical domain can complete neural network training directly within the optical domain, avoiding energy consumption, losses, and latency caused by optoelectronic conversion. Meanwhile, by fully exploiting the multi-dimensional characteristics of photons, such as wavelength, phase, and polarization, efficient coding schemes can be designed. Through the collaborative design of hardware and algorithms, the parallelism and high-speed characteristics of photons can be utilized to significantly enhance algorithm performance, improve algorithm generality and scalability, and enable more complex computing tasks such as image recognition, object detection, object tracking, large-scale model training, and reinforcement learning, breaking through the performance bottlenecks of traditional algorithms. [277]

In terms of architectures, current photonic computing systems still rely on certain electronic components (such as nonlinear activation layers and pooling layers), resulting in frequent optoelectronic/electro-optical conversions and analog-to-digital/digital-to-analog conversions. These processes severely limit the performance of current photonic computing systems, introducing additional latency and power consumption. Therefore, it is urgent to explore novel optoelectronic integrated computing architectures and conduct comprehensive collaborative optimization at both the hardware and algorithm levels. On one hand, developing novel nonlinear optical materials and devices can replace traditional electronic activation layers, or pooling operations can be realized through photonic methods to reduce dependence on electronic components. On the other hand, heterogeneous integration technologies can integrate photonic computing units and electronic control units at the chip level. This optoelectronic integrated architecture combines the flexibility of electronic computing with the high bandwidth and ultra-high-speed advantages of photonic computing, enabling efficient collaborative operation between photonic neuromorphic chips and traditional electronic systems. It provides more flexible and efficient customized computing solutions for scenarios such as large-scale parallel processing of massive data in data centers and real-time decision-making in autonomous driving.

The collaborative innovation of algorithms and architectures represents a crucial breakthrough in the development of photonic computing. Algorithm optimization should closely follow the characteristics of hardware architectures. For example, parallel computing strategies can be designed according to the optical path layout and signal processing flow of photonic computing chips. In addition, architectural innovations should be guided by algorithm requirements. For instance, the integration of photonic devices can be improved to meet the requirements of specific algorithms for optical signal modulation and demodulation. By establishing a feedback mechanism between algorithms and architectures during the design phase, performance degradation caused by the disconnection between algorithms and architectures can be effectively avoided, thus fully unleashing the technical potential of photonic computing and accelerating the transformation of photonic computing technology from theoretical research to practical applications.

**(3) Large-scale integration and standardized interfaces of optical computing chips**

The large-scale commercialization of photonic chips urgently requires solutions to the problems of high-density integration and standardized interfaces. Currently, there is a significant gap in the integration density of photonic chips compared to mature large-scale integrated circuits. In the future, researchers will focus on the advanced integration and packaging technologies, aiming to integrate core photonic functional devices such as light sources, modulators, detectors, and optical waveguides onto a single chip.

From the perspective of chip development, silicon-based photonic integrated chips, large-scale programmable chips, heterogeneous integrated chips, optoelectronic hybrid integrated chips, and optoelectronic fusion integrated chips are important research directions.[278] Among them, the silicon photonics platform has become the mainstream development direction for photonic computing systems due to its good compatibility with semiconductor processes, high integration density, and low cost advantages. By integrating various photonic devices such as waveguides, resonators, and modulators on the silicon-based platform, not only large-scale photonic integration has been achieved, but also successful applications in constructing compact and efficient PNNs and optical matrix computing, effectively enhancing system performance and stability. In the future, silicon-based photonic integrated chips will evolve towards higher integration density and programmability, further unleashing the potential of photonic computing.[279]

Heterogeneous integration technology based on silicon photonics has also attracted significant attention. This technology can integrate photonic devices made of different materials, combining light sources, information processing units, and detector arrays on a single chip to achieve complex functions while reducing system volume and power consumption. For example, by integrating III-V compound semiconductor lasers with high gain characteristics onto silicon-based waveguides through heterogeneous integration technology, the inherent defect of low luminous efficiency of silicon materials can be effectively addressed, providing high-performance light sources for optical computing chips and promoting the development of optical chips towards higher integration density and performance.

In the field of packaging technology, hybrid integration has become the key to enhancing the performance of photonic computing chips. 2.5D, 3D packaging, and co-packaged optics (CPO) technologies achieve efficient collaboration between optoelectronic devices and electronic devices in different ways.[1] 2.5D packaging realizes lateral integration of photonic and electronic chips through an interposer, significantly improving inter-chip communication speed and reducing signal latency. 3D packaging further increases integration density by vertically stacking chips, shortening the distance between chips and optimizing data transmission efficiency and heat dissipation performance. The CPO technology directly packages optical devices with electronic chips, greatly shortening the optical signal transmission path, effectively reducing power consumption and latency, and enhancing the overall system performance, especially suitable for scenarios with high bandwidth and energy efficiency requirements such as data centers.

Photonic-electronic convergent integration represents the development direction of photonic computing technology, further deepening the collaboration between photonics and electronics. By integrating photonic devices such as waveguides, modulators, and detectors with CMOS circuits through a silicon photonics integration platform and exploring optoelectronic hybrid neural network architectures, this technology fully leverages the high-speed multi-core parallel processing advantages of photonic computing and the flexible control capabilities of electronic computing. For example, in matrix operations, photons are responsible for efficient data transmission and parallel computing, while electronic devices handle complex logical control and nonlinear processing, achieving deep collaborative optimization of optoelectronic devices and providing more efficient computing solutions for fields such as AI and autonomous driving.

Despite the great potential demonstrated by heterogeneous integration, hybrid integration, and photonic-electronic convergent integration technologies, the development of photonic computing chips still faces challenges such as complex processes and low yield rates. In the future, continuous development is needed to optimize manufacturing processes, improve yield rates, and achieve cost reduction and efficiency enhancement through large-scale production, thereby promoting the transition of photonic computing chips from the laboratory to the market and truly unleashing their application value in various fields.

**(4) Performance improvement and application expansion**

Photonics neuromorphic computing, as a revolutionary frontier technology, showcases broad application prospects in AI, big data analysis, autonomous driving, medical diagnosis, optical communication signal processing, edge computing, human-computer interaction, and embodied intelligence.

In the field of data center AI, photonic neuromorphic chips have become an ideal choice for training large AI models due to their ultra-high energy efficiency ratio, providing core computing power support for intelligent computing centers and cloud computing platforms. Their unique high-speed parallel processing architecture can efficiently handle the real-time analysis requirements of massive data, significantly improving data processing efficiency and optimizing the training and inference processes of machine learning and deep learning. In the field of optical communication networks, the integration of photonics neuromorphic computing technology with optical communication systems has revolutionized optical signal processing and transmission modes. By enabling fast signal modulation and demodulation, channel equalization, and nonlinear compensation at optical communication nodes, it effectively improves the transmission rate and stability of optical communication networks, laying a technical foundation for the construction of high-speed and reliable optical communication networks. In edge computing scenarios such as autonomous driving, the low latency of photonics neuromorphic computing is of vital importance.[280] Autonomous vehicles need to process massive data collected from sensors (such as cameras and lidars) in real-time to achieve environmental perception and rapid decision-making. With its extremely low response time, photonics neuromorphic computing technology can significantly enhance the safety and reliability of autonomous driving systems. Especially in complex traffic scenarios, it can efficiently integrate multi-source data, optimize path planning, and improve decision-making strategies, promoting the development of autonomous driving towards a higher level of intelligence. In the fields of embodied intelligence and real-time human-computer interaction, the parallel processing advantages of photonics neuromorphic computing are fully utilized. By greatly reducing data transmission latency, it significantly improves the system response speed, providing an efficient and real-time feedback mechanism for robot dynamic control and human-computer interaction, effectively enhancing system performance.

With continuous innovation efforts from global research institutions and enterprises, through multi-dimensional breakthroughs in material optimization, device research and development, architectural innovation, algorithm modification, and integrated chip performance enhancement, photonics neuromorphic computing technology is transitioning from the exploratory stage to mature application. It is expected that around 2030, this technology will become the core pillar of intelligent computing, leading human society into a new era driven by photonics.

## ACKNOWLEDGMENTS

This work was supported in part by the National Key Research and Development Program of China under Grants 2021YFB2801900, 2021YFB2801901, 2021YFB2801902, and 2021YFB2801904, in part by the Fundamental Research Funds for the Central Universities under Grant QTZX23041, in part by the National Science Foundation for Young Scientists of China under Grant 62404173

## AUTHOR DECLARATIONS

### Conflict of Interest

The authors have no conflicts to disclose.

### Author Contributions

Shuiying Xiang and Chengyang Yu contributed equally to this paper.

Shuiying Xiang: Conceptualization (lead); Funding acquisition (lead); Writing–original draft (equal); Writing – review & editing (equal). Chengyang Yu: Writing–original draft (equal); Writing – review & editing (equal). Yuna Zhang: Writing–original draft (equal). Xintao Zeng: Writing–original draft (equal). Dianzhuang Zheng: Writing–original draft (equal). Yizhi Wang: Writing–original draft (equal). Xinran Niu: Writing–original draft (equal). Haowen Zhao: Writing–original draft (equal). Hanxu Zhou: Writing–original draft (equal). Yanan Han: Writing–original draft (equal). Xingxing Guo: Writing–original draft (equal). Yahui Zhang: Writing–original draft (equal).Yue Hao: Supervision (lead).

## DATA AVAILABILITY

The data that support the findings of this study are available from the corresponding author upon reasonable request.

## REFERENCES

[1] S. Shekhar et al., "Roadmapping the next generation of silicon photonics," Nat. Commun. **15**, 751 (2024).

[2] G. Wetzstein et al., "Inference in artificial intelligence with deep optics and photonics," Nature **588**, 39-47 (2020).

[3] M. Reck, A. Zeilinger, H.J. Bernstein, and P. Bertani, "Experimental realization of any discrete unitary operator," Phys. Rev. Lett. **73**, 58–61 (1994).

[4] A. Ribeiro, A. Ruocco, L. Vanacker, and W. Bogaerts, "Demonstration of a 4×4-port universal linear circuit," Optica **3**, 1348 (2016).

[5] W.R. Clements, P.C. Humphreys, B.J. Metcalf, W.S. Kolthammer, and I.A. Walsmley, "Optimal design for universal multiport interferometers," Optica **3**, 1460 (2016).

[6] A. Annoni, E. Guglielmi, M. Carminati, G. Ferrari, M. Sampietro, D.A. Miller, A. Melloni, and F. Morichetti, "Unscrambling light-automatically undoing strong mixing between modes," Light Sci Appl **6**, e17110–e17110 (2017).

[7] D. Pérez, I. Gasulla, L. Crudgington, D.J. Thomson, A.Z. Khokhar, K. Li, W. Cao, G.Z. Mashanovich, and J. Capmany, "Multipurpose silicon photonics signal processor core," Nat Commun **8**, 636 (2017).

[8] S. Pai, B. Bartlett, O. Solgaard, and D.A.B. Miller, "Matrix optimization on universal unitary photonic devices," Phys. Rev. Applied **11**, 064044 (2019).

[9] G. Cong, N. Yamamoto, T. Inoue, M. Okano, Y. Maegami, M. Ohno, and K. Yamada, "Arbitrary reconfiguration of universal silicon photonic circuits by bacteria foraging algorithm to achieve reconfigurable photonic digital-to-analog conversion," Opt. Express **27**(18), 24914 (2019).

[10] F. Shokraneh, S. Geoffroy-gagnon, and O. Liboiron-Ladouceur, "The diamond mesh, a phase-error- and loss-tolerant field-programmable MZI-based optical processor for optical neural networks," Opt. Express **28**, 23495 (2020).

[11] L. Pei, Z. Xi, B. Bai, J. Wang, J. Zheng, J. Li, and T. Ning, "Joint device architecture algorithm codesign of the photonic neural processing unit," Adv. Photon. Nexus **2**, (2023).

[12] B. Wu, S. Liu, J. Cheng, W. Dong, H. Zhou, J. Dong, M. Li, and X. Zhang, "Real-valued optical matrix computing with simplified MZI sesh," Intell. Comput. **2**, 0047 (2023).

[13] G. Giamougiannis, A. Tsakyridis, M. Moralis-Pegios, A.R. Totovic, M. Kirtas, N. Passalis, A. Tefas, D. Lazovsky, and N. Pleros, "Universal linear optics revisited: new perspectives for neuromorphic computing with silicon photonics," IEEE J. Select. Topics Quantum Electron. **29**(2: Optical Computing), 1–16 (2023).

[14] A. Shafiee, S. Banerjee, K. Chakrabarty, S. Pasricha, and M. Nikdast, "Analysis of optical loss and crosstalk noise in MZI-Based coherent photonic

neural networks," J. Lightwave Technol. **42**, 4598–4613 (2024).
[15] J. Lin, K. Yang, Q. Fu, P. Wang, S. Dai, W. Chen, D. Kong, J. Li, T. Dai, and J. Yang, "A robust MZI-based optical neural network using QR decomposition," J. Lightwave Technol. **43**, 1024–1031 (2025).
[16] L. Yang, R. Ji, L. Zhang, J. Ding, and Q. Xu, "On-chip CMOS-compatible optical signal processor," Opt. Express **20**,13560-13565(2012).
[17] A.N. Tait, M.A. Nahmias, B.J. Shastri, and P.R. Prucnal, "Broadcast and weight: an integrated network for scalable photonic spike processing," J. Lightwave Technol. **32**, 4029–4041 (2014).
[18] A.N. Tait, J. Chang, B.J. Shastri, M.A. Nahmias, and P.R. Prucnal, "Demonstration of WDM weighted addition for principal component analysis," Opt. Express **23**, 12758 (2015).
[19] A.N. Tait, T. Ferreira De Lima, M.A. Nahmias, B.J. Shastri, and P.R. Prucnal, "Continuous calibration of microring weights for analog optical networks," IEEE Photon. Technol. Lett. **28**, 887–890 (2016).
[20] A.N. Tait, T.F. De Lima, M.A. Nahmias, B.J. Shastri, and P.R. Prucnal, "Multi-channel control for microring weight banks," Opt. Express **24**, 8895 (2016).
[21] A.N. Tait, A.X. Wu, T.F. De Lima, E. Zhou, B.J. Shastri, M.A. Nahmias, and P.R. Prucnal, "Microring weight banks," IEEE J. Select. Topics Quantum Electron. **22**, 312–325 (2016).
[22] A.N. Tait, T.F. De Lima, E. Zhou, A.X. Wu, M.A. Nahmias, B.J. Shastri, and P.R. Prucnal, "Neuromorphic photonic networks using silicon photonic weight banks," Sci Rep **7**, 7430 (2017).
[23] C. Huang, S. Bilodeau, T. Ferreira De Lima, A.N. Tait, P.Y. Ma, E.C. Blow, A. Jha, H.-T. Peng, B.J. Shastri, and P.R. Prucnal, "Demonstration of scalable microring weight bank control for large-scale photonic integrated circuits," APL Photonics **5**, 040803 (2020).
[24] P.Y. Ma, A.N. Tait, T.F. De Lima, C. Huang, B.J. Shastri, and P.R. Prucnal, "Photonic independent component analysis using an on-chip microring weight bank," Opt. Express **28**, 1827 (2020).
[25] S. Xu, J. Wang, and W. Zou, "Optical convolutional neural network with WDM-based optical patching and microring weighting banks," IEEE Photon. Technol. Lett. **33**, 89–92 (2021).
[26] J. Cheng, Y. Zhao, W. Zhang, H. Zhou, D. Huang, Q. Zhu, Y. Guo, B. Xu, J. Dong, and X. Zhang, "A small microring array that performs large complex-valued matrix-vector multiplication," Front. Optoelectron. **15**, 15 (2022).
[27] W. Zhang, C. Huang, H.-T. Peng, S. Bilodeau, A. Jha, E. Blow, T.F. De Lima, B.J. Shastri, and P. Prucnal, "Silicon microring synapses enable photonic deep learning beyond 9-bit precision," Optica **9**, 579 (2022).
[28] T. Ferreira De Lima, E.A. Doris, S. Bilodeau, W. Zhang, A. Jha, H.-T. Peng, E.C. Blow, C. Huang, A.N. Tait, B.J. Shastri, and P.R. Prucnal, "Design automation of photonic resonator weights," Nanophotonics **11**, 3805–3822 (2022).
[29] E.C. Blow, S. Bilodeau, W. Zhang, T. Ferreira De Lima, J.C. Lederman, B. Shastri, and P.R. Prucnal, "Radio-frequency linear analysis and optimization of silicon photonic neural networks," Adv. Photonics Res. **5**, 2300306 (2024).
[30] D. Jin, S. Ren, J. Hu, D. Huang, D.J. Moss, and J. Wu, "Modeling of complex integrated photonic resonators using the scattering matrix method," Photonics **11**, 1107 (2024).
[31] Z. Cheng, C. Ríos, W.H.P. Pernice, C.D. Wright, and H. Bhaskaran, "On-chip photonic synapse," Sci. Adv. **3**, e1700160 (2017).
[32] I. Chakraborty, G. Saha, and K. Roy, "A photonic in-memory computing primitive for spiking neural networks using phase-change materials," Phys. Rev. Applied **11**, 014063 (2019).
[33] J. Feldmann, N. Youngblood, C.D. Wright, H. Bhaskaran, and W.H.P. Pernice, "All-optical spiking neurosynaptic networks with self-learning capabilities," Nature **569**, 208–214 (2019).
[34] J. Feldmann, N. Youngblood, M. Karpov, H. Gehring, X. Li, M. Stappers, M. Le Gallo, X. Fu, A. Lukashchuk, A.S. Raja, J. Liu, C.D. Wright, A. Sebastian, T.J. Kippenberg, W.H.P. Pernice, and H. Bhaskaran, "Parallel convolutional processing using an integrated photonic tensor core," Nature **589**, 52–58 (2021).
[35] M. Miscuglio, and V.J. Sorger, "Photonic tensor cores for machine learning," Applied Physics Reviews **7**, 031404 (2020).
[36] Y. Zhang, D. Yao, Y. Liu, C. Fang, S. Wang, G. Wang, Y. Huang, X. Yu, G. Han, and Y. Hao, "All-optical synapse with directional coupler structure based on phase change material," IEEE Photonics J. **13**, 1–6 (2021).
[37] C. Wu, H. Yu, S. Lee, R. Peng, I. Takeuchi, and M. Li, "Programmable phase-change metasurfaces on waveguides for multimode photonic convolutional neural network," Nat. Commun. **12**, 96 (2021).
[38] W. Zhou, N. Farmakidis, J. Feldmann, X. Li, J. Tan, Y. He, C.D. Wright, W.H.P. Pernice, and H. Bhaskaran, "Phase-change materials for energy-efficient photonic memory and computing," MRS Bulletin **47**, 502–510 (2022).
[39] W. Zhou, B. Dong, N. Farmakidis, X. Li, N. Youngblood, K. Huang, Y. He, C. David Wright, W.H.P. Pernice, and H. Bhaskaran, "In-memory photonic dot-product engine with electrically programmable weight banks," Nat Commun **14**, 2887 (2023).
[40] Y. Zhang, S. Xiang, Y. Han, X. Guo, Y. Zhang, Y. Shi, and Y. Hao, "Supervised learning and pattern recognition in photonic spiking neural networks based on MRR and phase-change materials," Opt. Commun. **549**, 129870 (2023).
[41] A. Lugnan, S. Aggarwal, F. Brückerhoff-Plückelmann, C.D. Wright, W.H.P. Pernice, H. Bhaskaran, and P. Bienstman, "Emergent self-adaptation in an integrated photonic neural network for backpropagation-free learning," Adv. Sci. **12**, 2404920 (2025).
[42] A.H.A. Nohoji, P. Keshavarzi, and M. Danaie, "A photonic crystal waveguide intersection using phase change material for optical neuromorphic synapses," Opt. Mater. **151**, 115372 (2024).
[43] M.P. Fok, Y. Tian, D. Rosenbluth, and P.R. Prucnal, "Pulse lead/lag timing detection for adaptive feedback and control based on optical spike-timing-dependent plasticity," Opt. Lett. **38**, 419 (2013).
[44] R. Toole, and M.P. Fok, "Photonic implementation of a neuronal algorithm applicable towards angle of arrival detection and localization," Opt. Express **23**, 16133 (2015).
[45] R. Toole, A.N. Tait, T. Ferreira De Lima, M.A. Nahmias, B.J. Shastri, P.R. Prucnal, and M.P. Fok, "Photonic implementation of spike-timing-dependent plasticity and learning algorithms of biological neural systems," J. Lightwave Technol. **34**, 470–476 (2016).
[46] Q. Ren, Y. Zhang, R. Wang, and J. Zhao, "Optical spike-timing-dependent plasticity with weight-dependent learning window and reward modulation," Opt. Express **23**, 25247 (2015).
[47] Q. Li, Z. Wang, Y. Le, C. Sun, X. Song, and C. Wu, "Optical implementation of neural learning algorithms based on cross-gain modulation in a semiconductor optical amplifier," edited by X. Zhang, B. Li, and C. Yu, (Beijing, China, 2016), p. 100190E.
[48] S. Xiang, J. Gong, Y. Zhang, X. Guo, Y. Han, A. Wen, and Y. Hao, "Numerical implementation of wavelength-dependent photonic spike timing dependent plasticity based on VCSOA," IEEE J. Quantum Electron. **54**, 1–7 (2018).
[49] Z. Song, S. Xiang, X. Cao, S. Zhao, and Y. Hao, "Experimental demonstration of photonic spike-timing-dependent plasticity based on a VCSOA," Sci. China Inf. Sci. **65**, 182401 (2022).
[50] Y. Zhang, S. Xiang, X. Cao, X. Guo, G. Han, and Y. Hao, "Plastic photonic synapse based on VCSOA for self-learning in photonic spiking neural network," J. Lightwave Technol. **41**, 1759–1767 (2023).
[51] D. Zheng, S. Xiang, X. Guo, Y. Zhang, X. Zeng, X. Zhu, Y. Shi, X. Chen, and Y. Hao, "Full-function Pavlov associative learning photonic neural networks based on SOA and DFB-SA," APL Photonics **9**, 026102 (2024).
[52] D. Zheng, S. Xiang, N. Li, Y. Zhang, X. Guo, L. Zhao, and Y. Hao, "Neuromorphic network with photonic weighting and photoelectronic nonlinear activation based on SOA and APD," ACS Photonics, acsphotonics.4c01083 (2024).

[53] D. Zheng, S. Xiang, N. Li, Y. Zhang, X. Guo, X. Zhu, X. Chen, Y. Shi, and Y. Hao, "The hybrid photonic convolutional neural networks based on SOA and FP-SA," J. Lightwave Technol. **42**, 8819–8825 (2024).
[54] B. Shi, N. Calabretta, and R. Stabile, "Deep neural network through an InP SOA-based photonic integrated cross-connect," IEEE J. Select. Topics Quantum Electron. **26**, 1–11 (2020).
[55] J.A. Alanis, J. Robertson, M. Hejda, and A. Hurtado, "Weight adjustable photonic synapse by nonlinear gain in a vertical cavity semiconductor optical amplifier," Applied Physics Letters **119**, 201104 (2021).
[56] T. Tian, Z. Wu, X. Lin, X. Tang, Z. Gao, M. Ni, G. Xia, H. Chen, and T. Deng, "Photonic implementation of spike timing dependent plasticity with weight-dependent learning window based on VCSOA," Laser Phys. **32**, 016201 (2022).
[57] S. Xu, J. Wang, H. Shu, Z. Zhang, S. Yi, B. Bai, X. Wang, J. Liu, and W. Zou, "Optical coherent dot-product chip for sophisticated deep learning regression," Light Sci Appl **10**, 221 (2021).
[58] N. Youngblood, "Coherent photonic crossbar arrays for large-scale matrix-matrix multiplication," IEEE J. Select. Topics Quantum Electron. **29**(2: Optical Computing), 1–11 (2023).
[59] Z. Zhu, A. Fardoost, F.G. Vanani, A.B. Klein, G. Li, and S.S. Pang, "Coherent general-purpose photonic matrix processor," ACS Photonics **11**, 1189–1196 (2024).
[60] M. Moralis-Pegios, G. Giamougiannis, A. Tsakyridis, D. Lazovsky, and N. Pleros, "Perfect linear optics using silicon photonics," Nat Commun **15**, 5468 (2024).
[61] B. Dong, F. Brückerhoff-Plückelmann, L. Meyer, J. Dijkstra, I. Bente, D. Wendland, A. Varri, S. Aggarwal, N. Farmakidis, M. Wang, G. Yang, J.S. Lee, Y. He, E. Gooskens, D.-L. Kwong, P. Bienstman, W.H.P. Pernice, and H. Bhaskaran, "Partial coherence enhances parallelized photonic computing," Nature **632**, 55–62 (2024).
[62] S. R. Kari, N.A. Nobile, D. Pantin, V. Shah, and N. Youngblood, "Realization of an integrated coherent photonic platform for scalable matrix operations," Optica **11**, 542 (2024).
[63] A. Hurtado, I. D. Henning, and M. J. Adams, "Optical neuron using polarisation switching in a 1550 nm - VCSEL," Opt. Express **18**, 25170–25176 (2010).
[64] B. J. Shastri et al., "Simulations of a graphene excitable laser for spike processing," Opt. Quantum Electron. **46**, 1353-1358 (2014).
[65] B. J. Shastri et al., "Spike processing with a graphene excitable laser," Sci. Rep. 6, 1-12 (2016).
[66] P. Y. Ma, B. J. Shastri, T. F. De Lima et al., "All-optical digital-to-spike conversion using a graphene excitable laser," Opt. Express **25**, 33504-33513 (2017).
[67] S. Xiang, A. Wen, W. Pan, "Emulation of spiking response and spiking frequency property in VCSEL-based photonic neuron," IEEE Photonics J. **8**: 1504109 (2016).
[68] Y. Zhang, S. Xiang, X. Cao, S. Zhao, X. Guo, A. Wen, Y. Hao, "Experimental demonstration of pyramidal neuron-like dynamics dominated by dendritic action potentials based on a VCSEL for all-optical XOR classification task," Photonics Res. **9**, 1055–1061 (2021).
[69] Y. Zhang, J. Robertson, S. Xiang, M. Hejda, J. Bueno, A. Hurtado, "All-optical neuromorphic binary convolution with a spiking VCSEL neuron for image gradient magnitudes," Photonics Res. **9**, B201-B209 (2021).
[70] S. Gao et al., "Hardware implementation of ultra-fast obstacle avoidance based on a single photonic spiking neuron," Laser Photonics Rev. **17**, 2300424 (2023).
[71] Y. Zhang et al., "Photonic neuromorphic pattern recognition with a spiking DFB-SA laser subject to incoherent optical injection," Laser Photonics Rev. **25**, 2400482 (2025).
[72] M. A. Nahmias et al., "A leaky integrate-and-fire laser neuron for ultrafast cognitive computing," IEEE J. Sel. Top. Quantum Electron. **19**, 1-12 (2013).
[73] Q. Li et al., "Simulating the spiking response of VCSEL-based optical spiking neuron," Opt. Commun. 407, 327-332 (2018).
[74] Y. Zhang et al., "An enhanced photonic spiking neural network based on the VCSEL-SA for recognition and classification," J. Lightwave Technol. **24**, (2024).
[75] Y. Zhang et al., "Spike encoding and storage properties in mutually coupled vertical-cavity surface-emitting lasers subject to optical pulse injection," Appl. Opt. **57**, 1731 (2018).
[76] Y. H. Zhang et al., "Polarization-resolved and polarization-multiplexed spike encoding properties in photonic neuron based on VCSEL-SA," Sci. Rep. **8**, 16095 (2018).
[77] Y. Zhang et al., "All-optical inhibitory dynamics in photonic neuron based on polarization mode competition in a VCSEL with an embedded saturable absorber," Opt. Lett. **44**, 1548-1551 (2019).
[78] S. Y. Xiang et al., "All-optical neuromorphic XOR operation with inhibitory dynamics of a single photonic spiking neuron based on a VCSEL-SA," Opt. Lett. **45**, 1104-1107 (2020).
[79] F. Selmi et al., "Relative refractory period in an excitable semiconductor laser," Phys. Rev. Lett. **112**, 183902 (2014).
[80] F. Selmi et al., "Temporal summation in a neuromimetic micropillar laser," Opt. Lett. **40**, 5690-5693 (2015).
[81] F. Selmi et al., "Spike latency and response properties of an excitable micropillar laser," Phys. Rev. E **94**, 042219 (2016).
[82] V. A. Pammi et al., "Photonic computing with single and coupled spiking micropillar lasers," IEEE J. Sel. Top. Quantum Electron. **26**, 1-7 (2019).
[83] S. Terrien et al., "Merging and disconnecting resonance tongues in a pulsing excitable microlaser with delayed optical feedback," Chaos **33**, (2023).
[84] D. Zheng et al., "Experimental demonstration of coherent photonic neural computing based on a Fabry-Perot laser with a saturable absorber," Photonics Res. **11**, 65-71 (2022).
[85] Z. W. Song et al., "Nonlinear neural computation in an integrated FP-SA spiking neuron subject to incoherent dual-wavelength optical pulse injections," Sci. China Inf. Sci. **66**, 229405 (2023).
[86] S. Y. Xiang et al., "Hardware-algorithm collaborative computing with photonic spiking neuron chip based on an integrated Fabry-Perot laser with a saturable absorber," Optica **10**, 162-171 (2023).
[87] X. X. Guo et al., "Hardware implementation of multi-layer photonic spiking neural network with three cascaded photonic spiking neurons," J. Lightwave Technol. **41**, 6533-6541 (2023).
[88] Y. N. Han et al., "Experimental demonstration of delay-weight learning and pattern classification with a FP-SA-based photonic spiking neuron chip," J. Lightwave Technol. **42**, 1497-1503 (2024).
[89] Y. N. Zhang et al., "Evolution of neuron-like spiking response and spike-based all-optical XOR operation in a DFB with saturable absorber," J. Lightwave Technol. **42**, 2026-2035 (2024).
[90] C. Y. Yu et al., "Neuromorphic convolution with a spiking DFB-SA laser neuron based on rate coding," Opt. Express **31**, 43698-43711 (2023).
[91] Y. N. Han et al., "Pattern recognition in multi-synaptic photonic spiking neural networks based on a DFB-SA chip," Opto-Electron. Sci. **2**, 230021-230021 (2023).
[92] S. Y. Xiang et al., "Photonic integrated neuro-synaptic core for convolutional spiking neural network," Opto-Electron. Adv. **6**, 230140 (2023).
[93] X. Zeng et al., "Photonic spiking neural network based on DML and DFB-SA laser chip for pattern classification," Opt. Express **33**, 12045-12058 (2025).
[94] T. Van Vaerenbergh et al., "Cascadable excitability in microrings," Opt. Express **20**, 20292-20308 (2012).
[95] J. Xiang et al., "All-optical spiking neuron based on passive microresonator," J. Lightwave Technol. **38**, 4019-4029 (2020).
[96] J. Xiang et al., "All-optical silicon microring spiking neuron," Photonics Res. **10**, 939-946 (2022).

[97] J. Xiang, Y. Zhao, A. He et al., “Photonic neuromorphic processing with on-chip electrically-driven microring spiking neuron,” Laser Photonics Rev. **19**, 2400604 (2025).
[98] A. Jha et al., “Photonic spiking neural networks and graphene-on-silicon spiking neurons,” J. Lightwave Technol. **40**, 2901-2914 (2022).
[99] I. Chakraborty et al. “Toward fast neural computing using all-photonic phase change spiking neurons,” Sci. Rep. **8**, 12980 (2018).
[100] Q. Zhang et al., “On-chip spiking neural networks based on add-drop ring microresonators and electrically reconfigurable phase-change material photonic switches,” Photonics Res. **12**, 755-766 (2024).
[101] J. K. George et al., “Neuromorphic photonics with electro-absorption modulators,” Opt. Express **27**, 5181-5191 (2019).
[102] R. Amin et al., “ITO-based electro absorption modulator for photonic neural activation function,” APL Mater. **7**, 081112 (2019).
[103] I. A. D. Williamson et al., “Reprogrammable electro-optic nonlinear activation functions for optical neural networks,” IEEE J. Sel. Top. Quantum Electron. **26**, 1-12 (2019).
[104] Z. Xu et al., “Reconfigurable nonlinear photonic activation function for photonic neural network based on non-volatile opto-resistive RAM switch,” Light: Sci. Appl. **11**, 288 (2022).
[105] F. Ashtiani, A. J. Geers, F. Aflatouni, “An on-chip photonic deep neural network for image classification,” Nature **606**, 501-506 (2022).
[106] A. Jha, C. Huang, P. R. Prucnal, “Reconfigurable all-optical nonlinear activation functions for neuromorphic photonics,” Opt. Lett. **45**, 4819-4822 (2020).
[107] Z. Fu et al., “Programmable low-power consumption all-optical nonlinear activation functions using a micro-ring resonator with phase-change materials,” Opt. Express **30**, 44943-44953 (2022).
[108] G. Mourgias-Alexandris et al., “An all-optical neuron with sigmoid activation function,” Opt. Express **27**, 9620-9630 (2019).
[109] B. Shi et al., “Lossless monolithically integrated photonic InP neuron for all-optical computation,” in Optical Fiber Communication Conference, Optica Publishing Group, 2020, p. W2A. 12.
[110] M. Miscuglio et al., “All-optical nonlinear activation function for photonic neural networks,” Opt. Mater. Express **8**, 3851-3863 (2018).
[111] C. Chen et al., “Ultra-broadband all-optical nonlinear activation function enabled by $MoTe_2$/optical waveguide integrated devices,” Nat. Commun. **15**, 9047 (2024).
[112] Y. Tian et al., “Photonic neural networks with Kramers-Kronig activation,” Adv. Photonics Res. 4, 2300062 (2023).
[113] B. Wu et al., “Low-threshold all-optical nonlinear activation function based on a Ge/Si hybrid structure in a microring resonator,” Opt. Mater. Express **12**, 970-980 (2022).
[114] Y. Shi et al., “Nonlinear germanium-silicon photodiode for activation and monitoring in photonic neuromorphic networks,” Nat. Commun. **13**, 6048 (2022).
[115] H. Li et al., “All-optical nonlinear activation function based on germanium silicon hybrid asymmetric coupler,” IEEE J. Sel. Top. Quantum Electron. **29**, 1-6 (2022).
[116] B. Zhao et al., “Cascadable optical nonlinear activation function based on Ge-Si,” Opt. Lett. **49**, 6149-6152 (2024).
[117] Z. Zhou et al., “Ultrafast silicon/graphene optical nonlinear activator for neuromorphic computing,” Adv. Opt. Mater. **12**, 2401686 (2024).
[118] M. Ruiz-Llata, L. Horacio, and W. Cardinal, “Fully interconnected neural network system based on an optical broadcast,” Algor. Syst. Opt. Inf. Process. **4471**, 159-166 (2001).
[119] Y. Shen et al., “Deep learning with coherent nanophotonic circuits,” Nat. Photonics **11**, 441-446 (2017).
[120] C. Huang et al., “Demonstration of photonic neural network for fiber nonlinearity compensation in long-haul transmission systems,” 2020 Opt. Fiber Commun. Conf. Exhib. (OFC), IEEE, 1-3 (2020).
[121] C. Huang et al., “A silicon photonic–electronic neural network for fibre nonlinearity compensation,” Nat. Electronics **4**, 837-844(2021).
[122] K. Mizutani et al., “OPTWEB: a lightweight fully connected inter-FPGA network for efficient collectives,” IEEE Trans. Comput. **70**, 849-862(2021).
[123] H. Zhang et al., “An optical neural chip for implementing complex-valued neural network,” Nat. commun. **12**, 457 (2021).
[124] G. Mourgias-Alexandris et al., “Noise-resilient and high-speed deep learning with coherent silicon photonics,” Nat. Commun. **13**, 5572(2022).
[125] M. Moralis-Pegios et al., “Neuromorphic silicon photonics and hardware-aware deep learning for high-speed inference,” J. Lightwave Technol. **40**, 3243-3254(2022).
[126] C. Feng et al., “A compact butterfly-style silicon photonic–electronic neural chip for hardware-efficient deep learning,” ACS Photonics **9**, 3906-3916 (2022).
[127] S. Ohno et al., “Si microring resonator crossbar array for on-chip inference and training of the optical neural network,” ACS Photonics **9**, 2614-2622(2022).
[128] H. Zhang et al., “Molecular property prediction with photonic chip-based machine learning,” Laser Photonics Rev. **17**, 2200698 (2023).
[129] K. Lu, and X. Guo, “Efficient training of unitary optical neural networks,” Opt. Express **31**, 39616-39623 (2023).
[130] T. Xu et al., “Control-free and efficient integrated photonic neural networks via hardware-aware training and pruning,” Optica **11**, 1039-1049(2024).
[131] Z. Xu et al., “Large-scale photonic chiplet Taichi empowers 160-TOPS/W artificial general intelligence,” Science **384**, 202-209 (2024).
[132] A. Mehrabian, Y. Al-Kabani, V. J. Sorger, and T. El-Ghazawi, “PCNNA: a photonic convolutional neural network accelerator,” in Proc. 2018 31st IEEE International System-on-Chip Conference (SOCC), 2018, pp. 169-173.
[133] V. Bangari, B. A. Marquez, H. Miller, A. N. Tait, M. A. Nahmias, T. Ferreira de Lima, H.-T. Peng, P. R. Prucnal, and B. J. Shastri, “Digital electronics and analog photonics for convolutional neural networks (DEAP-CNNs),” IEEE J. Sel. Top. Quantum Electron. **26**, 7701213 (2020).
[134] X. Xu, M. Tan, B. Corcoran, J. Wu, A. Boes, T. G. Nguyen, S. T. Chu, B. E. Little, D. G. Hicks, R. Morandotti, A. Mitchell, and D. J. Moss, “11 TOPS photonic convolutional accelerator for optical neural networks,” Nature **589**, 7840 (2021).
[135] S. Xu, J. Wang, S. Yi, and W. Zou, “High-order tensor flow processing using integrated photonic circuits,” Nat. Commun. **13**, 7970 (2022).
[136] J. Cheng, Y. Xie, Y. Liu, J. Song, X. Liu, Z. He, W. Zhang, X. Han, H. Zhou, K. Zhou, H. Zhou, J. Dong, and X. Zhang, “Human emotion recognition with a microcomb-enabled integrated optical neural network,” Nanophotonics **12**, 3883–3894 (2023).
[137] B. Bai, Q. Yang, H. Shu, L. Chang, F. Yang, B. Shen, Z. Tao, J. Wang, S. Xu, W. Xie, W. Zou, W. Hu, J. E. Bowers, and X. Wang, “Microcomb-based integrated photonic processing unit,” Nat. Commun. **14**, 66 (2023).
[138] S. Chen, Y. Zheng, Y. Xu, X. Zhu, S. Huang, S. Wang, X. Xu, C. Xia, Z. Liu, C. Huang, R. Morandotti, S. T. Chu, B. E. Little, Y. Liu, Y. Bai, D. J. Moss, X. Xu, and K. Xu, “High-bit-efficiency TOPS optical tensor convolutional accelerator using microcombs,” Laser Photon. Rev. **19**, 2300123 (2025).
[139] Y. Bai, Y. Xu, S. Chen, X. Zhu, S. Wang, S. Huang, Y. Song, Y. Zheng, Z. Liu, S. Tan, R. Morandotti, S. T. Chu, B. E. Little, D. J. Moss, X. Xu, and K. Xu, “TOPS-speed complex-valued convolutional accelerator for feature extraction and inference,” Nat. Commun. **16**, 292 (2025).
[140] J. K. George, H. Nejadriahi, and V. J. Sorger, “Towards on-chip optical FFTs for convolutional neural networks,” arXiv:1708.09534 (2017).
[141] Y. Tian, Y. Zhao, S. Liu, Q. Li, W. Wang, J. Feng, and J. Guo, “Scalable and compact photonic neural chip with low learning-capability-loss," Nanophotonics **11**, 329-344 (2021).
[142] C. D. Wright, Y. Liu, K. I. Kohary, M. M. Aziz, and R. J. Hicken, “Arithmetic and biologically-inspired computing using phase-change materials,” Adv. Mater. **23**, 3408–3413 (2011).
[143] B. Dong, S. Aggarwal, W. Zhou, U. E. Ali, N. Farmakidis, J. S. Lee, Y. He, X. Li, D.-L. Kwong, C. D. Wright, W. H. P. Pernice, and H. Bhaskaran, “Higher-dimensional processing using a photonic tensor core with continuous-time data,” Nat. Photonics **17**, 1080–1088 (2023).
[144] X. Meng, G. Zhang, N. Shi, G. Li, J. Azaña, J. Capmany, J. Yao, Y. Shen, W. Li, N. Zhu, and M. Li, “Compact optical convolution processing unit

based on multimode interference," Nat. Commun. **14**, 3000 (2023).
[145] J. Li, X. Meng, J. Zhang, W. Li, N. Shi, M. Li, "High-efficiency reconfigurable optical convolutional neural network," J. Lightwave Technol. **43**, 4076-4085 (2025).
[146] B. Shi, N. Calabretta, and R. Stabile, "Parallel photonic convolutional processing on-chip with cross-connect architecture and cyclic AWGs," IEEE J. Sel. Top. Quantum Electron. **29**, 7400310 (2023).
[147] J. Cheng, C. Li, J. Dai, Y. Chu, X. Niu, X. Dong, and J.-J. He, "Direct optical convolution computing based on arrayed waveguide grating router," Laser Photonics Rev. **18**, 2301221 (2024).
[148] D. Yi, C. Zhao, Z. Zhang, H. Xu, and H. K. Tsang, "Accelerating convolutional processing by harnessing channel shifts in arrayed waveguide gratings," Laser Photonics Rev. **19**, 2400435 (2025).
[149] S. Zhang, H. Zhou, B. Wu, X. Jiang, D. Gao, J. Xu, and J. Dong, "Redundancy-free integrated optical convolver for optical neural networks based on arrayed waveguide grating," Nanophotonics **13**, 19-28 (2024).
[150] C. Pappas, T. Moschos, A. Prapas, M. Kirtas, M. Moralis-Pegios, A. Tsakyridis, O. Asimopoulos, N. Passalis, A. Tefas, and N. Pleros, "Reaching the peta-computing: 163.8 TOPS through multidimensional AWGR-based accelerators," J. Lightwave Technol., **43**, 1773-1785 (2025).
[151] W. Maass, "Networks of spiking neurons: the third generation of neural network models," Neural Netw. **10**, 1659–1671 (1997).
[152] Y. Han, S. Xiang, Y. Zhang *et al.*, "An all-MRR-based photonic spiking neural network for spike sequence learning," Photonics **9**, 120 (2022).
[153] S. Xiang, Y. Zhang, J. Gong *et al.*, "STDP-based unsupervised spike pattern learning in a photonic spiking neural network with VCSELs and VCSOAs," IEEE J. Sel. Top. Quantum Electron. **25**, 1–9 (2019).
[154] S. Xiang, Z. Ren, Z. Song *et al.*, "Computing primitive of fully VCSEL-based all-optical spiking neural network for supervised learning and pattern classification," IEEE Trans. Neural Netw. Learn. Syst. **32**, 2494–2505 (2020).
[155] S. Xiang, Z. Ren, Y. Zhang *et al.*, "Training a multi-layer photonic spiking neural network with modified supervised learning algorithm based on photonic STDP," IEEE J. Sel. Top. Quantum Electron. **27**, 1–9 (2020).
[156] Z. W. Song, S. Y. Xiang, Z. X. Ren *et al.*, "Photonic spiking neural network based on excitable VCSELs-SA for sound azimuth detection," Opt. Express **28**, 1561–1573 (2020).
[157] Y. Han, S. Xiang, Z. Song, *et al.*, "Noisy image segmentation based on synchronous dynamics of coupled photonic spiking neurons," Opt. Express **31**, 35484–35492 (2023).
[158] D. Owen-Newns, J. Robertson, M. Hejda, *et al.*, "GHz rate neuromorphic photonic spiking neural network with a single vertical-cavity surface-emitting laser (VCSEL)," IEEE J. Sel. Top. Quantum Electron. **29**, 1–10 (2022).
[159] D. Owen-Newns, J. Robertson, M. Hejda, *et al.*, "Photonic spiking neural networks with highly efficient training protocols for ultrafast neuromorphic computing systems," Intell. Comput. **2**, 0031 (2023).
[160] J. Robertson, P. Kirkland, G. Di Caterina, *et al.*, "VCSEL-based photonic spiking neural networks for ultrafast detection and tracking," *Neuromorphic Comput. Eng.* **4**, 014010 (2024).
[161] D. Owen-Newns, L. Jaurigue, J. Robertson, *et al.*, "Photonic spiking neural network built with a single VCSEL for high-speed time series prediction," *Commun. Phys.* **8**, 110 (2025).
[162] Y. Lu, W. Zhang, B. Fu, *et al.*, "Synaptic delay plasticity based on frequency-switched VCSELs for optical delay-weight spiking neural networks," *Opt. Lett.* **47**, 5587–5590 (2022).
[163] Y. J. Lee, M. B. On, X. Xiao *et al.*, "Photonic spiking neural networks with event-driven femtojoule optoelectronic neurons based on Izhikevich-inspired model," *Opt. Express*, **30**, 19360–19389 (2022).
[164] L. El Srouji, Y. J. Lee, M. B. On *et al.*, "Scalable nanophotonic-electronic spiking neural networks," *IEEE J. Sel. Top. Quantum Electron.* **29**, 1–13 (2022).
[165] Y. J. Lee, M. B. On, L. El Srouji *et al.*, "Demonstration of programmable brain-inspired optoelectronic neuron in photonic spiking neural network with neural heterogeneity," *J. Lightwave Technol.* (2024).
[166] Z. Song, S. Xiang, S. Zhao *et al.*, "A hybrid-integrated photonic spiking neural network framework based on an MZI array and VCSELs-SA," *IEEE J. Sel. Top. Quantum Electron.*, **29**, 1–11 (2022).
[167] Y. Han, S. Xiang, Z. Ren *et al.*, "Delay-weight plasticity-based supervised learning in optical spiking neural networks," *Photon. Res.* **9**, B119–B127 (2021).
[168] S. Xiang, T. Zhang, Y. Han *et al.*, "Neuromorphic speech recognition with photonic convolutional spiking neural networks," *IEEE J. Sel. Top. Quantum Electron.* **29**, 1–7 (2023).
[169] Y. Han, S. Xiang, T. Zhang, *et al.*, "Conversion of a single-layer ANN to photonic SNN for pattern recognition," *Sci. China Inf. Sci.* **67**, 112403 (2024).
[170] T. Inagaki, K. Inaba, T. Leleu *et al.*, "Collective and synchronous dynamics of photonic spiking neurons," *Nat. Commun.* **12**, 2325 (2021).
[171] B. Ma, J. Zhang, X. Li, *et al.*, "Stochastic photonic spiking neuron for Bayesian inference with unsupervised learning," *Opt. Lett.* **48**, 1411–1414 (2023).
[172] M. A. Nahmias, H. T. Peng, T. F. de Lima, *et al.*, "A laser spiking neuron in a photonic integrated circuit," arXiv preprint arXiv:2012.08516 (2020).
[173] Y. Zhang, S. Xiang, X. Guo *et al.*, "Spiking information processing in a single photonic spiking neuron chip with double integrated electronic dendrites," *Photon. Res.* **11**, 2033–2041 (2023).
[174] S. Gao, S. Xiang, Z. Song, *et al.*, "Hardware implementation of ultra-fast obstacle avoidance based on a single photonic spiking neuron," *Laser Photonics Rev.* **17**, 2300424 (2023).
[175] Y. Shi, S. Xiang, X. Guo *et al.*, "Photonic integrated spiking neuron chip based on a self-pulsating DFB laser with a saturable absorber," *Photonics Res.* **11**, 1382–1389 (2023).
[176] D. Zheng, S. Xiang, Y. Zhang *et al.*, "Photonics neural networks for multimodal recognition based on the self-activated MAC function of DFB - SA," *ACS Photonics*, 2025.
[177] J. Hu, D. Mengu, D. C. Tzarouchis *et al.*, "Diffractive optical computing in free space," *Nat. Commun.* **15**, 1525 (2024).
[178] Z. Xue, T. Zhou, Z. Xu *et al.*, "Fully forward mode training for optical neural networks," *Nature*, **632**, 280–286 (2024).
[179] X. Yuan, Y. Wang, Z. Xu *et al.*, "Training large-scale optoelectronic neural networks with dual-neuron optical-artificial learning," *Nat. Commun.* **14**, p. 7110 (2023).
[180] T. Zhou, X. Lin, J. Wu *et al.*, "Large-scale neuromorphic optoelectronic computing with a reconfigurable diffractive processing unit," *Nat. Photonics* **15**, 367–373 (2021).
[181] J. R. Ong, C. C. Ooi, T. Y. L. Ang *et al.*, "Photonic convolutional neural networks using integrated diffractive optics," *IEEE J. Sel. Top. Quantum Electron.* **26**, 1–8 (2020).
[182] H. H. Zhu, J. Zou, H. Zhang *et al.*, "Space-efficient optical computing with an integrated chip diffractive neural network," *Nat. Commun.* **13**, 1044 (2022).
[183] S. Zarei, A. Khavasi, "Realization of optical logic gates using on-chip diffractive optical neural networks," *Sci. Rep.*, **12**, 15747 (2022).
[184] T. Fu, Y. Zang, Y. Huang *et al.*, "Photonic machine learning with on-chip diffractive optics," *Nat. Commun.* **14**, 70 (2023).
[185] Y. Huang, T. Fu, H. Huang *et al.*, "Sophisticated deep learning with on-chip optical diffractive tensor processing," Photon. Res.**11**, 1125–1138 (2023).
[186] J. Cheng, C. Huang, J. Zhang *et al.*, "Multimodal deep learning using on-chip diffractive optics with in situ training capability," Nat. Commun. vol. 15, no. 1, p. 6189 (2024).

[187] W. Liu, T. Fu, Y. Huang *et al.*, "C-DONN: compact diffractive optical neural network with deep learning regression," *Opt. Express*, vol. 31, no. 13, pp. 22127–22143 (2023).
[188] D. Verstraeten et al., "An experimental unification of reservoir computing methods," Neural Netw. **20**, 391–403 (2007).
[189] K. Vandoorne et al., "Toward optical signal processing using photonic reservoir computing," Opt. Express **16**, 11182–11192 (2008).
[190] K. Vandoorne et al., "Parallel reservoir computing using optical amplifiers," IEEE Trans. Neural Netw. **22**, 1469–1481 (2011).
[191] K. Vandoorne et al., "Experimental demonstration of reservoir computing on a silicon photonics chip," Nat. Commun. **5**, 3541 (2014).
[192] A. Katumba et al., "Low-loss photonic reservoir computing with multimode photonic integrated circuits," Sci. Rep. **8**, 2653 (2018).
[193] D. Brunner and I. Fischer, "Reconfigurable semiconductor laser networks based on diffractive coupling," Opt. Lett. **40**, 3854–3857 (2015).
[194] M. Nakajima et al., "Scalable reservoir computing on coherent linear photonic processor," Commun. Phys. **4**, 20 (2021).
[195] E. Gooskens et al., "Wavelength dimension in waveguide-based photonic reservoir computing," Opt. Express **30**, 15634–15647 (2022).
[196] C. Ma et al., "Integrated photonic reservoir computing with an all-optical readout," Opt. Express **31**, 34843–34854 (2023).
[197] X. Zuo et al., "Integrated silicon photonic reservoir computing with PSO training algorithm for fiber communication channel equalization," J. Lightwave Technol. **41**, 5841–5850 (2023).
[198] D. Wang et al., "Ultrafast silicon photonic reservoir computing engine delivering over 200 TOPS," Nat. Commun. **15**, 10841 (2024).
[199] S. Heinsalu et al., "Silicon loop-type multimode waveguide structure with fan-out output for photonic reservoir computing," J. Lightwave Technol. **42**, 7321-7329, (2024).
[200] S. Biasi et al., "An array of micro resonators as a photonic extreme learning machine," APL Photonics **8**, 9 (2023).
[201] Z. Li et al., "Packet header recognition utilizing an all-optical reservoir based on reinforcement-learning-optimized double-ring resonators," IEEE J. Sel. Top. Quantum Electron. **29**, 1–8 (2023).
[202] C. Gao et al., "Reservoir computing using arrayed waveguide grating," in 2023 Opto-Electronics and Communications Conference (OECC) (2023).
[203] Y. Nie et al., "Integrated laser graded neuron enabling high-speed reservoir computing without a feedback loop," Optica **11**, 1690–1699 (2024).
[204] G. Donati et al., "Time delay reservoir computing with a silicon microring resonator and a fiber-based optical feedback loop," Opt. Express **32**, 13419–13437 (2024).
[205] D. Brunner et al., "Parallel photonic information processing at gigabyte per second data rates using transient states," Nat. Commun. **4**, 1364 (2013).
[206] J. Nakayama et al., "Laser dynamical reservoir computing with consistency: an approach of a chaos mask signal," Opt. Express **24**, 8679–8692 (2016).
[207] A. Argyris et al., "Fast physical repetitive patterns generation for masking in time-delay reservoir computing," Sci. Rep. **11**, 6701 (2021).
[208] X. Guo et al., "Photonic implementation of the input and reservoir layers for a reservoir computing system based on a single VCSEL with two Mach-Zehnder modulators: erratum," Opt. Express **33**, 1197–1197 (2025).
[209] R. M. Nguimdo et al., "Simultaneous computation of two independent tasks using reservoir computing based on a single photonic nonlinear node with optical feedback," IEEE Trans. Neural Netw. Learn. Syst. **26**, 3301–3307 (2015).
[210] J. Bueno et al., "Conditions for reservoir computing performance using semiconductor lasers with delayed optical feedback," Opt. Express **25**, 2401–2412 (2017).
[211] J. Vatin et al., "Enhanced performance of a reservoir computer using polarization dynamics in VCSELs," Opt. Lett. **43**, 4497–4500 (2018).
[212] G. Q. Xia et al., "Prediction performance of reservoir computing using a semiconductor laser with double optical feedback," in 2018 Conference on Lasers and Electro-Optics Pacific Rim (CLEO-PR).
[213] X. X. Guo et al., "Polarization multiplexing reservoir computing based on a VCSEL with polarized optical feedback," IEEE J. Sel. Top. Quantum Electron. **25**, 1–9 (2019).
[214] X. X. Guo et al., "Four-channels reservoir computing based on polarization dynamics in mutually coupled VCSELs system," Opt. Express **27**, 23293–23306 (2019).
[215] X. X. Guo et al., "Enhanced prediction performance of a neuromorphic reservoir computing system using a semiconductor nanolaser with double phase conjugate feedbacks," J. Lightwave Technol. **39**, 129–135 (2021).
[216] J. Y. Tang et al., "Parallel time-delay reservoir computing with quantum dot lasers," IEEE J. Quantum Electron. **58**, 1–9 (2022).
[217] Y. W. Shen et al., "Deep photonic reservoir computing recurrent network," Optica **10**, 1745–1751 (2023).
[218] X. Guo et al., "Photonic reservoir computing system for pattern recognition based on an array of four distributed feedback lasers," ACS Photonics **11**, 1327–1334 (2024).
[219] X. Guo et al., "Experimental demonstration of a photonic reservoir computing system based on Fabry Perot laser for multiple tasks processing," Nanophotonics **13**, 1569–1580 (2024).
[220] R. Zhang et al., "Deep tree photonic reservoir computing with parallel architecture," in 2024 Asia Communications and Photonics Conference (ACP) and International Conference on Information Photonics and Optical Communications (IPOC) (IEEE, 2024).
[221] L. Y. Zhang et al., "Heterogeneous forecasting of chaotic dynamics in vertical-cavity surface-emitting lasers with knowledge-based photonic reservoir computing," Photon. Res. **13**, 728–736 (2025).
[222] C. D. Zhou et al., "Photonic deep residual time-delay reservoir computing," Neural Netw. **178**, 106575 (2024).
[223] C. D. Zhou et al., "Streamlined photonic reservoir computer with augmented memory capabilities," Opto-Electron. Adv. **8**, 240135 (2025).
[224] Z. W. Dai et al., "Photonic spiking reservoir computing system based on a DFB-SA laser for pattern recognition," ACS Photonics **12**, 989-996 (2025).
[225] J. Jin et al., "Adaptive time-delayed photonic reservoir computing based on Kalman-filter training," Opt. Express **30**, 13647–13658 (2022).
[226] G. O. Danilenko et al., "Impact of filtering on photonic time-delay reservoir computing," Chaos **33**, 013116 (2023).
[227] Y. Tanaka and H. Tamukoh, "Self-organizing multiple readouts for reservoir computing," IEEE Access **11**, 138839–138849 (2023).
[228] S. Sackesyn et al., "Experimental realization of integrated photonic reservoir computing for nonlinear fiber distortion compensation," Opt. Express **29**, 30991–30997 (2021).
[229] X. Guo et al., "Short-term prediction for chaotic time series based on photonic reservoir computing using VCSEL with a feedback loop," Photon. Res. **12**, 1222–1230 (2024).
[230] T. Wang et al., "Reservoir computing-based advance warning of extreme events," Chaos Solitons Fractals **178**, 114673 (2024).
[231] S. R. Ahmed, R. Baghdadi, M. Bernadskiy et al. "Universal photonic artificial intelligence acceleration," Nature **640**, 368–374 (2025).
[232] S. Hua, E. Divita, S. Yu et al. "An integrated large-scale photonic accelerator with ultralow latency," Nature **640**, 361–367 (2025).
[233] Z. Chen et al., "Deep learning with coherent VCSEL neural networks," *Nat. Photon.* **17**, 723–730 (2023).
[234] S. Afifi et al., "TRON: transformer neural network acceleration with non-coherent silicon Photonics," in *Proceedings of the Great Lakes Symposium on VLSI 2023 (GLSVLSI '23)*, 15–21 (2023).
[235] H. Zhu et al., "Lightening-transformer: a dynamically-operated optically-interconnected photonic transformer accelerator," in *2024 IEEE International Symposium on High-Performance Computer Architecture (HPCA)*, 686–703 (2024).
[236] T.-C. Hsueh et al., "Optical comb-based monolithic photonic-electronic accelerators for self-attention computation," *IEEE J. Sel. Top. Quantum Electron.* **30**, 3500417 (2024).
[237] H. Sha et al., "Vision transformer with photonic integrated circuits," in *Conference 13236*, Paper 13236-23 (2024).
[238] Y. Ma et al., "Time-delay signature concealment of chaos and ultrafast decision making in mutually coupled semiconductor lasers with a phase-modulated Sagnac loop," *Opt. Express* **28**, 1665–1678 (2020).
[239] M. Naruse et al., "Decision making photonics: solving bandit problems using photons," *IEEE J. Sel. Top. Quantum Electron.* **26**, 7700210 (2020).

[240] B. Shen et al., "Harnessing microcomb-based parallel chaos for random number generation and optical decision making," *Nat. Commun.* **14**, 4590 (2023).
[241] K. Morijiri et al., "Parallel photonic accelerator for decision making using optical spatiotemporal chaos," *Optica* **10**, 339–348 (2023).
[242] M. Ishikawa et al., "Experimental studies on learning capabilities of optical associative memory," *Appl. Opt.* **29**, 289–295 (1990).
[243] S. Wang et al., "Photonic associative learning neural network based on VCSELs and STDP," *J. Lightwave Technol.* **38**, 4691–4698 (2020).
[244] J. Y. S. Tan, "All-optical associative learning element," PhD Thesis, University of Oxford (2021).
[245] J. Y. S. Tan et al., "Monadic pavlovian associative learning in a backpropagation-free photonic network," *Optica* **9**, 792–802 (2022).
[246] J. Ouyang et al., "16-channel photonic solver for optimization problems on a silicon chip," *Chip* **4**, 100117 (2025).
[247] M. Naruse et al., "Ultrafast photonic reinforcement learning based on laser chaos," *Sci. Rep.* **7**, 8772 (2017).
[248] Y. Tang et al., "Device-system end-to-end design of photonic neuromorphic processor using reinforcement learning," *Laser Photonics Rev.* **17**, 2200223 (2023).
[249] Z. Yang et al., "Deep reinforcement learning based on optical neural networks in path planning," in *2023 Asia Communications and Photonics Conference/2023 International Photonics and Optoelectronics Meetings (ACP/POEM)*, 1–4 (2023).
[250] Z. Yang et al., "Tunable-bias based optical neural network for reinforcement learning in path planning," *Opt. Express* **32**, 18099–18112 (2024).
[251] X. Li et al., "High-efficiency reinforcement learning with hybrid architecture photonic integrated circuit," *Nat. Commun.* **15**, 1044 (2024).
[252] B. Wu, W. Zhang, S. Zhang et al., "A monolithically integrated optical Ising machine," Nat. Commun. **16**, 4296 (2025).
[253] B. Wu, H. Zhou, J. Cheng et al., "Monolithically integrated asynchronous optical recurrent accelerator," eLight **5**, 7 (2025).
[254] J. Gu et al., in "ROQ: A noise-aware quantization scheme towards robust optical neural networks with low-bit controls," 2020 (IEEE), p. 1586-1589.
[255] M. Kirtas et al., "Quantization-aware training for low precision photonic neural networks," Neural Netw. **155**, 561-573 (2022).
[256] M. Kirtas et al., in "Normalized post-training quantization for photonic neural networks," 2022 (IEEE), p. 657-663.
[257] R. Shao, G. Zhang, and X. Gong, "Generalized robust training scheme using genetic algorithm for optical neural networks with imprecise components," Photonics Res. **10**, 1868 (2022).
[258] J. Spall, X. Guo, and A. I. Lvovsky, "Hybrid training of optical neural networks," Optica **9**, 803-811 (2022).
[259] Y. Zhan et al., "Physics-aware analytic-gradient training of photonic neural networks," Laser Photon. Rev. **18** (2024).
[260] Y. Wang et al., "Asymmetrical estimator for training encapsulated deep photonic neural networks," Nat. Commun. **16** (2025).
[261] T. W. Hughes, M. Minkov, Y. Shi, and S. Fan, "Training of photonic neural networks through in situ backpropagation and gradient measurement," Optica **5**, 864 (2018).
[262] T. Zhou et al., "In situ optical backpropagation training of diffractive optical neural networks," Photonics Res. **8**, 940-953 (2020).
[263] H. Zhou et al., "Self-configuring and reconfigurable silicon photonic signal processor," ACS Photonics **7**, 792-799 (2020).
[264] H. Zhang et al., "Efficient on-chip training of optical neural networks using genetic algorithm," ACS Photonics **8**, 1662-1672 (2021).
[265] M. J. Filipovich et al., "Silicon photonic architecture for training deep neural networks with direct feedback alignment," Optica **9**, 1323-1332 (2022).
[266] W. Zhang et al., in "Online training and pruning of photonic neural networks," 2023 (IEEE), p. 1-2.
[267] S. Pai et al., "Experimentally realized in situ backpropagation for deep learning in photonic neural networks," Science **380**, 398-404 (2023).
[268] Y. Wan et al., "Efficient stochastic parallel gradient descent training for on-chip optical processor," Opto-Electron. Adv. **7**, 230182 (2024).
[269] J. Spall, X. Guo, and A. I. Lvovsky, "Training neural networks with end-to-end optical backpropagation," Adv. Photonics **7**, 16004 (2025).
[270] H. Zhou et al., "Photonic matrix multiplication lights up photonic accelerator and beyond," Light Sci. Appl. Light: Sci. Appl. **11**, 30 (2022).
[271] C. Huang et al., "Prospects and applications of photonic neural networks," Adv. Phys.: X **7**, 1981155 (2022).
[272] K. Liao et al., "Integrated photonic neural networks: opportunities and challenges," ACS Photonics **10**, 2001-2010 (2023).
[273] N. Farmakidis et al., "Integrated photonic neuromorphic computing: opportunities and challenges," Nat. Rev. Electr. Eng. **1**, 358-373 (2024).
[274] R. Li et al., "Photonics for neuromorphic computing: fundamentals, devices, and opportunities," Adv. Mater. **36**, 2312825 (2024).
[275] X. Guo et al., "Integrated neuromorphic photonics: synapses, neurons, and neural networks," Adv. Photonics Res. 2, 2000212 (2021).
[276] P.L. McMahon, "The physics of optical computing," Nat. Rev. Phys. **3**, 717-734 (2021).
[277] Y. Bai et al., "Photonic multiplexing techniques for neuromorphic computing," Nanophotonics **12**, 795-817 (2023).
[278] B.J. Shastri et al., "Photonics for artificial intelligence and neuromorphic computing," Nat. Photonics **15**, 102-114 (2021).
[279] C.D. Schuman et al., "Opportunities for neuromorphic computing algorithms and applications," Nat. Comput. Sci. **2**, 10-19 (2022).
[280] D. Brunner et al., "Roadmap on neuromorphic photonics," J. Phys.: Photonics **3**, 031001 (2021).